\documentclass[twocolumn]{aastex62}

\DeclareMathAlphabet{\mathitbf}{OML}{cmm}{b}{it}
\newcommand{\Jvec}{\mathitbf{J}}
\newcommand{\Bvec}{\mathitbf{B}}

\usepackage{color}
\usepackage{graphicx}

\shorttitle{Observations of a Footpoint Drift of an Erupting Flux Rope} \shortauthors{Zemanov\'a et al.}
\begin{document}

\title{Observations of a Footpoint Drift of an Erupting Flux Rope}

\author[0000-0002-7565-5437]{Alena Zemanov\'a}
\affil{Astronomical Institute of the Academy of Sciences of the
Czech Republic, CZ-25135 Ond\v{r}ejov, Czech Republic}
 \email{alena.zemanova@asu.cas.cz}
 
\author[0000-0003-1308-7427]{Jaroslav Dud\'{\i}k}
\affil{Astronomical Institute of the Academy of Sciences of the
Czech Republic, CZ-25135 Ond\v{r}ejov, Czech Republic}

\author[0000-0001-5810-1566]{Guillaume Aulanier}
\affiliation{LESIA, Observatoire de Paris, Psl Research University, CNRS, Sorbonne Universit\'es, UPMC Univ. Paris 06, Univ. Paris Diderot, Sorbonne Paris Cité, 5 place Jules Janssen, F-92195 Meudon, France}

\author[0000-0001-8985-2549]{Julia K. Thalmann}
\affil{University of Graz, IGAM/Institute of Physics, A-8010 Graz, Austria}

\author[0000-0002-0473-4103]{Peter G\"{o}m\"{o}ry}
\affil{Astronomical Institute, Slovak Academy of Sciences, 05960 Tatransk\'a Lomnica, Slovakia}

\begin{abstract}
We analyze the imaging observations of an M-class eruptive flare of 2015 November, 4. The pre-eruptive H$\alpha$ filament was modelled by the non-linear force free field model, which showed that it consisted of two helical systems. Tether-cutting reconnection involving these two systems led to the formation of a hot sigmoidal loop structure rooted in a small hook that formed at the end of the flare ribbon. Subsequently, the hot loops started to slip away form the small hook until it disappeared. The loops continued slipping and the ribbon elongated itself by several tens of arc seconds. A new and larger hook then appeared at the end of elongated ribbon with hot and twisted loops rooted there. After the eruption of these hot loops, the ribbon hook expanded and later contracted. 
We interpret these observations in the framework of the recent three dimensional (3D) extensions to the standard solar flare model, which predict the drift of the flux rope footpoints. The hot sigmoidal loop is interpreted as the flux rope, whose footpoints drift during the eruption. While the deformation and drift of the new hook can be described by the model, the displacement of the flux rope footpoint from the filament to that of the erupting flux rope indicate that the hook evolution can be more complex than those captured by the model.

\end{abstract}


\section{INTRODUCTION}
Eruptive flares are one of the most geo-effective manifestations of solar activity. Their observed characteristics are widely explained using the standard two dimensional (2D) model of eruptive flares. This so called CSHKP model \citep{Carmichael1964, Sturrock1966, Hirayama1974, Kopp1976} is a phenomenological model. According to it a flare occurs beneath rising and erupting prominence/filament \cite{Carmichael1964} used the concept of the 2D reconnection process suggested by \cite{Petschek1964} to explain the flare with two parallel chromospheric ribbons situated on each side of magnetic polarity inversion line (PIL). In accordance with imaging observations available at that time it was suggested that a vertical current sheet with a X-type magnetic null point is located above the flare loop system rooted in these bright ribbons. The reconnection of the stretched (open) field lines with anti-parallel magnetic orientation below the rising filament then produces the flare loops on one side \citep{Kopp1976} and twisted field lines contributing to the flux rope on another side \citep{Shibata1995}. As the flare proceeds, the observed H$\alpha$ ribbons move apart because the X-point is moving upward \citep{Hirayama1974} creating new, hot and higher flare loops. 

During the era of modern soft X-ray and EUV/UV imaging observations this model proved to be consistent with some 
observed phenomena of eruptive flares but it does not explain the evolution of the flux rope. Neither its formation and destabilization is accounted for, nor the location of flux rope (filament/prominence) footpoints and linkage between the shape and overall evolution of flare ribbons in connection to flare loops and expelled coronal mass ejection (CME) are considered. The progress in flare observations, and more than 20 years of computational modelling of twisted  magnetic structures in 3D, turned the CSHKP model into a snapshot of the eruptive flare in preferred plane of symmetry within a possibly complex 3D magnetic structure. 

\begin{figure}
         \includegraphics[width=8.5cm,clip]{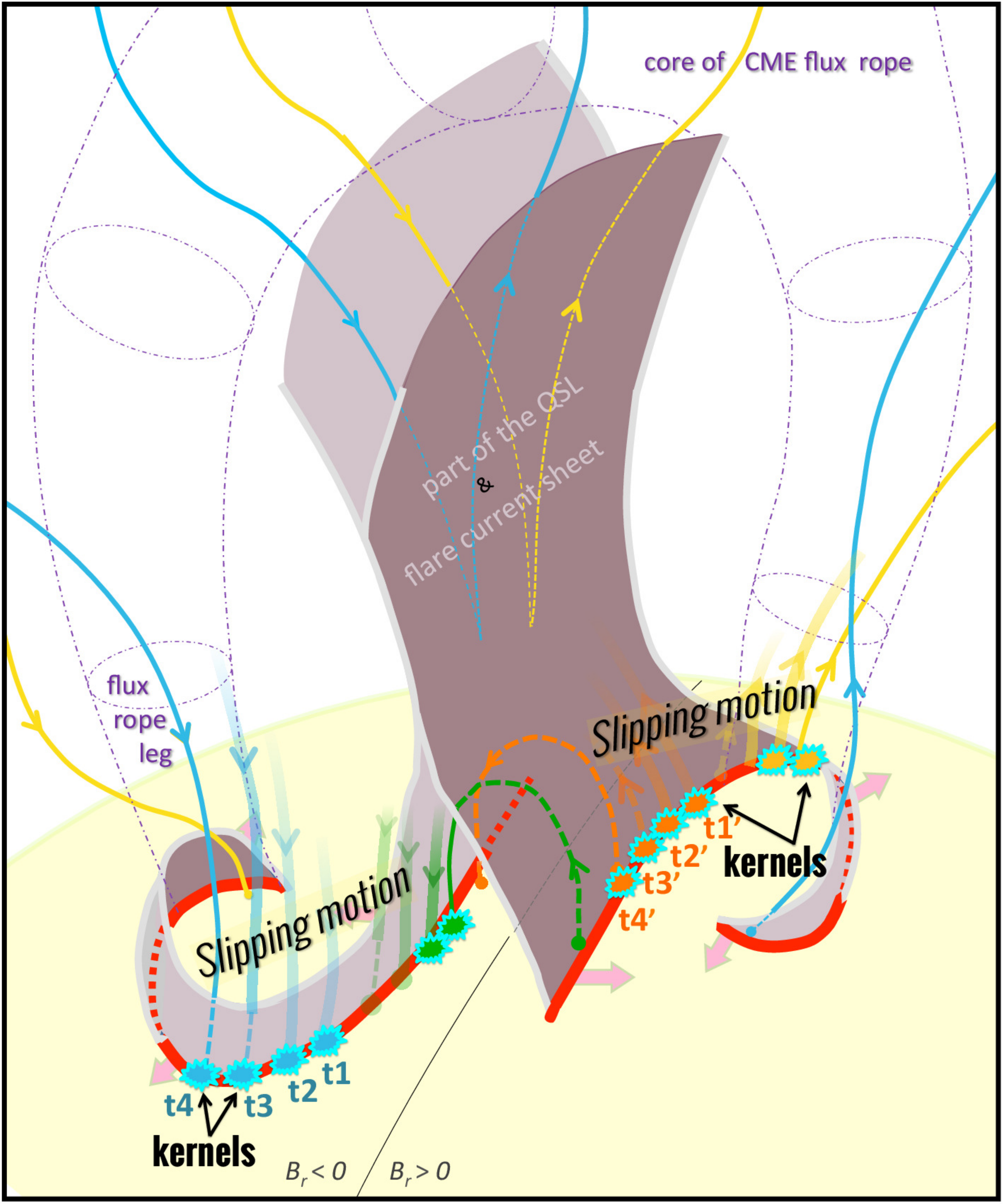}
          \caption{A cartoon showing the standard model of solar (eruptive) flare in 3D. Red lines mark J-shaped (flare/current) ribbons which are footprints of QSLs (grey shaded regions). Only the parts of QSLs' volume containing the strongest currents are depicted. The dash-dotted grey lines represent the core of the flux rope, legs of which are anchored within the hooked parts of the J-shaped ribbons. The individual field lines are shown in colours. Flare loops, in green and orange, are shown at the straight part of J-shaped ribbons. Blue and yellow field lines contribute to flux rope envelope and are anchored in ribbon hooks. Slipping motion of reconnecting field lines is depicted by kernels located at flare ribbons with the sense of motion denoted by t1--t4. Thin black line marks PIL. (From \cite{Dudik2016}, reproduced by permission of the AAS.)} \label{fig_3Dmodel}
\end{figure}

The so called standard flare model in 3D has been introduced by \cite{Aulanier2012_I, Aulanier2013_II} and \cite{Janvier2013_III, Janvier2014}. This set of papers attempted to introduce a compact overview of phenomena usually observed in flares based on numerical MHD model in 3D. The standard model of an eruptive flare in 3D now includes twisted structure, a flux rope, which is an essential ingredient of this kind of a flare \citep{Janvier2015}. A simplified cartoon denoting some of the main features of the model is shown in Figure~\ref{fig_3Dmodel}. 
         
\cite{Demoulin1996} studied the continuous magnetic field line linkage in twisted magnetic field configurations. They showed that when field line linkage is continuous, specific locations within a 3D domain may exhibit strong gradients in the associated field line mapping. The physical properties at these locations may further be similar to that in separatrices. These locations form 3D layers, so called Quasi-separatrix layers (QSLs, grey shaded regions in Figure~\ref{fig_3Dmodel}), and their cross-section with the bottom boundary of computational domain (i.e. QSL traces  or footprints) create a pair of J-shaped ribbons (red lines in Figure~\ref{fig_3Dmodel}). \cite{Aulanier2010, Aulanier2012_I} performed an analysis of 3D MHD simulation of an asymmetric solar eruption and identified the spatial distribution of electric currents. Thin and narrow current layers identified in the model form a J-shape pattern at the cross-section with the photosphere (bottom boundary of the simulation domain) representing the flare ribbons. The straight part of the J-ribbon, parallel to the PIL (black line in Figure~\ref{fig_3Dmodel}), possesses direct currents (j$_z$/B$_z>0$) only and corresponds to the footpoints of cusp loops (green and orange loops in Figure~\ref{fig_3Dmodel}) located below the vertical current sheet developed in a wake of the CME. The hooked part of the J-ribbon surrounds the legs of the expanding flux rope/CME (grey dash-dotted lines in Figure~\ref{fig_3Dmodel}) and involves both direct (j$_z$/B$_z>0$) and return currents (j$_z$/B$_z<0$). \cite{Janvier2014} quantified the evolution of photospheric currents during an X-class flare and confirmed the predictions of the 3D numeric MHD simulations of \cite{Aulanier2010, Aulanier2012_I}.
These predictions have been supported also by detailed comparison of {\it SDO}/AIA and {\it Hinode}/XRT (imaging) observations to nonlinear force-free field (NLFFF) modelling \citep{Savcheva2015, Savcheva2016, Zhao2016}. This 3D MHD model is also capable to explain the evolution of strong-to-weak transition in the flare loops.  

In the standard solar flare model in 3D, the magnetic reconnection is also a 3D phenomenon. \cite{Priest2003} and \cite{Priest2016} showed that 3D reconnection can be crucially different from that in 2D. The existence of QSLs enables magnetic reconnection to occur without null points \citep{Priest1995, Priest2003}. 

The magnetic reconnection in QSLs occurs via flipping of magnetic field lines when they pass such a narrow 
current layer \citep{Priest1995}. \cite{Aulanier2006} showed that reconnecting field lines gradually slip along one another. The slipping does not correspond to real bulk motion but to rearrangements of magnetic field lines due to reconnection. The slipping footpoints (kernels t1--t4 and t1'--t4' in Figure~\ref{fig_3Dmodel}) move along the arc-shaped trajectories on both sides of the inversion line which correspond to the intersections of QSLs with the line-tied boundary. Apparent slipping motion of loops has been observed by several authors, for coronal loops in a non-flaring active region \citep{Aulanier2007,Testa2013}, and also for flare loops \citep{Dudik2014, Li2015, Dudik2016}, where slipping velocities on the order of 10 $-$ 100 km\,s$^{-1}$ are common. \cite{Aulanier2007} and \cite{Dudik2016} have also found evidence of apparent slipping motions of the loops in both directions.

\cite{Janvier2013_III, Savcheva2016} reported that QSLs and their footprints connected to flux rope evolve in time during the flare. The temporal evolution of the flux rope footpoints is related to coronal dimmings \citep[see e.g. review by][]{Webb2000}. These areas of diminished intensity appear usually during the eruption and evolve in time \citep{Dissauer2018a}. They are often divided to core and secondary dimmings \citep{Mandrini2007, Dissauer2018a, Dissauer2018b, Veronig2019}. Core dimmings are areas of strongly reduced EUV emission, presented close to the eruption site and localized in opposite magnetic polarity regions. Secondary dimmings are widespread and develop away from the eruption site e.g. due to expansion and interaction of the erupting field with neighbouring magnetic regions \citep{Mandrini2007}.

\begin{figure*}
\centering
         \includegraphics[width=8.3cm,clip]{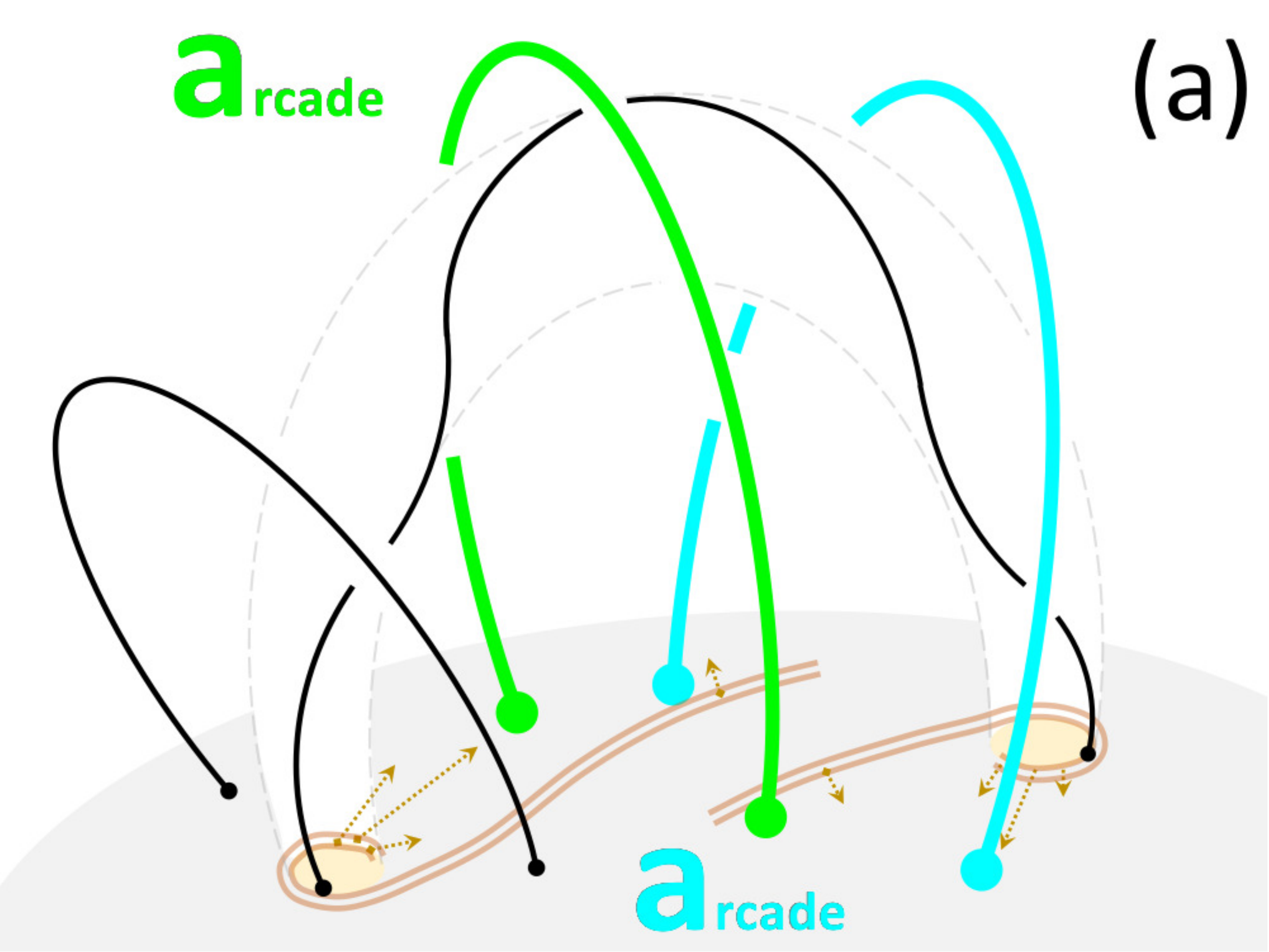}
         \setlength{\columnsep}{0.3cm}
         \includegraphics[width=8.3cm,clip,]{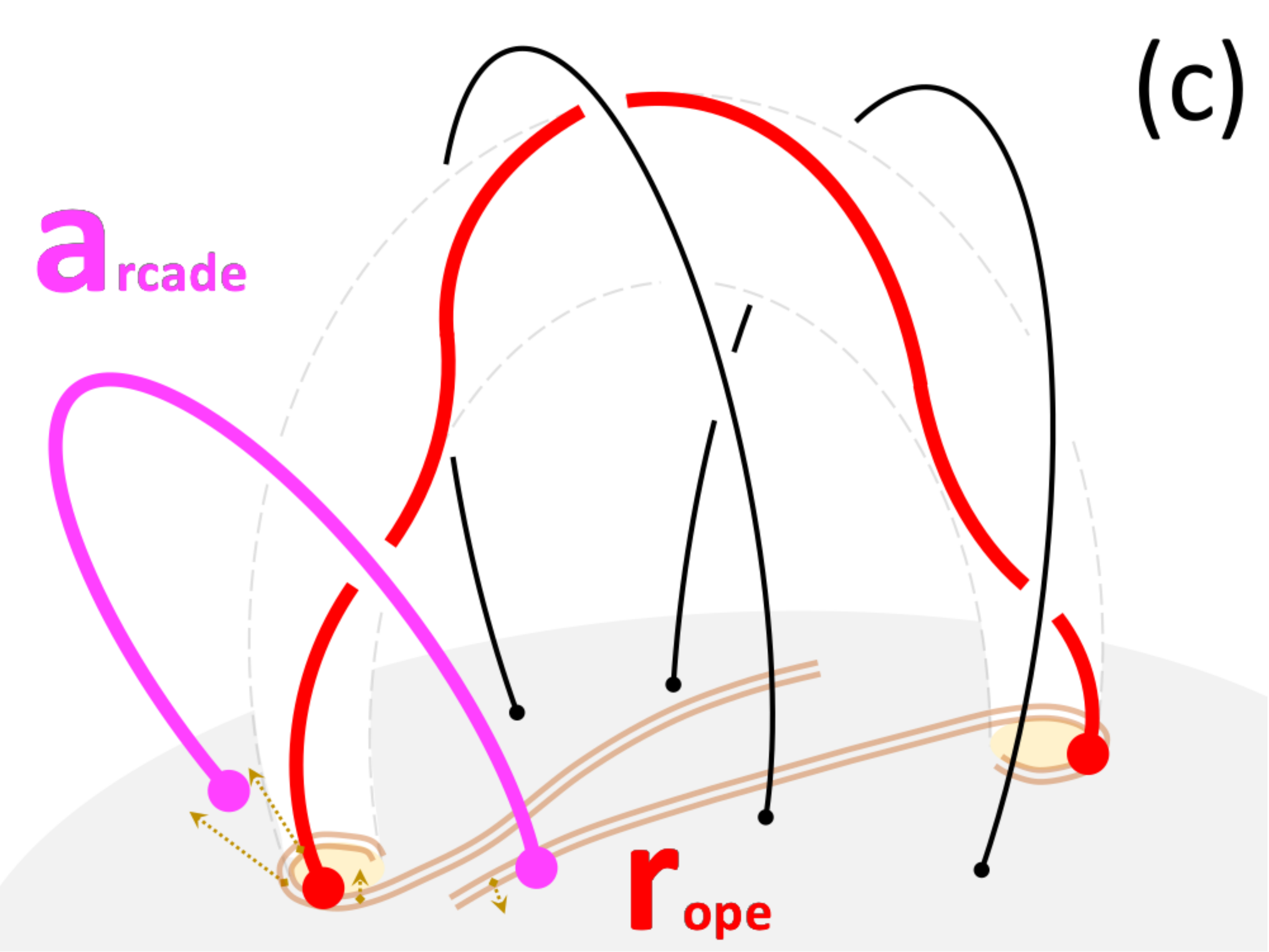}
         
         \includegraphics[width=8.3cm,clip]{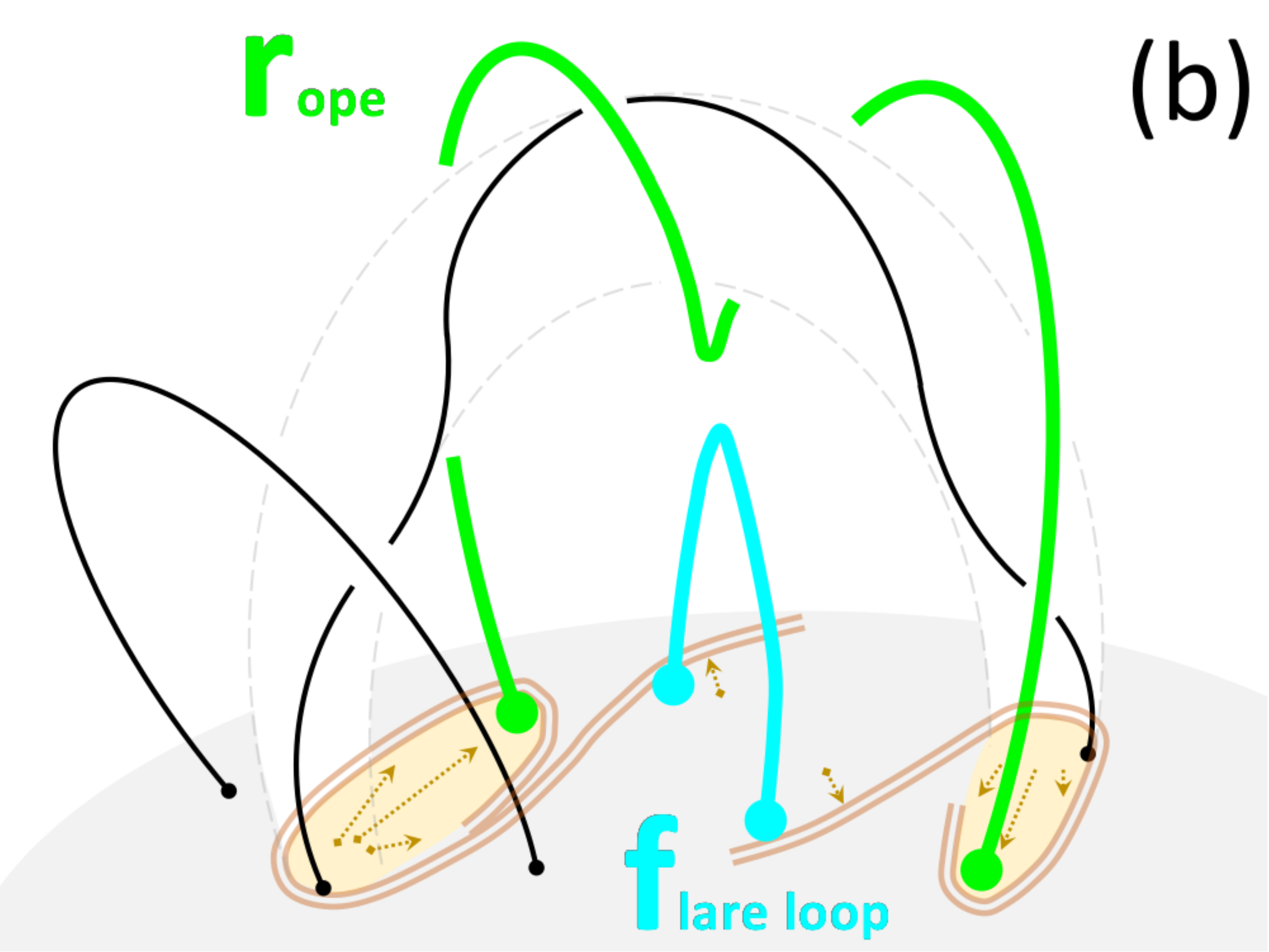}
         \setlength{\columnsep}{0.3cm}
         \includegraphics[width=8.3cm,clip,]{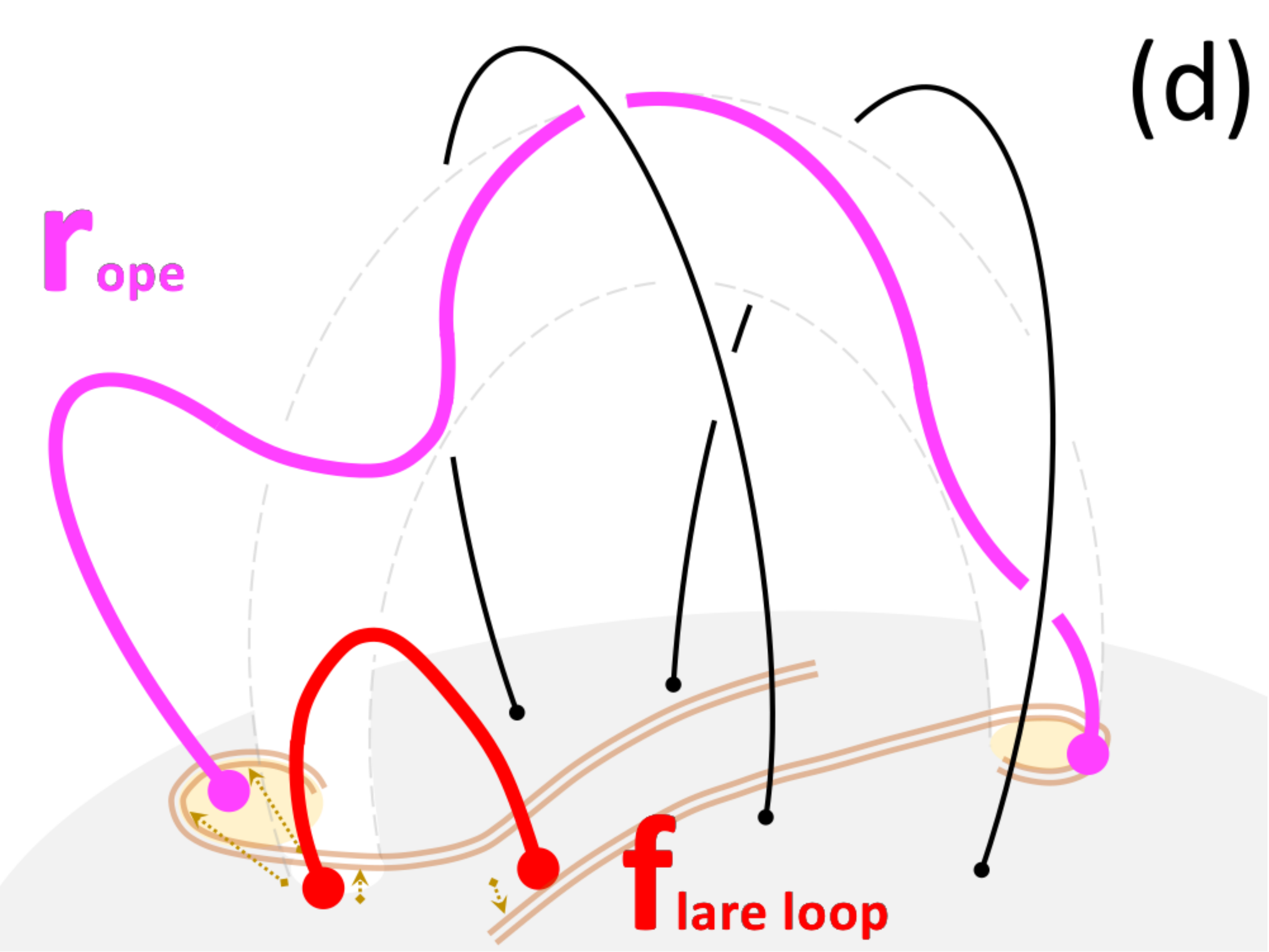}
         \caption{Cartoons showing the aa--rf (a--b) and ar--rf (c--d) reconnection geometries and the associated ribbon/QSLs deformation and drift. Grey dashed lines show the envelope of a flux rope at the onset of the eruption. J-shaped ribbons are plotted in brown together with arrows showing their evolution. Yellow areas denote footpoints of flux rope field lines. (a) Two arcade field lines `a' (green and cyan) reconnect to produce (b) a flux rope field line `r' (green) and a flare loop `f' (cyan). The straight part of J-shaped ribbons move away {\bf from} PIL while hooked parts of the J's expand as a result. (c) An inclined arcade `a' (pink) and a flux rope field line `r' (red) reconnect to produce (d) a new flux rope field line `r' (pink) and a flare loop `f' (red). The hook of the J-shaped ribbon drifts away from the position {\bf it occupied} at the onset of the eruption (light grey dashed lines).} \label{fig_cartoon}
\end{figure*}

\begin{figure*}
\centering
        \includegraphics[width=8.4cm,clip,viewport=28 8 493 350]{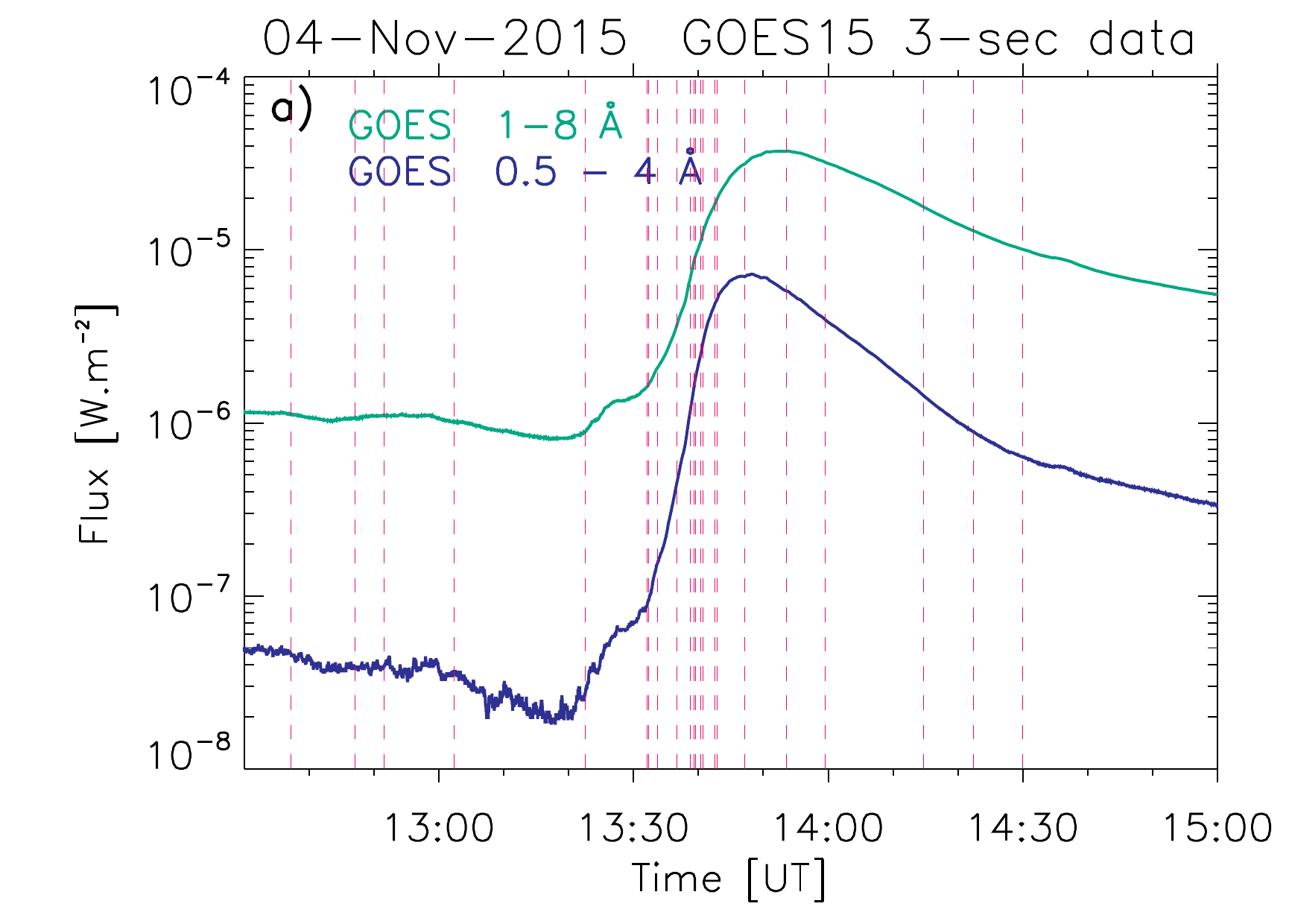}
        \includegraphics[width=8.6cm,clip,viewport=2 3 490 340]{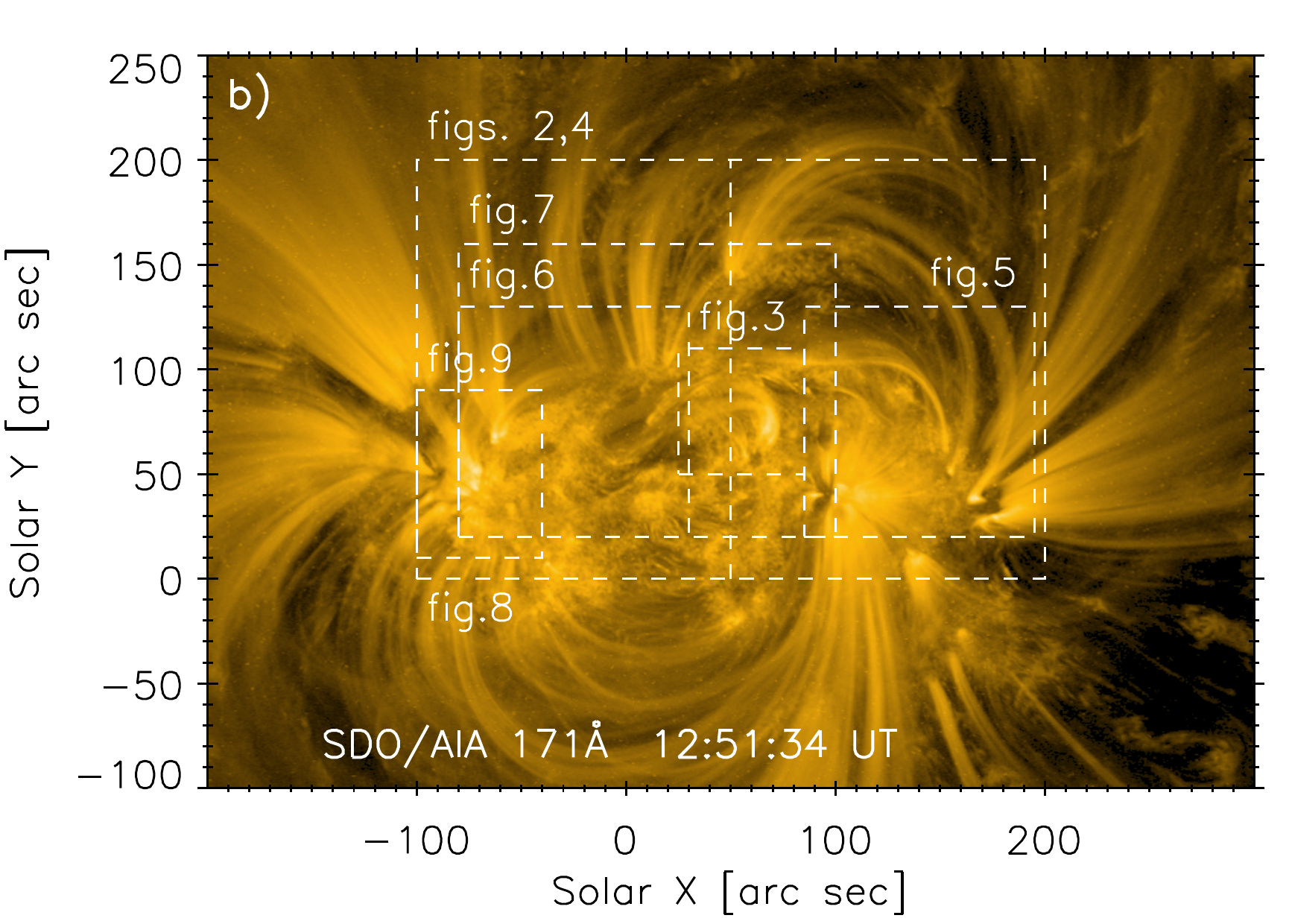}
         \caption{(a) Time evolution of the {\it GOES-15} soft X-ray flux during the M3.7 flare with red vertical dashed lines marking
        the times of {\it SDO}/AIA images used throughout the paper.
        (b) An overview image of NOAA AR 12 443 in {\it SDO}/AIA 171\,\AA~filter about 40 minutes before the flare. Dashed rectangles 
        show FOVs of subsequent individual figures.} \label{fig_goes}
\end{figure*}

\begin{figure*}[ht]
        \centering  
        \includegraphics[width=9.5cm,clip,viewport=8 50 490 340]{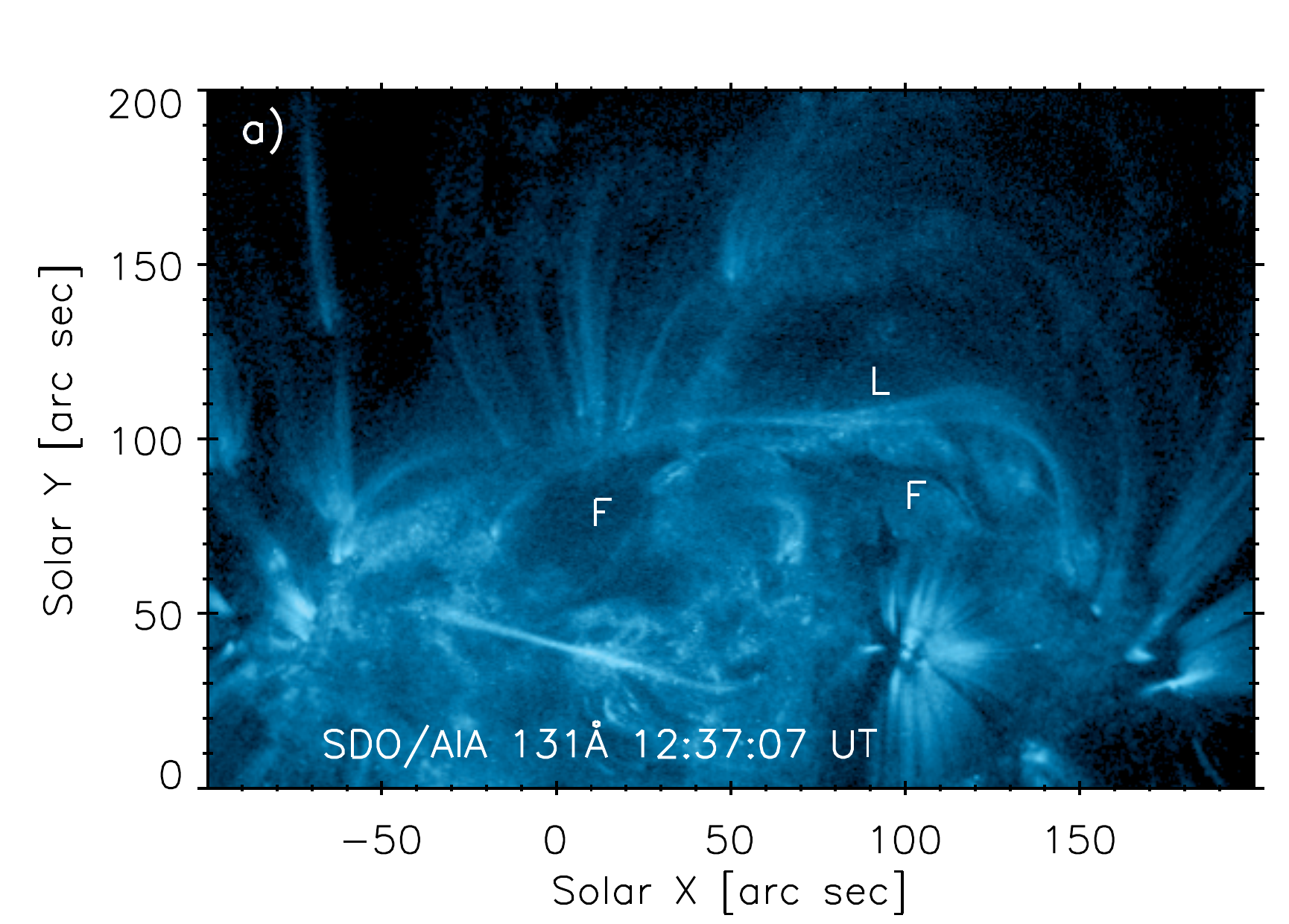}
        \includegraphics[width=8.13cm,clip,viewport=78 50 490 340]{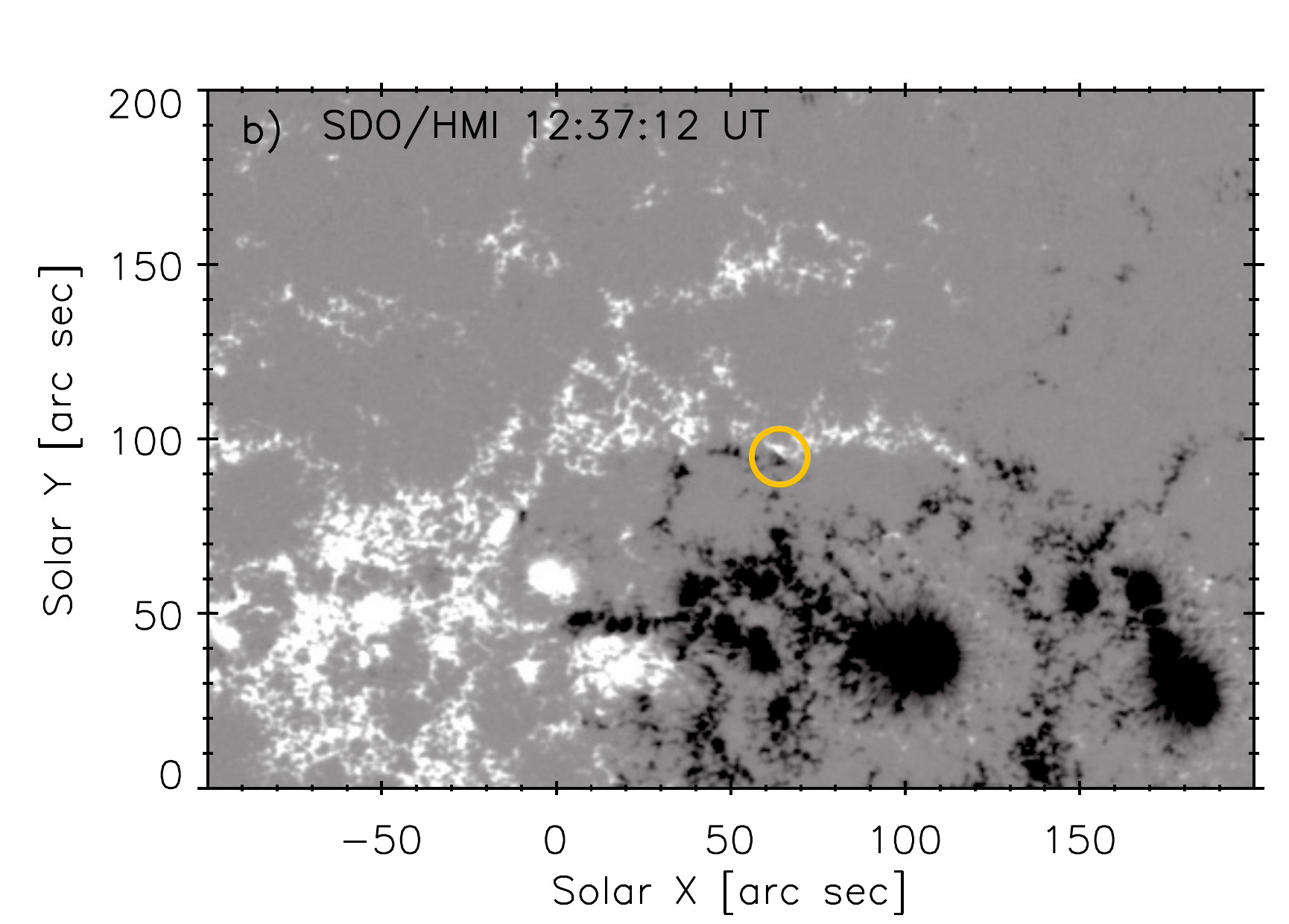}
        \includegraphics[width=9.5cm,clip,viewport=8 50 490 340]{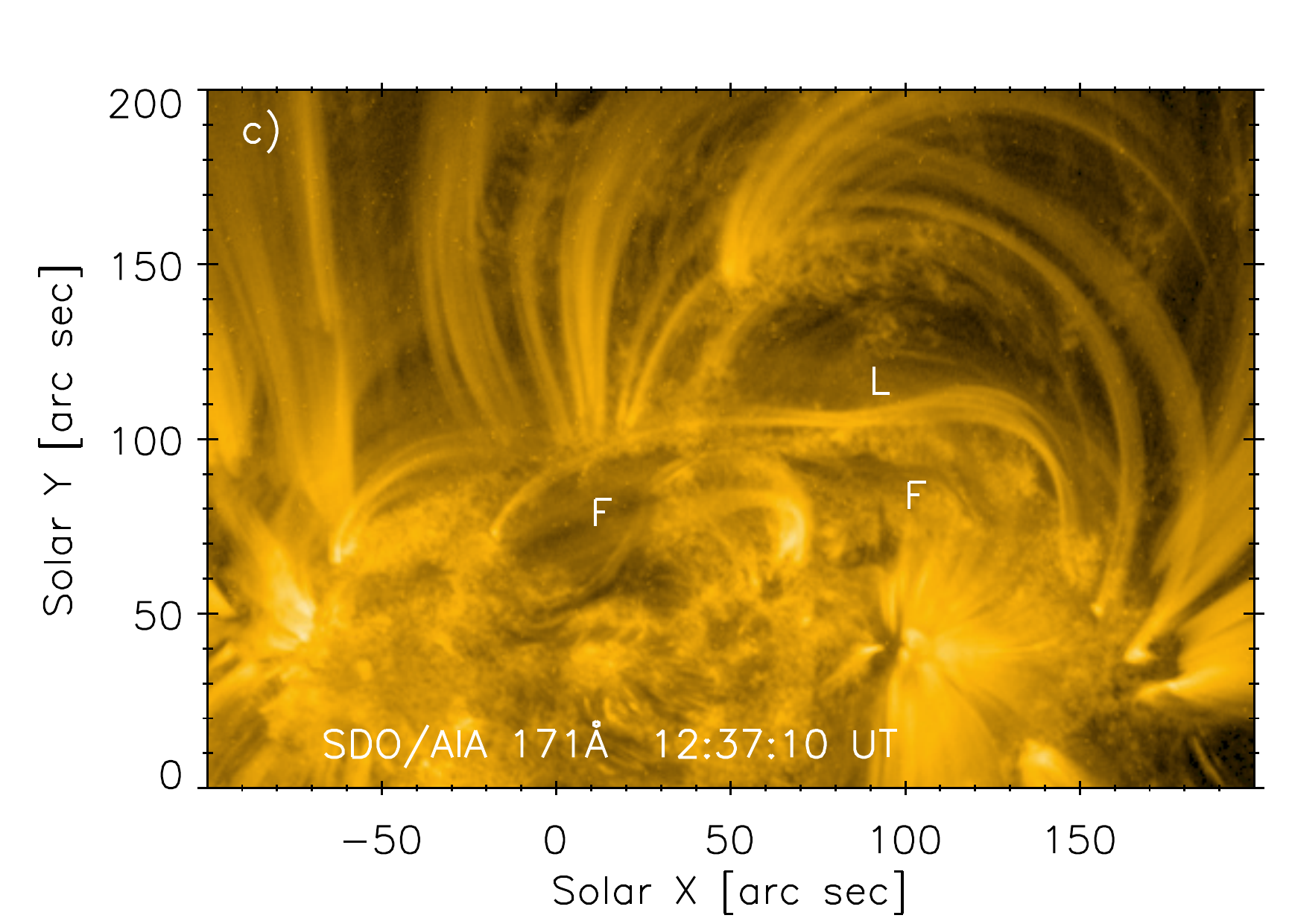}
        \includegraphics[width=8.13cm,clip,viewport=78 50 490 340]{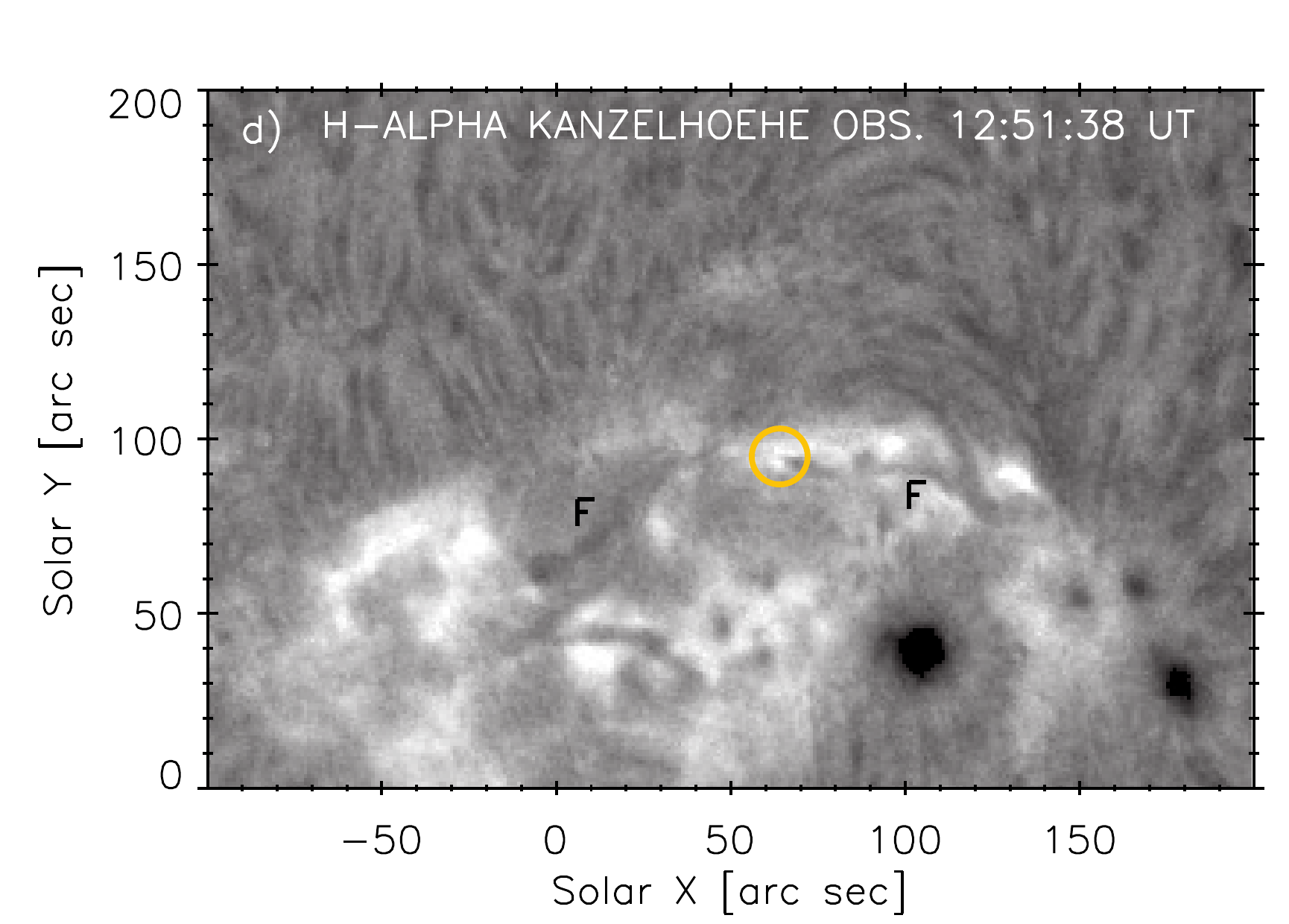}
        \includegraphics[width=9.5cm,clip,viewport=8 0 490 340]{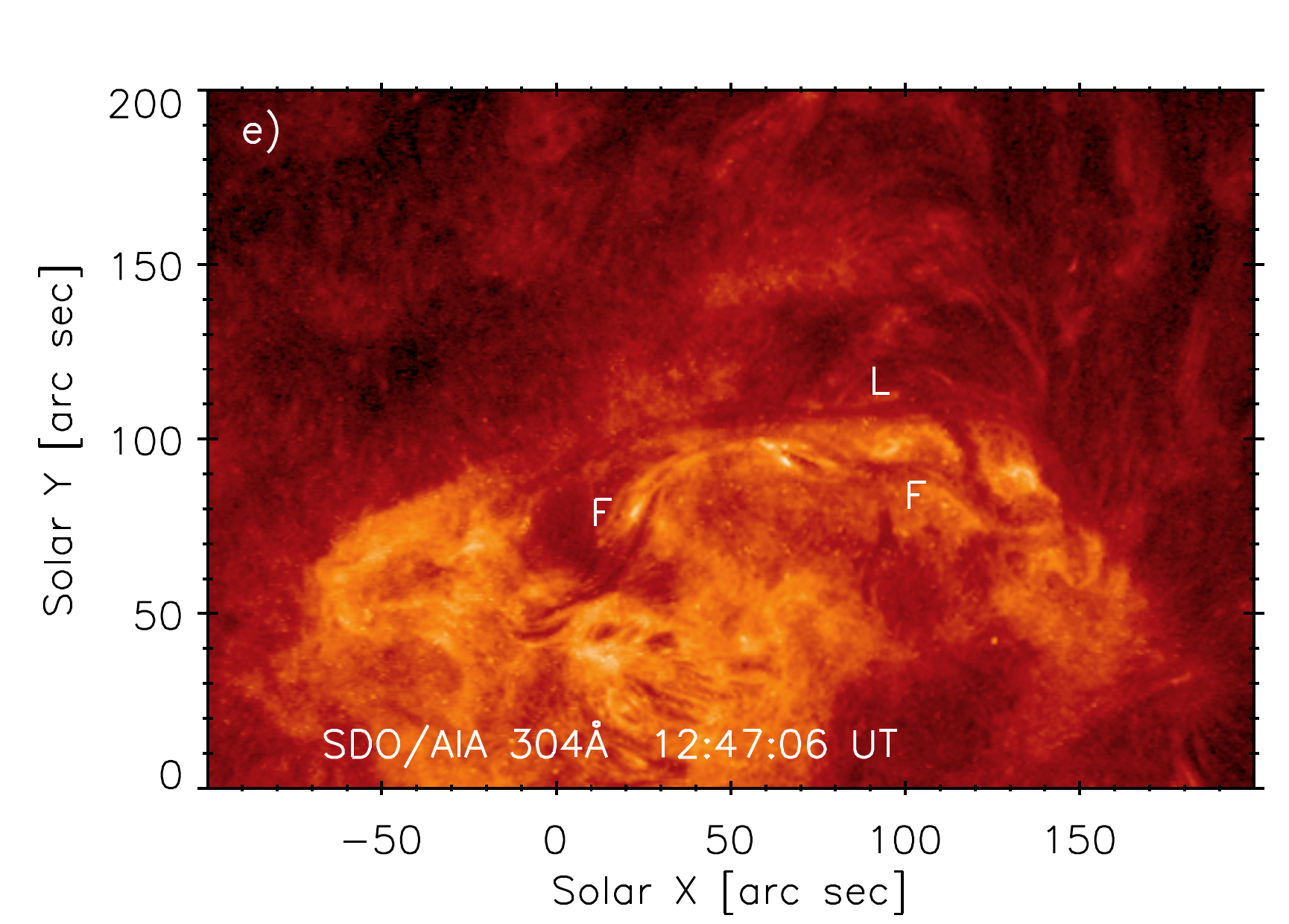}
        \includegraphics[width=8.03cm,clip,viewport=65 -10 488 325]{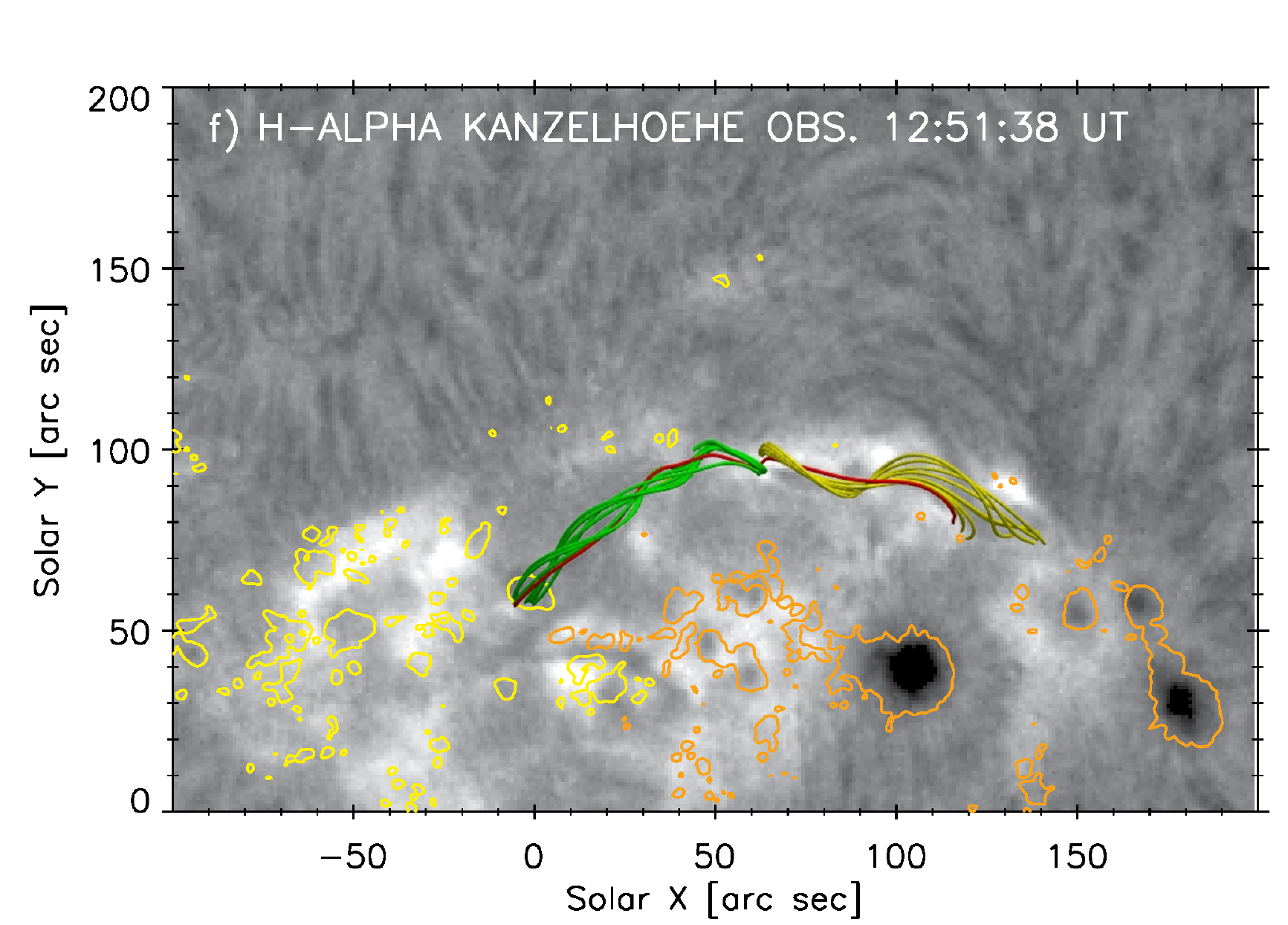}
\caption{Pre-flare situation. Left column shows {\it SDO}/AIA images: (a) 131\,\AA, (c) 171\,\AA, and (d) 304\,\,\AA. 
Right column shows: (b) line of sight (LOS) magnetic field from {\it SDO}/HMI (scaled $\pm$ 500\,G), (d) H$\alpha$
observation from Kanzelh\"{o}he Observatory and (f) NLFFF model of the filament overlaid over H$\alpha$ image.
Letter F on panels (a), (c) and (e) marks eastern/western parts of the filament as seen in H$\alpha$ (d) and L marks
the loop emitting in coronal temperatures and arching over the filament. The yellow circle depicted on (b) and (d) panels
shows the position of primary brightening.
Colour coded field lines on NLFFF model (f) depict two helical systems of magnetic field lines corresponding to eastern/green and
western/yellow parts of the filament F. The single red field line threads the body of the filament F from the positive polarity 
at east to the negative polarity at west. Yellow/orange contours outline positive/negative vertical photospheric magnetic 
field and are drawn at $\pm 500$\,G.} 
\label{fig_model}
\end{figure*}

\cite{Dissauer2018b} provided a study of 62 disk flares where it was possible to follow the evolution of coronal dimmings. Based on the growth rate of dimmed areas they define the impulsive phase of dimmings and found that 60\% of the events showed core dimmings which created only 3-13\% of the total dimmed area. \cite{Vanninathan2018} performed the plasma diagnostics of six coronal dimmings events. The core dimmings showed much faster and deeper decrease in emission measure and density than secondary dimmings.
The steepest changes in plasma parameters occurred 20-30 minutes after the flare start while in secondary dimmings the minimum was reached after 30-90 minutes. \cite{Vanninathan2018} and \cite{Veronig2019} supported the idea that the core dimming regions correspond to the footpoints of erupting flux rope and magnetic field there can open to interplanetary space. In the framework of the 3D model, the core dimming regions, i.e. the footpoints of the erupting flux rope should be located within the hooked parts of the J-shaped ribbons. 

The latest extension of the standard solar flare model in 3D deals with the drifting of the line-tied footpoints of the flux rope \citet{Aulanier2019}. They used an MHD model of the erupting flux rope configuration \citep{Zuccarello2015} and calculated the QSLs at different times during the model evolution to show that while the straight part of flare ribbons move away from PIL (as in 2D), the ribbon hooks first expand and then contract. Some locations in the photosphere thus can be swept by the ribbon hook more than once. The authors interpreted this evolution as a natural consequence of magnetic reconnection present in their 3D model. They introduced three basic reconnection geometries acting in the model:
(1) reconnection of two arcade field lines which produces a new flux rope field line and a flare loop similarly as in CSHKP model, referred to as `aa--rf reconnection', where letter `a' denotes an arcade field line, `r' denotes a flux rope field line and `f' denotes a flare loop (see Figure 4 in \cite{Aulanier2019}; 
(2) reconnection between two flux rope field lines that produces another flux rope field line and a flare loop, referred to as `rr--rf reconnection' (see Figure 4 in \cite{Aulanier2019}; 
and (3) reconnection between an (inclined) arcade field line and a flux rope field line producing a new flux rope field line and a flare loop, referred to as `ar--rf reconnection' (see Figure 5, 6 in \cite{Aulanier2019}).
The last one is primarily involved in shifting of the flux rope footpoints. Note, that QSLs hooks do not drift as a rigid body. They expand and shrink, what can be explained by series of sequential reconnections of individual field lines. In addition to the theoretical study, \cite{Aulanier2019} added an example of two flares, during which they observed drift and deformation of the ribbon hooks. 

 Figure~\ref{fig_cartoon} schematically shows the difference between aa--rf and ar--rf reconnection geometries. Grey dashed lines in all panels of Figure~\ref{fig_cartoon} show a flux rope at the onset of the eruption. The panels a) and b) show the ribbon/QSLs deformation due to aa--rf reconnection geometry. In Figure~\ref{fig_cartoon}a two overlying arcades `a' (green and cyan), rooted from outside of the straight parts of J-shaped ribbons (brown), reconnect. The reconnection turned the green arcade into a flux rope field line `r' and cyan one became a flare loop `f' rooted inside the straight part of J-shaped ribbons (Figure~\ref{fig_cartoon}b). The straight parts of the ribbons move apart, similarly as in 2D CSHKP model, while the hooks (curved parts) of the J-shaped ribbons expand (yellow areas) as a consequence of enlargement of the flux rope envelope. 
The cartoons in Figure~\ref{fig_cartoon}c--d show the drift of hook (curved part) of the ribbon/QSLs due to ar--rf reconnection geometry. An inclined arcade `a' (pink) and a flux rope field line  (red) (Figure~\ref{fig_cartoon}c), reconnect to produce a new flux rope field line and a flare loop (pink and red, respectively in Figure~\ref{fig_cartoon}d). The hook of the J-shaped ribbon on the left (Figure~\ref{fig_cartoon}d) thus drifts away from the position of the flux rope at the onset of the eruption (light grey dashed lines).
Although the cartoons show the effects of these to reconnections separately, in reality, both can happen nearly simultaneously. The true deformation of ribbons/QSLs is then a combination of these two effects.

In this study we report on pronounced evolution of the ribbon hooks observed during the eruptive M-class flare of 2015 November 4. For the first time, provide an observational evidence of a drift of the flux rope footpoint over distances of several tens of arc seconds. This drift was observed to occur by slipping reconnection of hot loops. We argue that such phenomenon can be explained by standard model of eruptive flares in three dimensions when considering the 3D reconnection geometries recently introduced by \cite{Aulanier2019}. 

Throughout the paper we will refer to a flare, and its timing, as it is defined according to its soft X-ray (SXR) light curves observed by {\it GOES} satellites. The term eruption will be used to refer to the sudden rise and escape of hot EUV/SXR loops. The paper is organized as follows: Section 2 describes the data sets and observations, Section 3 contains the interpretation of observed phenomena and in Section 4 we present conclusions.

\section{DATA AND OBSERVATIONS}
On 2015 November 4, the active region NOAA AR 12443 has produced 5 flares and one halo CME. In this study we focus on evolution of the M3.7 flare associated with this halo CME. The flare occurred close to the solar disk center (N09W04), started at 13:31 UT, peaked at 13:52\,UT and ended at 14:13\,UT according to {\it GOES-15} observation (Figure~\ref{fig_goes}a). The whole event was observed by Atmospheric Imaging Assembly \citep[AIA,][]{Lemen2012} and Helioseismic and Magnetic Imager \citep[HMI,][]{Hoeksema2014} instruments on board {\it Solar Dynamic Observatory} \citep[{\it SDO},][]{Pesnell2012} and by the X-ray Solar Telescope \citep[XRT,][]{Golub2007} on board {\it Hinode} \citep{Kosugi2007}. The beginning of the flare was also observed in H$\alpha$ by Kanzelh\"{o}he Solar Observatory, University of Graz.

The {\it SDO}/AIA images and HMI line of sight magnetograms have been processed by standard SSW \emph{aia\_prep.pro} routine. The HMI full-disk vector magnetic field data (hmi.B 720\,s data series) has been used for NLFFF modelling of pre-flare coronal magnetic field. {\it Hinode}/XRT data were processed using standard SSW \emph{xrt\_prep.pro} routine and then manually co-aligned with AIA 94\,\AA~filter.

\begin{figure*}[ht]
        \centering
        \includegraphics[width=6.63cm,clip,viewport=15 5 380 380]{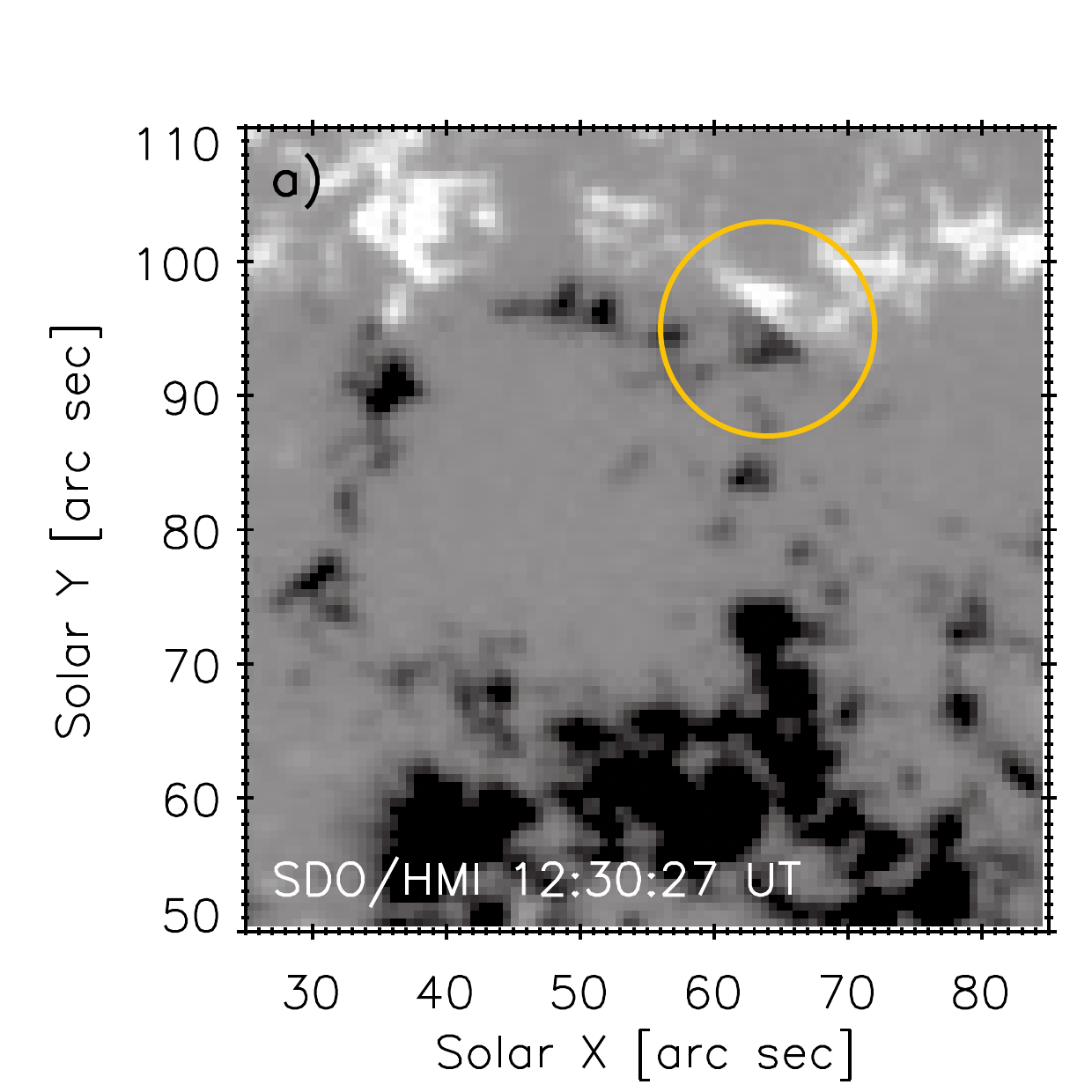}
       \includegraphics[width=5.53cm,clip,viewport=75 5 380 380]{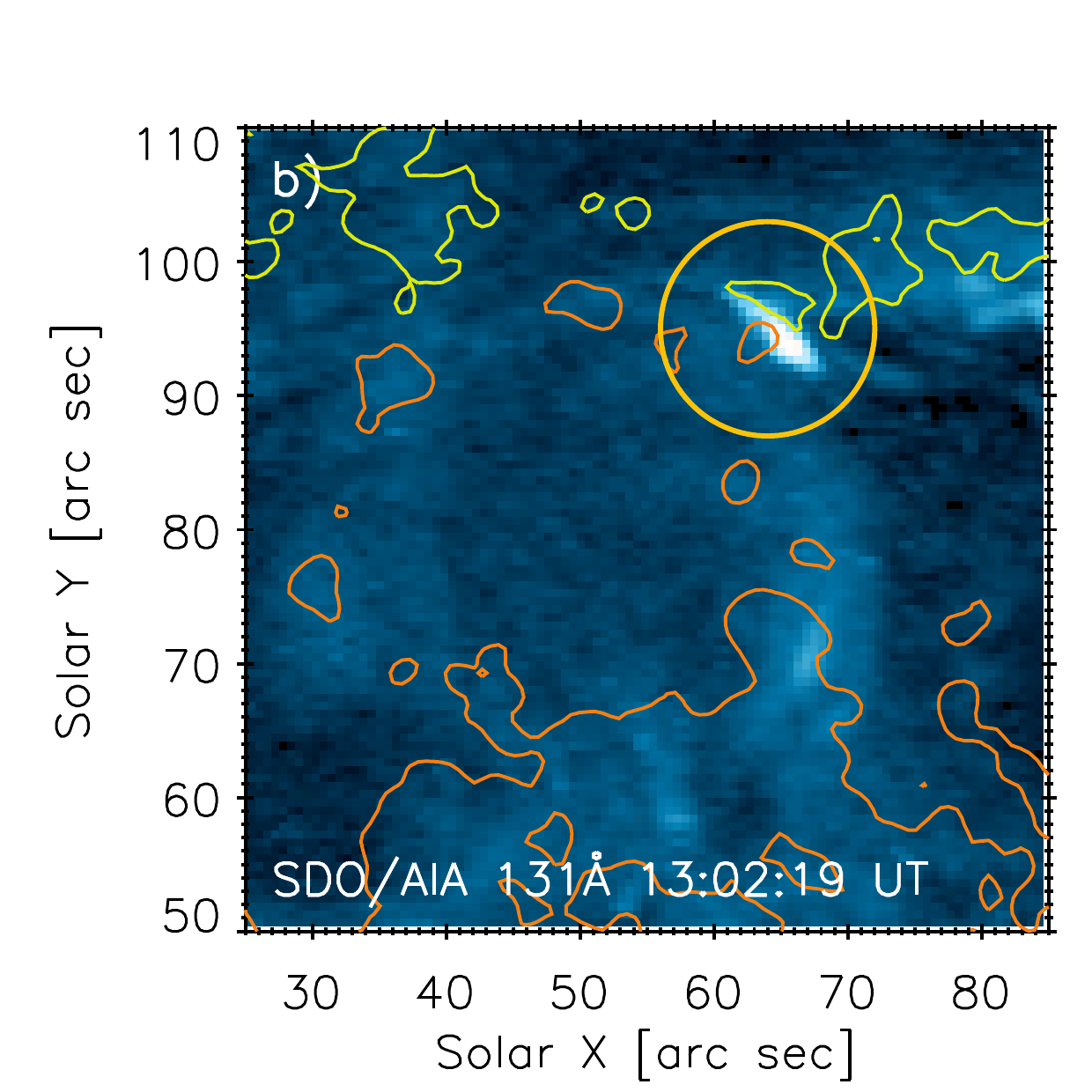} 
        \includegraphics[width=5.53cm,clip,viewport=75 5 380 380]{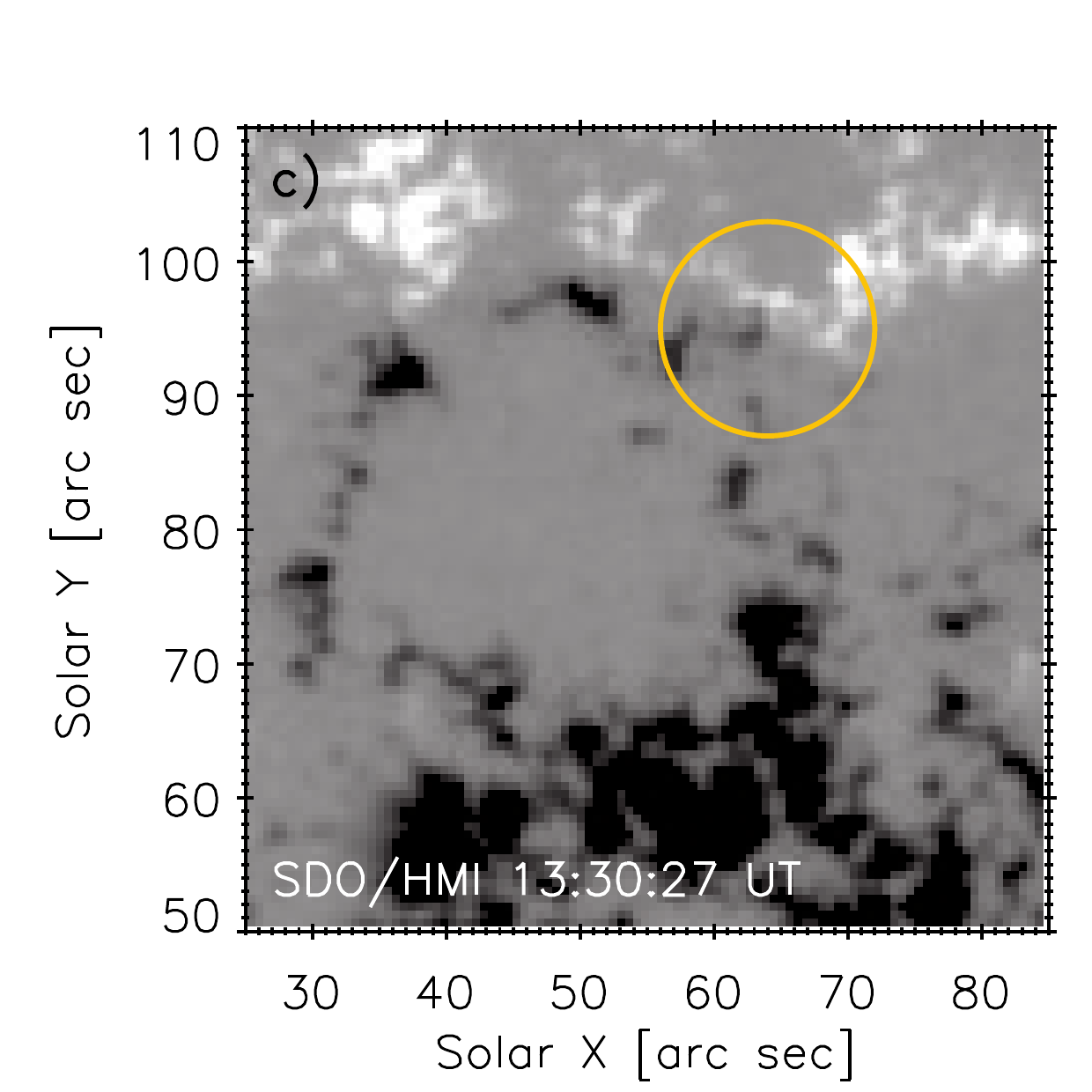}
        \caption{View of the negative supergranule showing the magnetic flux cancellation area within the circle 
(identical to that in Figures~\ref{fig_model} and \ref{fig_flevol}). (a) and (c) show {\it SDO}/HMI LOS 
magnetograms (scaled at $\pm$ 500\,G) with small bipolar patch in the middle of the circle. (b) shows {\it SDO}/AIA 
131\,\AA~image with primary brightening overlaid with LOS magnetic field contours at $\pm$ 200\,G (yellow/orange, resp.).}\label{fig_mf_can}
\end{figure*}

\subsection{Pre-flare conditions}

The HMI magnetogram in Figure~\ref{fig_model}b shows the photospheric magnetic field of the NOAA AR 12443. It consists of several small trailing spots of positive polarity and four well developed spots of negative polarity. The active region also includes a well developed negative supergranule. The northern border of the supergranule neighbours with positive plage polarity and a small bipolar patch formed by these opposite polarity fields is shown within a yellow circle (Figure~\ref{fig_model}b).  

In H$\alpha$  at 12:51:38\,UT (Figure~\ref{fig_model}d) we observed two parts (eastern/western) of filament F, 
lying along the polarity inversion line of magnetic field. Left column of the Figure~\ref{fig_model}~shows a pre-flare structure of magnetic field in the corona. At about 12:37\,UT, the EUV filters 131\,\AA~and 171\,\AA~show a loop structure L, emitting at coronal temperatures, and overlying the filament F. The eastern and western parts of the filament are not clearly discernible in those filters but are visible about 10 minutes later in 304\,\AA~(Figure~\ref{fig_model}e). The L can be observed there as dark structure.

To investigate the structure of pre-flare coronal magnetic field we used a NLFFF model based on {\it SDO}/HMI full-disk vector magnetic field at 12:59\,UT (hmi.B 720s data series). From the total field and inclination and ambiguity-resolved azimuth\footnote{http://jsoc.stanford.edu/jsocwiki/FullDiskDisamb} 
we derive the local (heliographic) surface magnetic field, following \cite{Gary1990}. The resulting full-resolution photospheric magnetic field map with spatial resolution of $\approx$\,1\,$\arcsec$ is used as input to the method of \cite{Wiegelmann2010} to reconstruct the non-linear force free (NLFF) magnetic field in corona. We employ NLFF solutions based on different choices of the free model parameters and judge the relative success of the individual models based on their quality (in the form of dimensionless numbers), as well as their success to reproduce the coronal observations of the filament. For the dimensionless numbers which quantify 
the goodness of a NLFFF model, we find for the current-weighted average of the sine of the angle between the magnetic field, $\Bvec$, and electric current density, $\Jvec$,
CWsin\,=\,0.1, i.e., a mean current-weighted angle of  $\theta_j\simeq7^\circ$. For a perfectly force-free solution, CWsin\,=\,0 and $\theta_j=0^\circ$
\cite[for details see][]{Schrijver2006}. In addition, we quantify the degree to which $\nabla\cdot\Bvec=0$ is fulfilled,
using the fractional flux measure $\langle|f_i|\rangle$ \cite[][]{Wheatland2000}, where we find $\langle|f_i|\rangle\simeq3\times10^{-4}$, and the non-solenoidal contribution, $E_{\rm div}$ to the total energy, $E$, introduced in \cite{Valori2013}, where we find $|E_{\rm div}|/E\simeq0.01$. 

\begin{figure*}
        \centering
        \includegraphics[width=8.42cm,clip,viewport= 15 51 487 328]{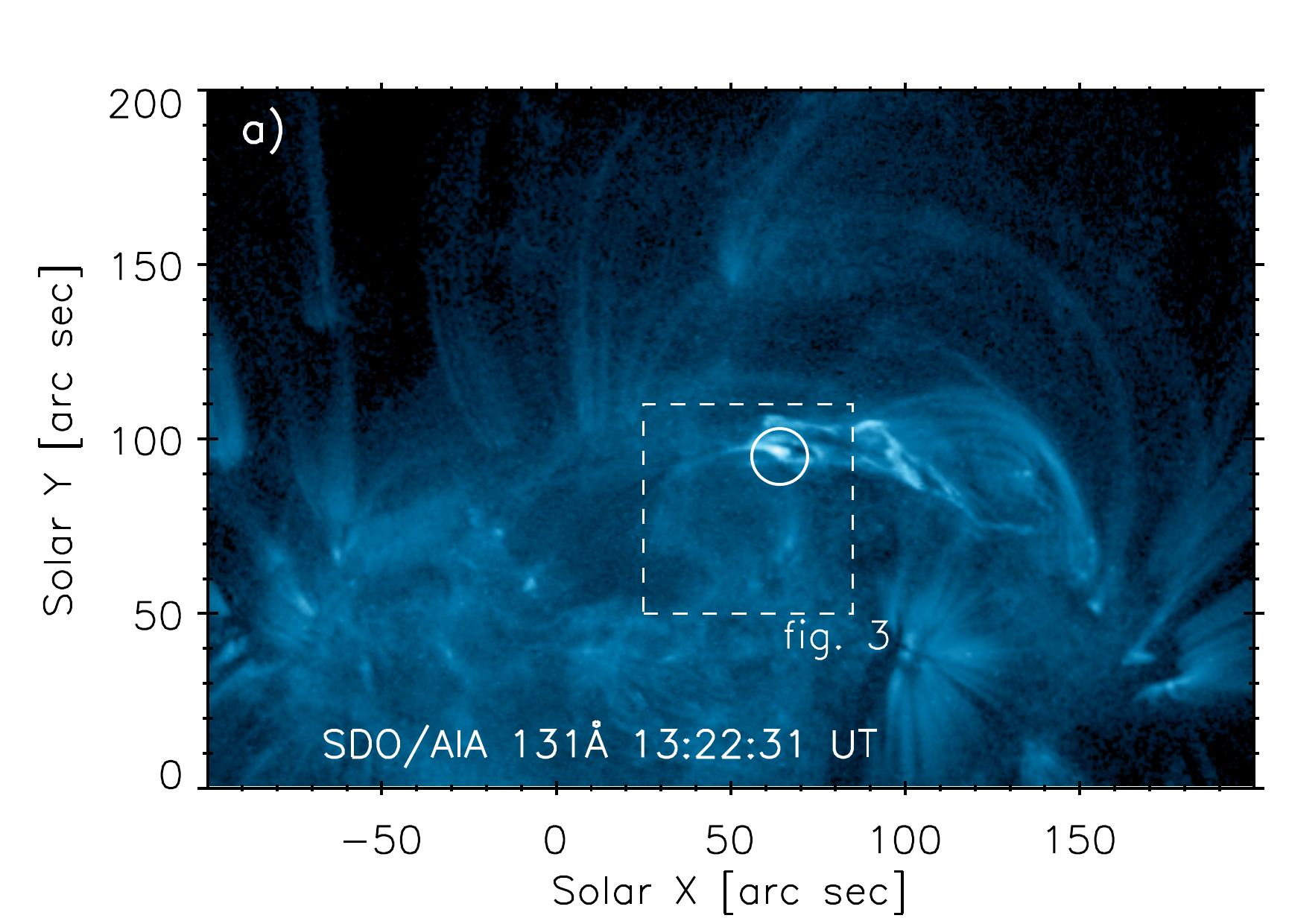}
        \includegraphics[width=7.32cm,clip,viewport=78 51 487 328]{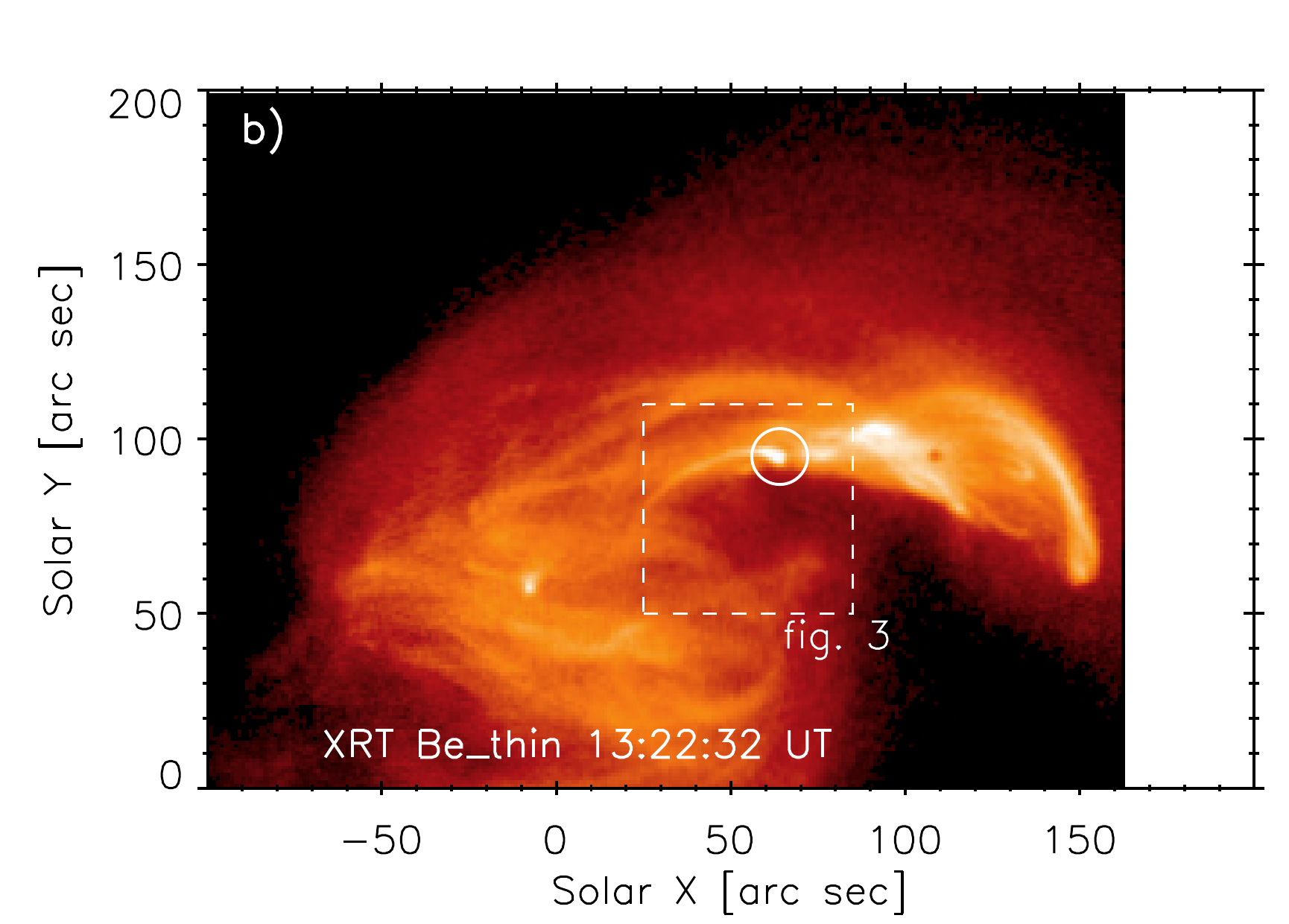}
      
        \includegraphics[width=8.42cm,clip,viewport= 15 51 487 328]{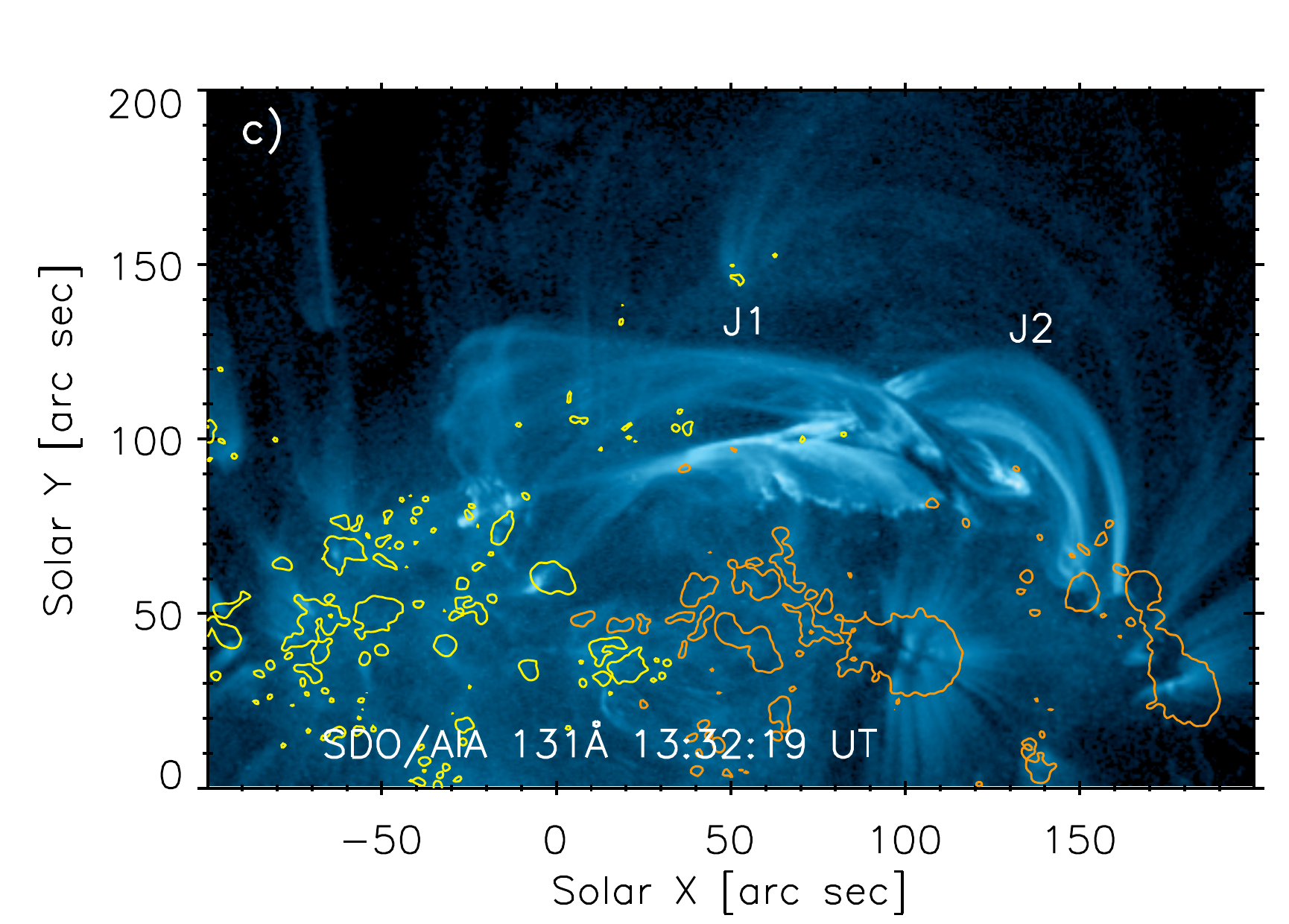}
        \includegraphics[width=7.32cm,clip,viewport=78 51 487 328]{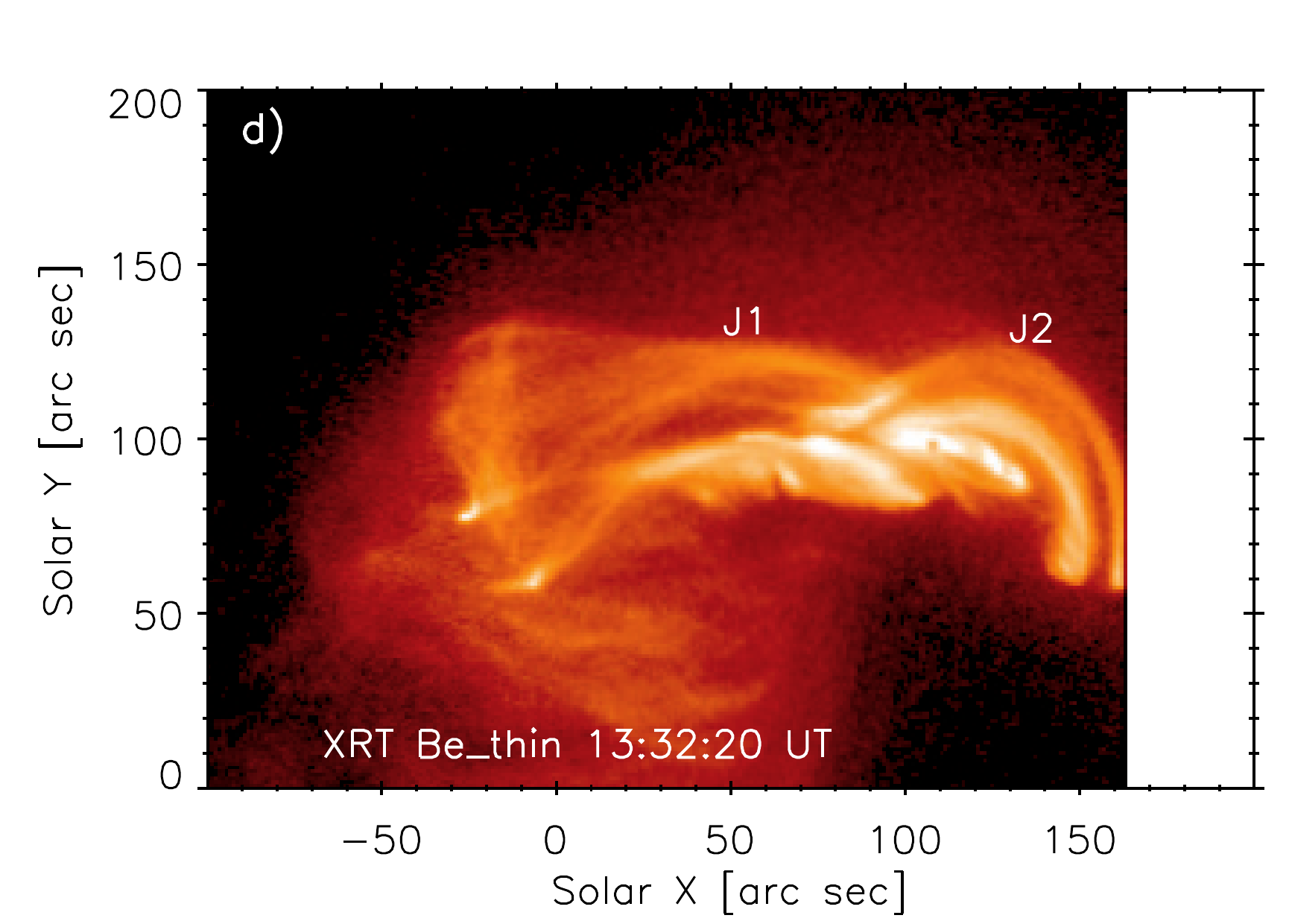}
        
        \includegraphics[width=8.42cm,clip,viewport= 15 51 487 328]{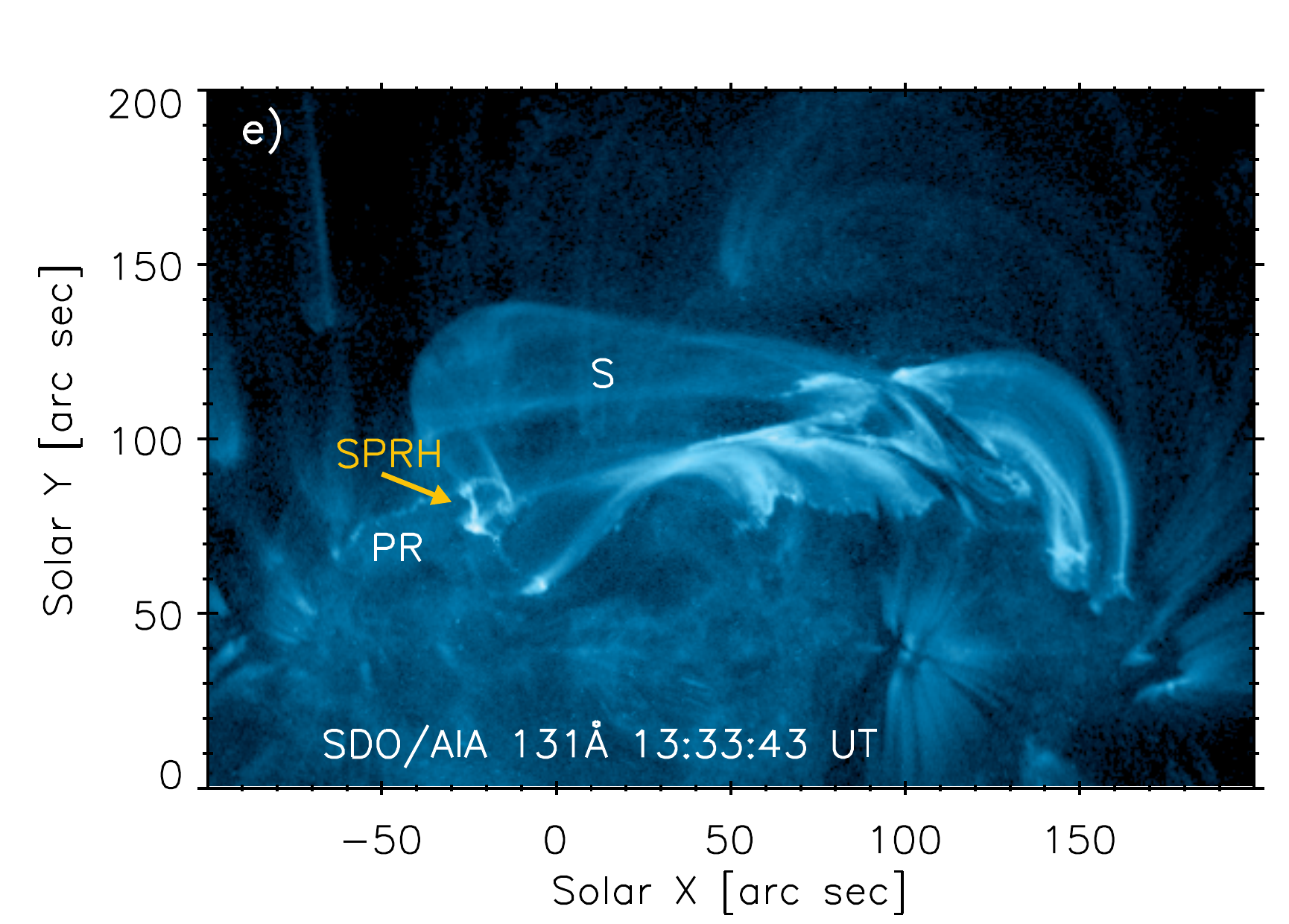}
        \includegraphics[width=7.32cm,clip,viewport=78 51 487 328]{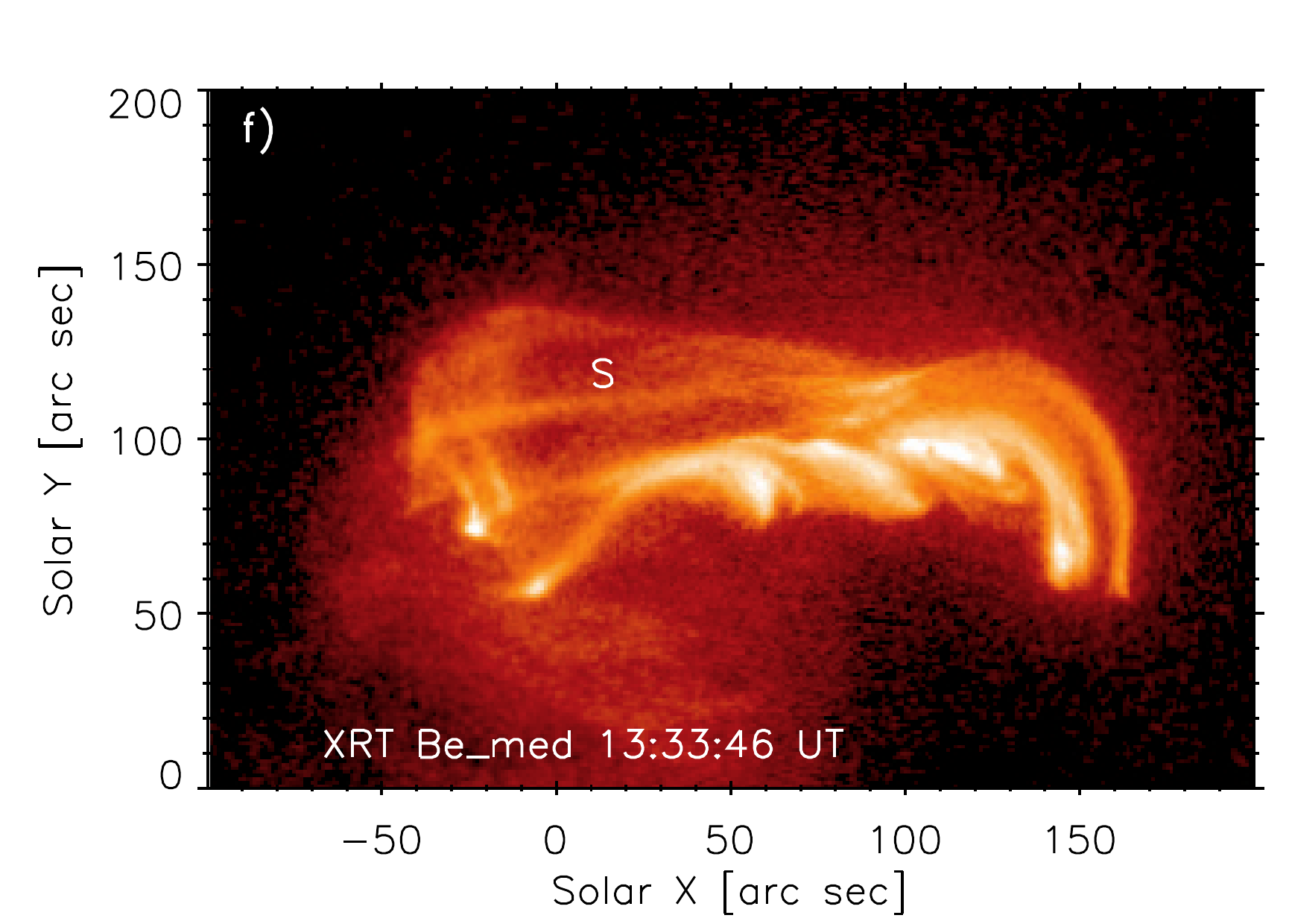}
        
        \includegraphics[width=8.42cm,clip,viewport= 15 4 487 328]{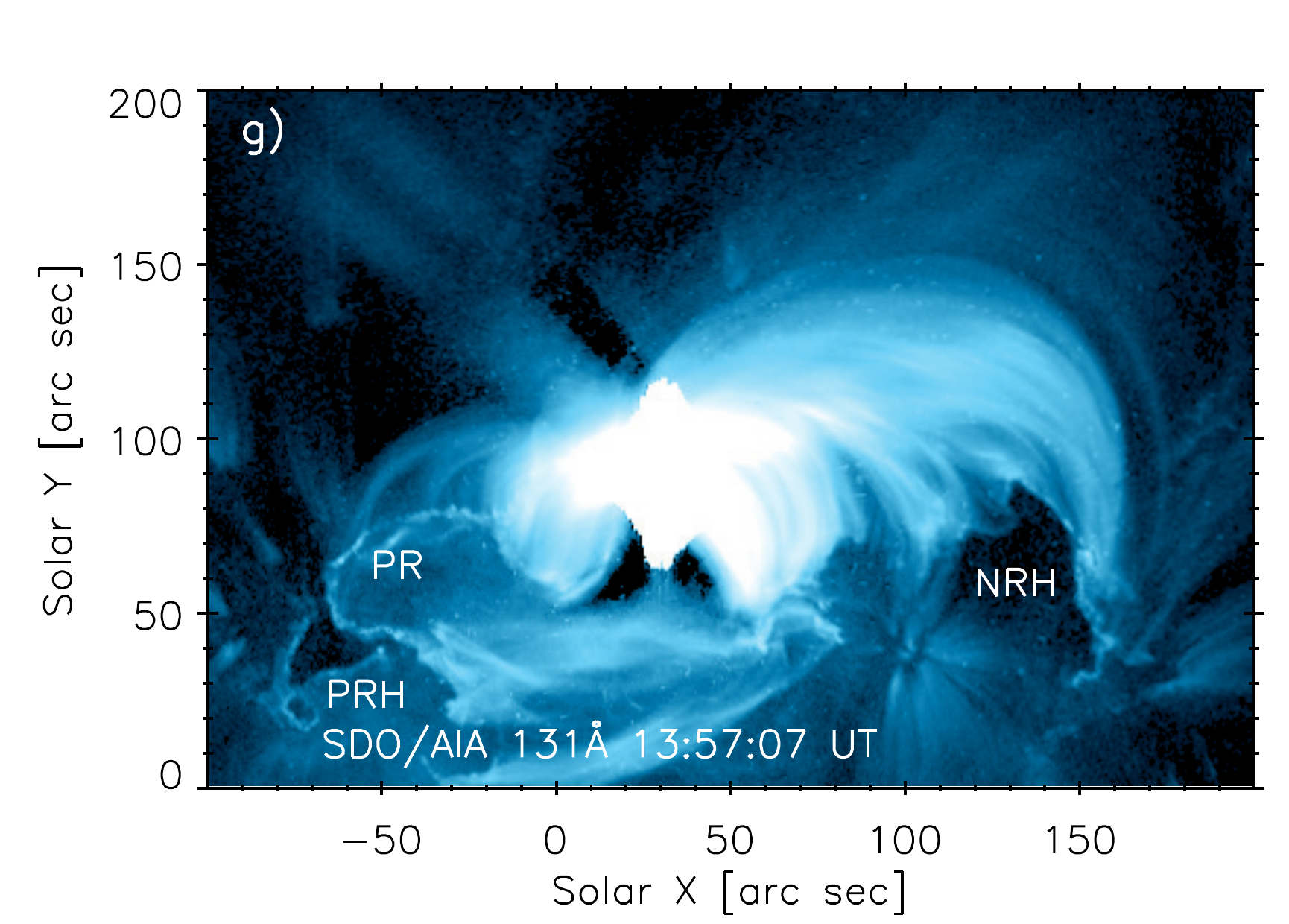}
        \includegraphics[width=7.32cm,clip,viewport=78 4 487 328]{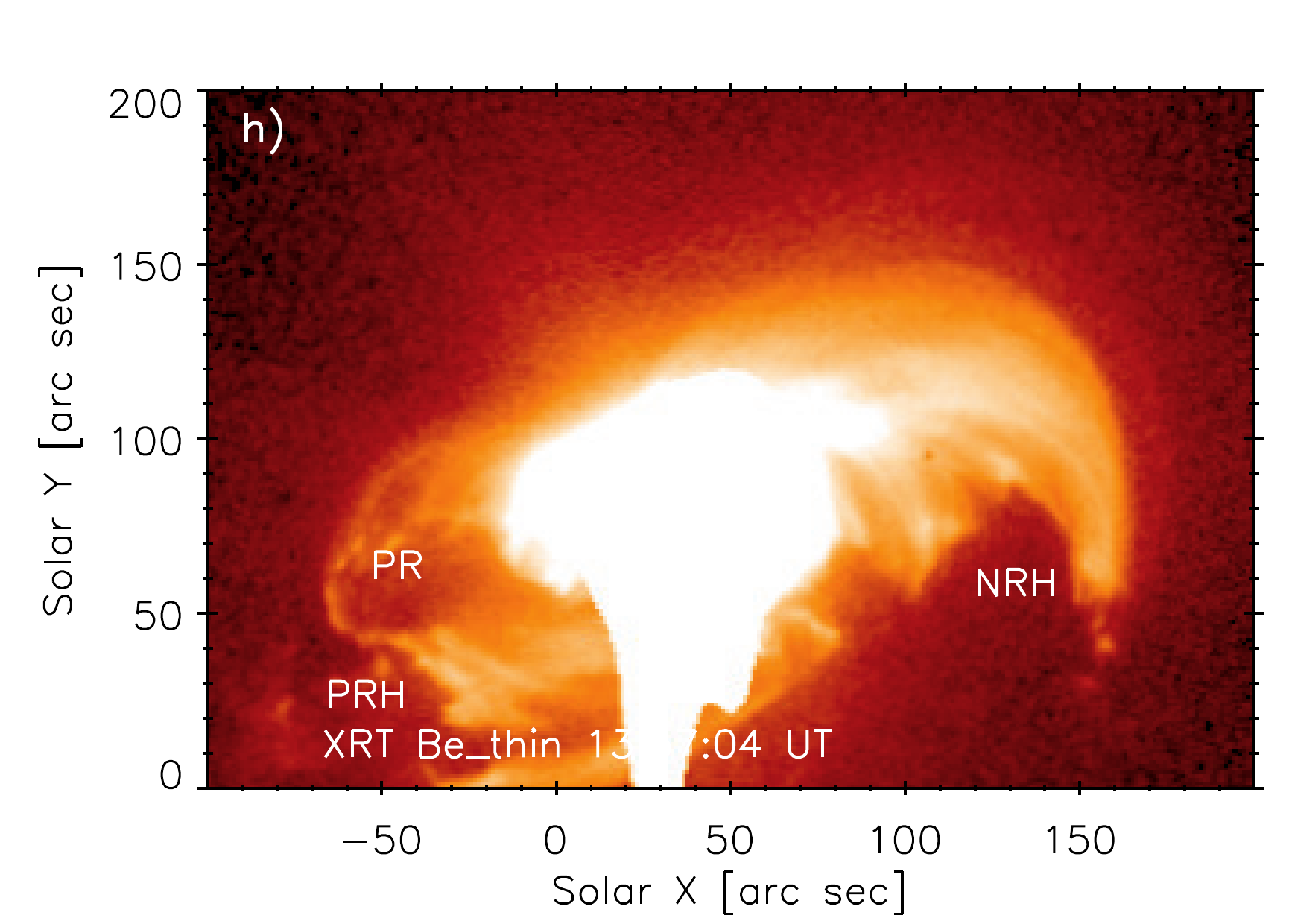}
        
\caption{An overview of flare evolution. White rectangles with a circle on (a) and (b) shows FOV of Figure~\ref{fig_mf_can}. 
Labels J1 and J2 on (c) and (d) show the J-shaped loop structures. The yellow/orange contours at (c) show LOS magnetic field at $\pm500G$.
(e) and (f) show the sigmoidal loop structure S in {\it SDO}/AIA 131\,\AA~and {\it Hinode}/XRT Be\_med filters, respectively. The yellow arrow on (e) shows the SPRH. 
Panels (g) and (h) show the elongated ribbon PR and positions of the hooks PRH and NRH after the eruption
of hot loops rooted in them. (A movie of {\it SDO}/AIA 131\AA~images is available (movie\_fl\_evol.mpg).)} \label{fig_flevol}
\end{figure*}

These values, together with the presence of modelled complex helical fields co-spatial with the observed filament, suggest that our NLFFF model may realistically approximate the pre-flare corona in the present case.

The result of NLFFF modelling is shown in Figure~\ref{fig_model}f. It shows two systems of helical magnetic field neighbouring each other. The green/yellow field lines are co-spatial, respectively, with the eastern/western parts of the filament F observed in H$\alpha$ (Figure~\ref{fig_model}d). The field lines of the green system were obtained by calculating NLFFF sample field lines starting from the location of the negative polarity patch in the middle of the yellow circle at Figure~\ref{fig_model}b. The yellow field lines are calculated from the positive polarity patch within this yellow circle. The single red field line is calculated from a random location along the north-south-aligned axis connecting the aforementioned two footpoint locations. It threads the bodies of both green and yellow helical systems, indicating that they indeed are parts of one twisted structure.

Figure~\ref{fig_mf_can} shows changes of the magnetic field close to the supergranule before the flare. Comparing Figures~\ref{fig_model}b, d and f with Figure~\ref{fig_mf_can} one can notice that green and yellow parts of the filament F are rooted in a small bipolar patch at which we observed cancellation of magnetic flux during the period 12:30--13:30\,UT. At about 13:02\,UT, a small brightening (Figure~\ref{fig_mf_can}b) appeared over this small bipolar patch. This brightening was visible in all {\it SDO}/AIA EUV filters and had elongated shape with a length of about 10\,$\arcsec$ (Figure~\ref{fig_mf_can}b). It was smaller and more circular in {\it SDO}/AIA 1600\AA~UV filter but there was no brightening visible in 1700\AA~UV filter. From this we deduce that it was emitted by coronal plasma at log\,T[K]\,$\gtrsim$\,4.7 \citep{Lemen2012}. The NLFFF model shows that it appeared in the area where foopoints of the loops with opposite magnetic orientation meet (Figure~\ref{fig_model}b, d and f).

\subsection{Overall evolution of the event}
Figure~\ref{fig_flevol} (and its accompanying movie) shows the overall evolution of the event. At 13:22\,UT the western part of the filament F got activated. The primary brightening at the cancellation site was still visible, and was located at negative footpoint of the eastern part of F (Figure~\ref{fig_flevol}a, b), i.e. at the negative footpoint of green field line system in Figure~\ref{fig_model}f. Both brightening and filament activation were observed prior the start of M3.7 flare. The {\it GOES-15} soft X-ray light curves showed a small bump during the time interval 13:20--13:30\,UT, followed since 13:31\,UT by the M3.7 class flare (Figure~\ref{fig_goes}a). 
 
At about 13:32\,UT, two hot J-shaped loop structures, J1 and J2, were present in AIA 131\,\AA~and XRT Be\_thin 
filters (Figure~\ref{fig_flevol}c, d). These structures were not exactly co-spatial with eastern/western parts of the observed H$\alpha$ filament F (Figure~\ref{fig_model}d, f) but they evolved due to their reconnection. The reconnection between J1 and J2 produced also a small arcade of hot flare loops below J1. The crossing area of J1 and J2 shows hot emission in XRT Be\_thin passband, contrary to AIA 131\,\AA, where it is partially obscured by dark threads. Later, the new loop structure S, seen almost edge-on, appeared due to ongoing reconnection between J1 and J2. It was also visible in XRT Be\_med filter (Figure~\ref{fig_flevol}f). Yellow arrow in Figure~\ref{fig_flevol}e shows the position of small positive ribbon hook SPRH which appeared at the north end of positive ribbon PR at the beginning of the flare. The S was rooted in the SPRH and belonged to the hot loops which formed over the flux cancelling location and primary brightening. During the impulsive phase of the flare the SPRH disappeared, followed by PR elongation, and formation of a new positive ribbon hook PRH at its end. The new PRH was located further away, i.e. $\approx$\,40\,$\arcsec$ to the south (along Y axis; Figure~\ref{fig_flevol}e and g), from the position of former SPRH.

\begin{figure*}
 \centering
        \includegraphics[width=5.2cm,clip,viewport=10 0 355 315]{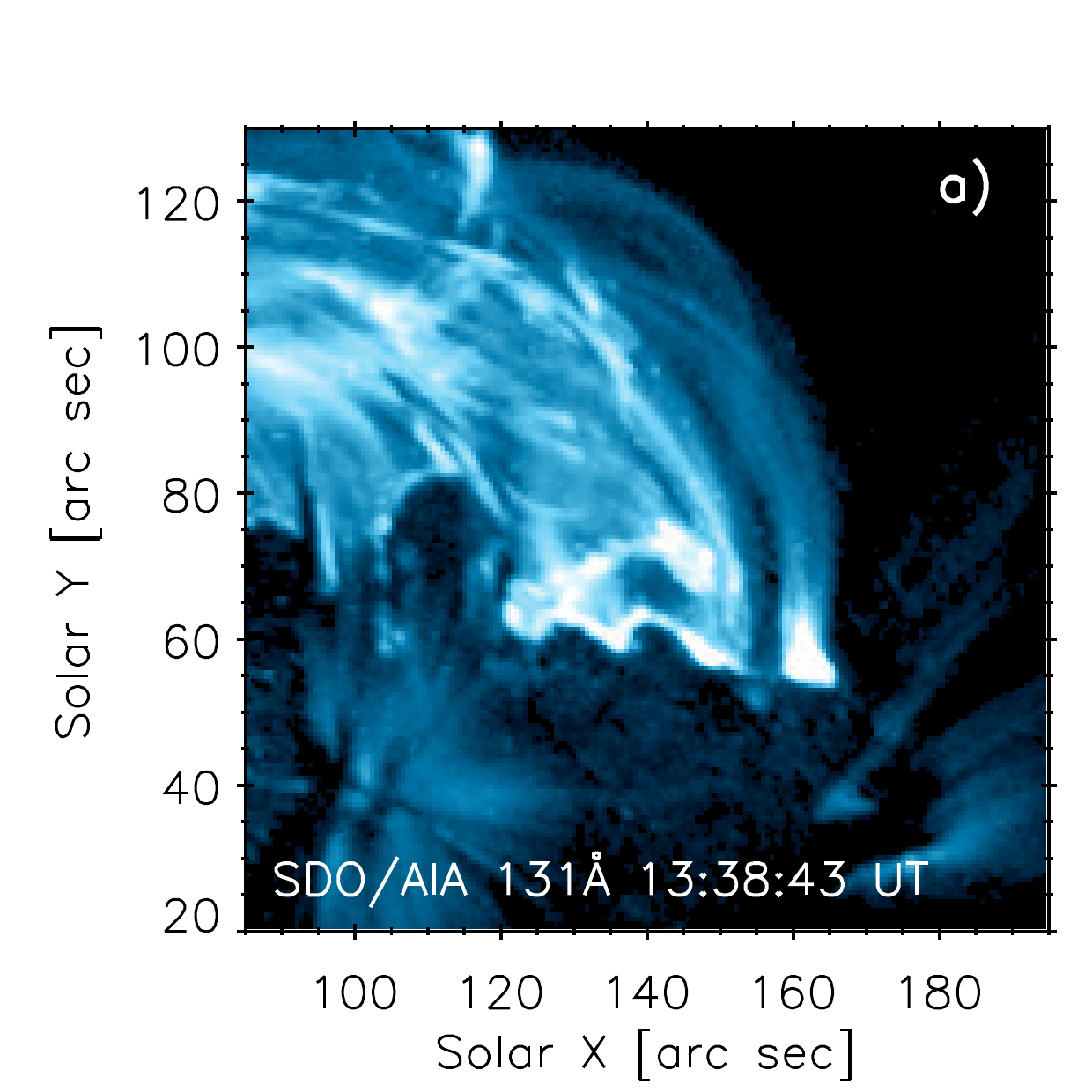}      
        \includegraphics[width=4.15cm,clip,viewport=78 0 355 315]{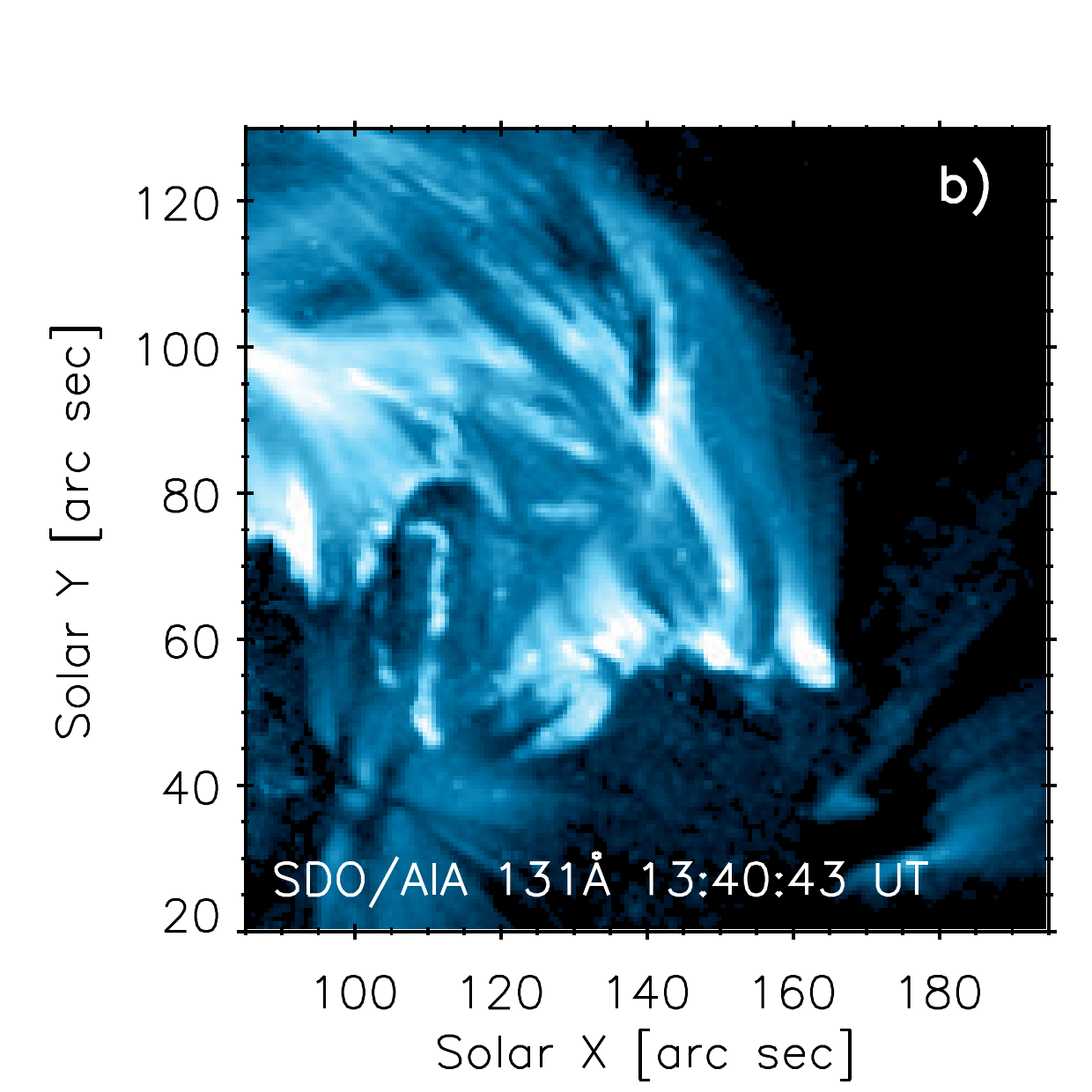}
        \includegraphics[width=4.15cm,clip,viewport=78 0 355 315]{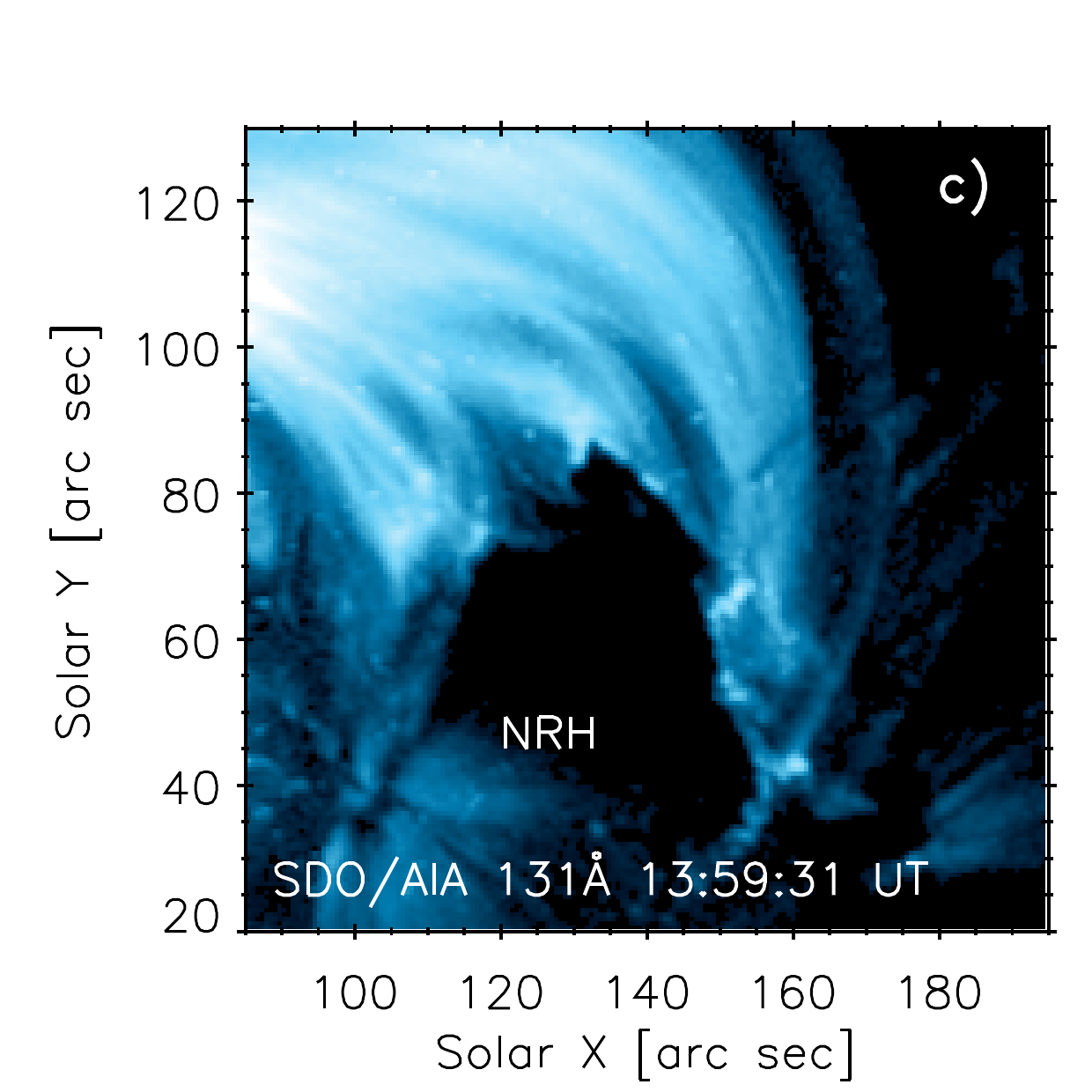}
        \includegraphics[width=4.15cm,clip,viewport=78 0 355 315]{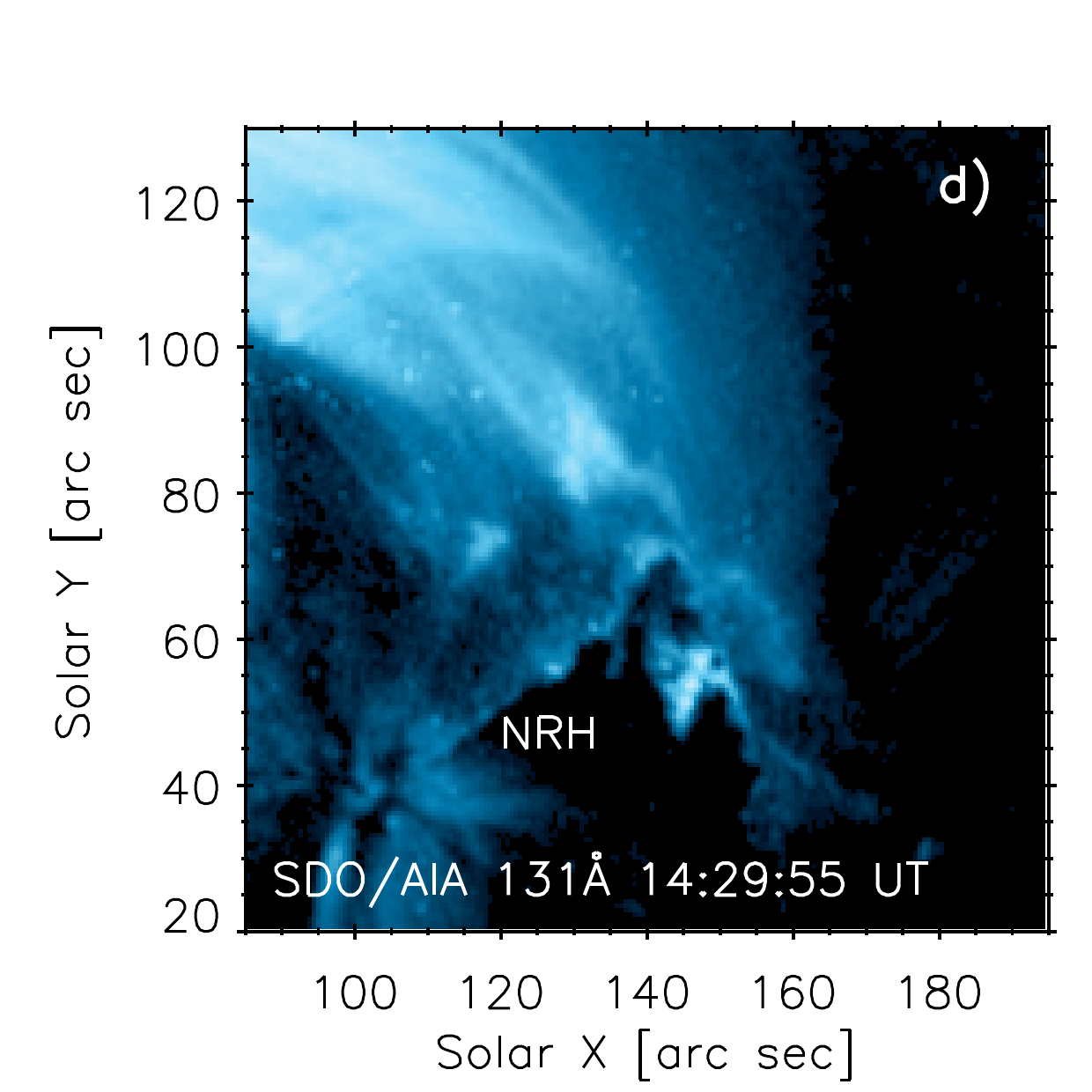}
        \caption{An illustration of NRH evolution. It was very complicated but since 13:50\,UT NRH had a clear hook shape. 
        (A movie of this figure is available (movie\_nrh.mpg).)} \label{fig_nrh}
\end{figure*}
The evolution of negative ribbon and its hook was complicated since the start of the flare (Figure~\ref{fig_nrh} and its accompanying movie). The negative ribbon hook NRH (Figure~\ref{fig_nrh}c) was formed later than PRH. It appeared few minutes after the eruption of hot loops rooted in PRH, which erupted at about 13:40\,UT. Since the eruption of hot loops widening of PRH, and later also of NRH  (Figure~\ref{fig_flevol}g, h), was observed and huge dimmed areas appeared within both ribbon hooks. Below the erupted hot loops, the growing arcade of hot flare loops was observed. During the gradual phase of the flare, the expansion of the hooks turned to their contraction. {\it SOHO}/LASCO coronograph C2 detected a halo CME\footnote{https://cdaw.gsfc.nasa.gov/CME\_list/} at 14:48\,UT.
\begin{figure*}
        \centering
        \includegraphics[width=5.2cm,clip,viewport=10 0 355 315]{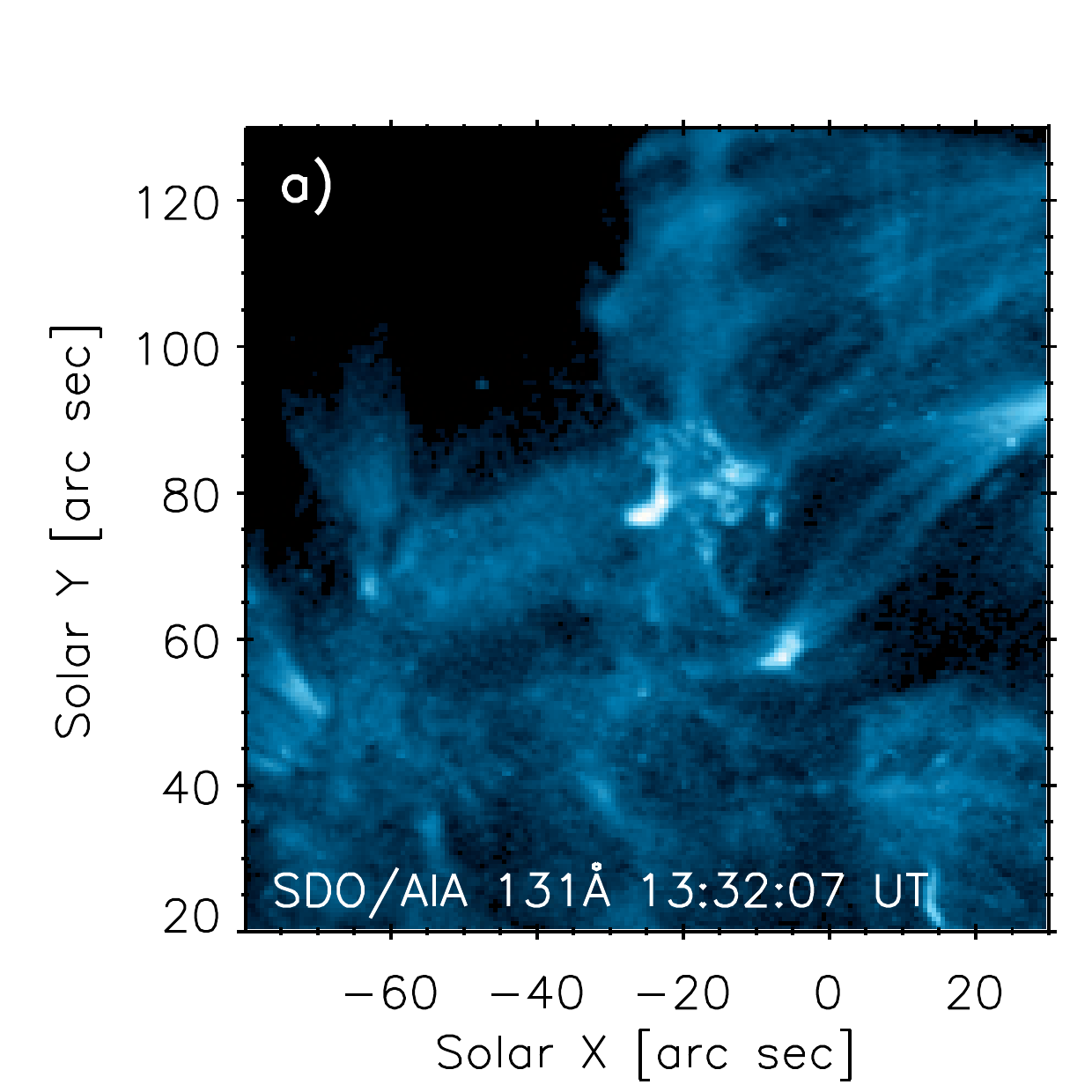}      
        \includegraphics[width=4.15cm,clip,viewport=78 0 355 315]{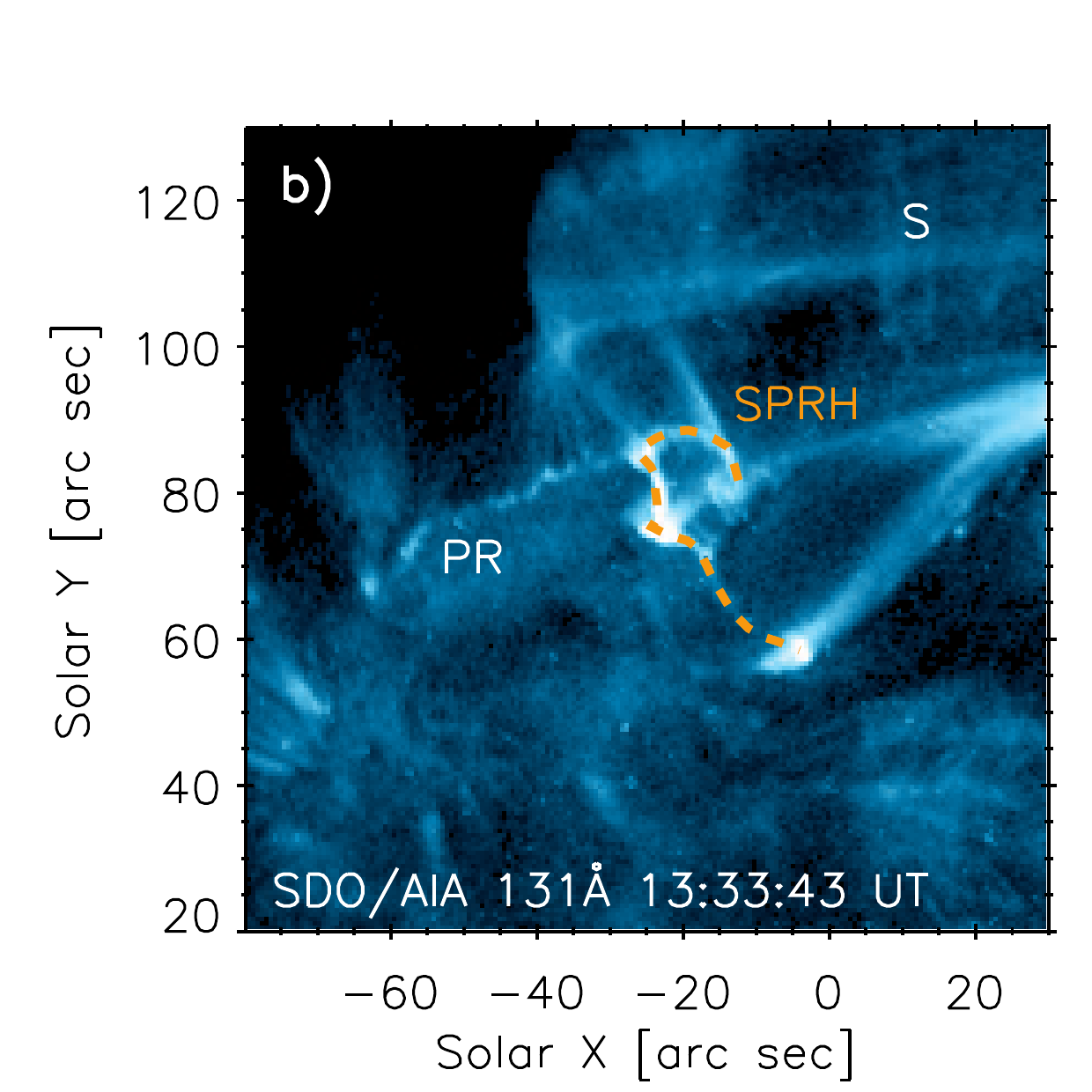}
        \includegraphics[width=4.15cm,clip,viewport=78 0 355 315]{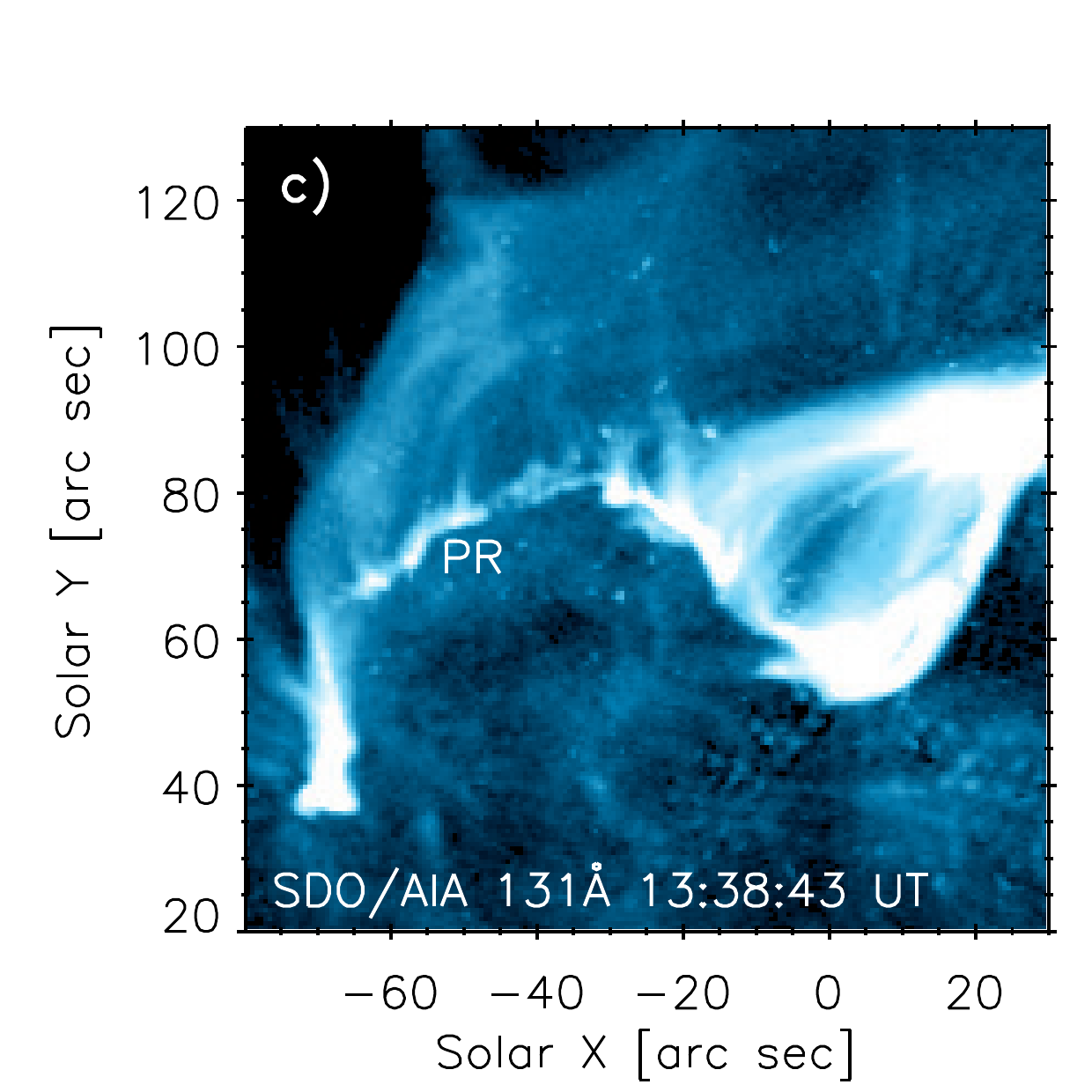}
        \includegraphics[width=4.15cm,clip,viewport=78 0 355 315]{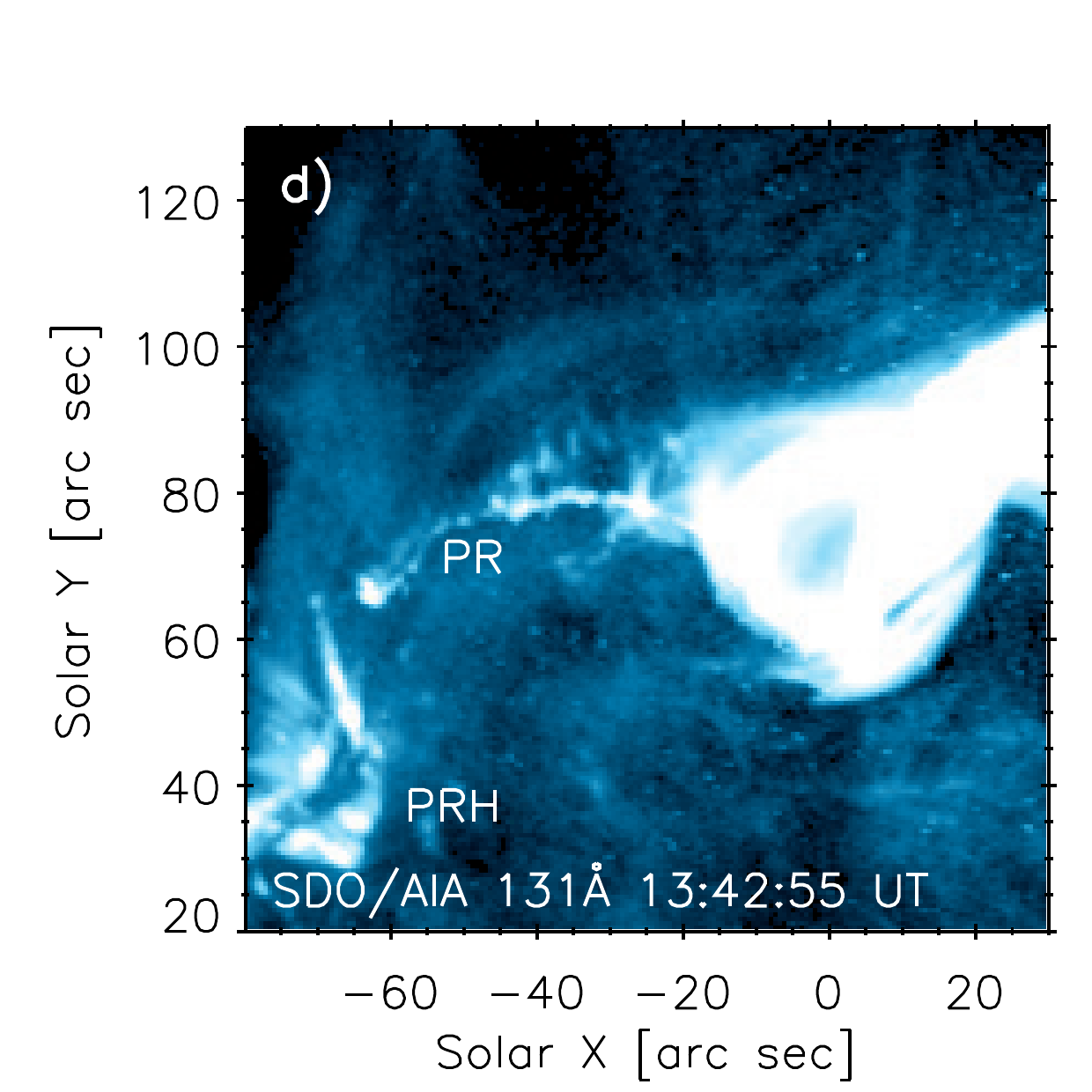}
        \caption{Evolution of SPRH and elongation of PR in 131\,\AA. Panels (a) and (b) show formation of SPRH at the north end 
of the original positive ribbon. Its straight part and SPRH are highlighted by dashed orange line in (b) and the sigmoidal 
loop structure rooted in SPRH is labelled by letter S. Panel (c) shows the elongated PR. SPRH disappeared and the elongated 
PR had an elbow at its place. (d) shows a new PRH which appeared at the end of elongated ribbon PR. (A movie of this Figure is available (movie\_sprh.mpg).)} \label{fig_sprh}
\end{figure*}

\subsection{Observation of the positive polarity ribbon elongation and its hook evolution.}
In this subsection we describe the observations of the disappearance of SPRH, the process of PR elongation and formation of new PRH. Figure~\ref{fig_sprh} (and its accompanying movie) shows the evolution of PR during the time interval 13:32--13:43\,UT in 131\,\AA. At 13:32:07\,UT, SPRH started to form at the north end of the original PR (Figure~\ref{fig_sprh}a). Few seconds later it became a hook of PR, and can be seen in Figure~\ref{fig_sprh}b, where the SPRH and the straight part of original PR are highlighted 
by orange dashed line. Simultaneously, to the east of SPRH, the new part of PR started to form. Within  approximately two minutes, SPRH disappeared and PR elongated further to south-east until about 13:40\,UT (Figure~\ref{fig_sprh}c). This elongation of the ribbon with `a hook-shaped segment' at its end, was for this flare also reported by \cite{Li2017}, who found 120 km\,s$^{-1}$ to be the elongation velocity. 
\begin{figure*}
        \centering
        \includegraphics[width=9.625cm,clip,viewport=15 50 470 330]{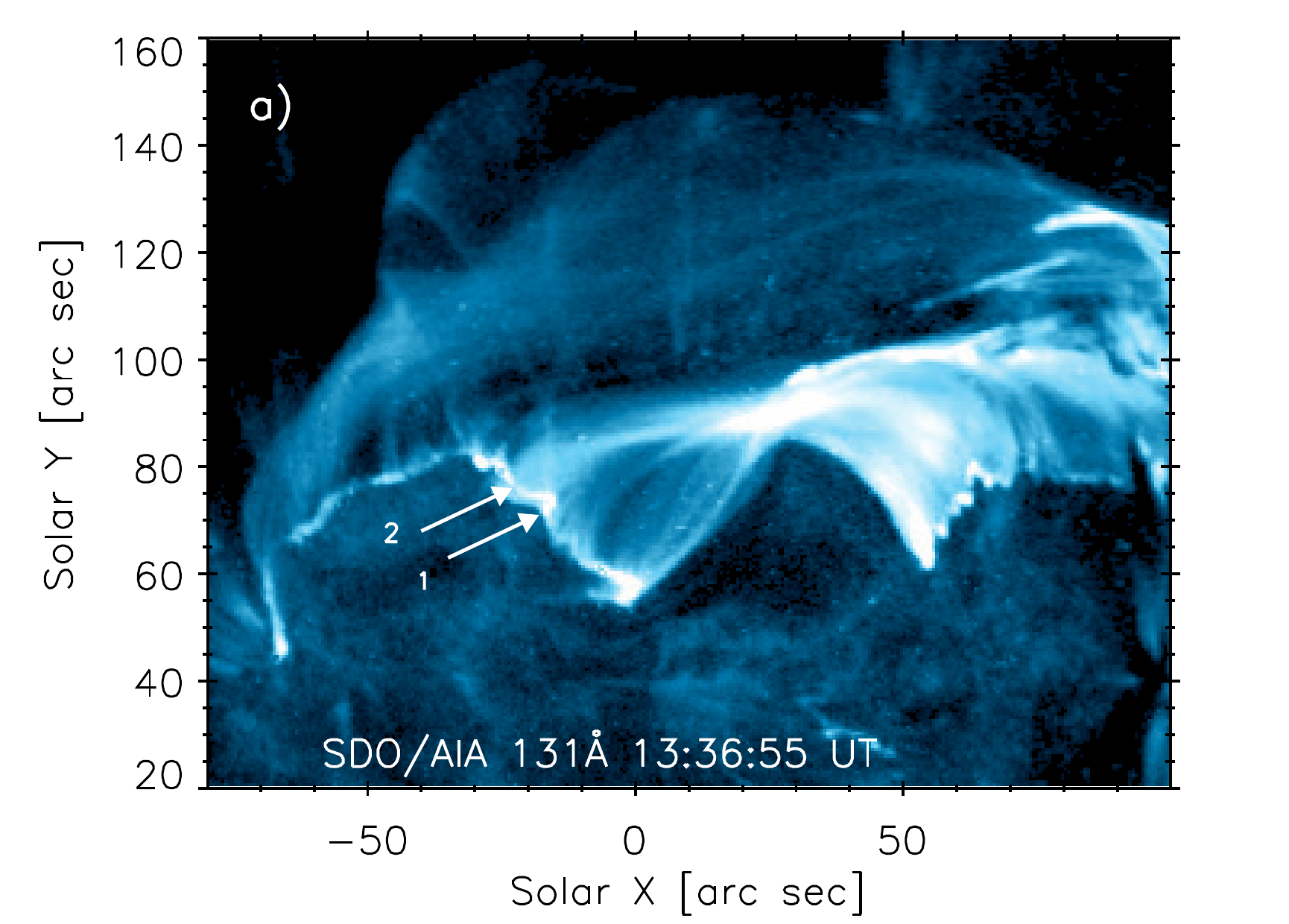} 
        \includegraphics[width=8.29cm,clip,viewport=78 50 470 330]{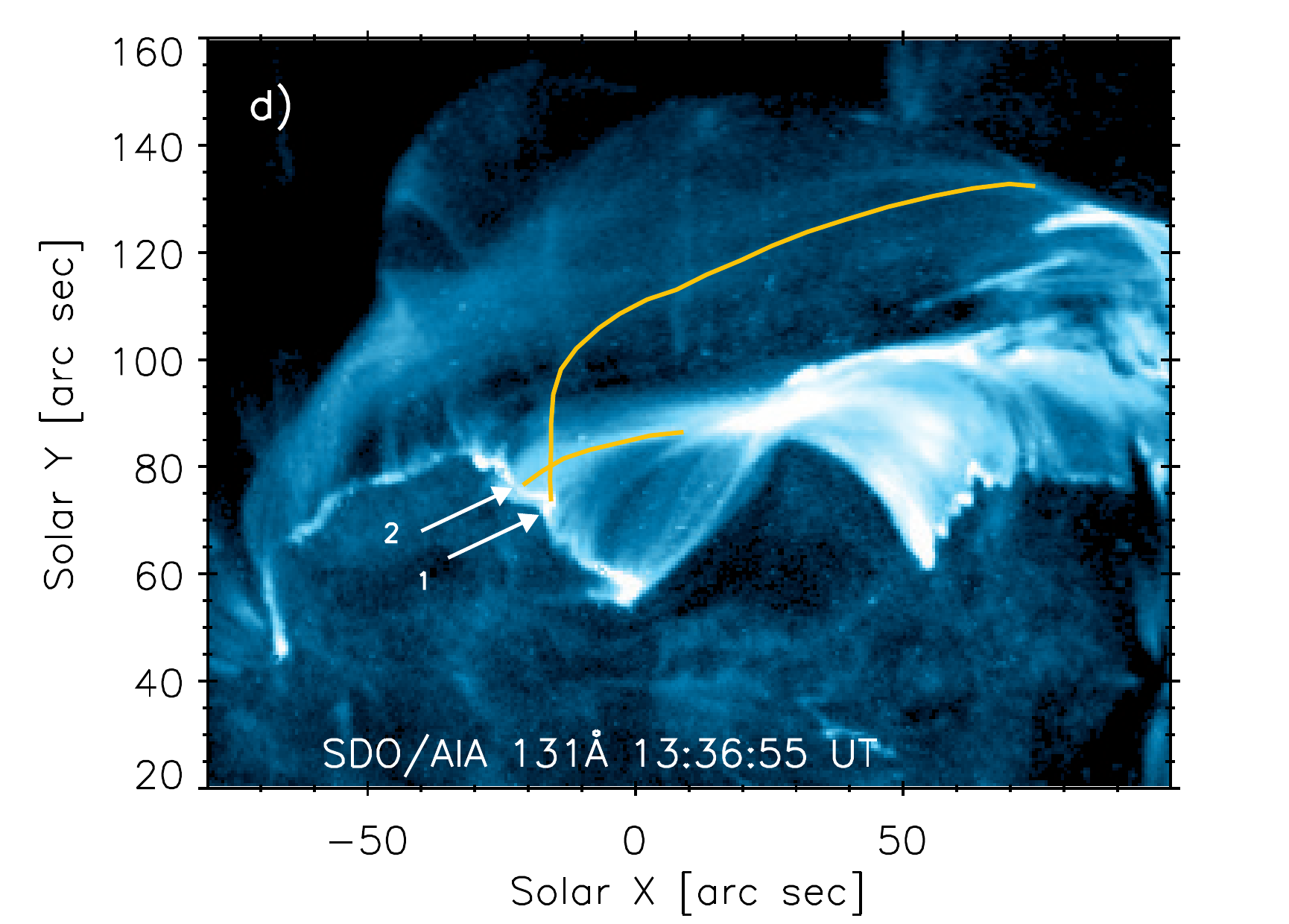}  
        
        \includegraphics[width=9.625cm,clip,viewport=15 50 470 330]{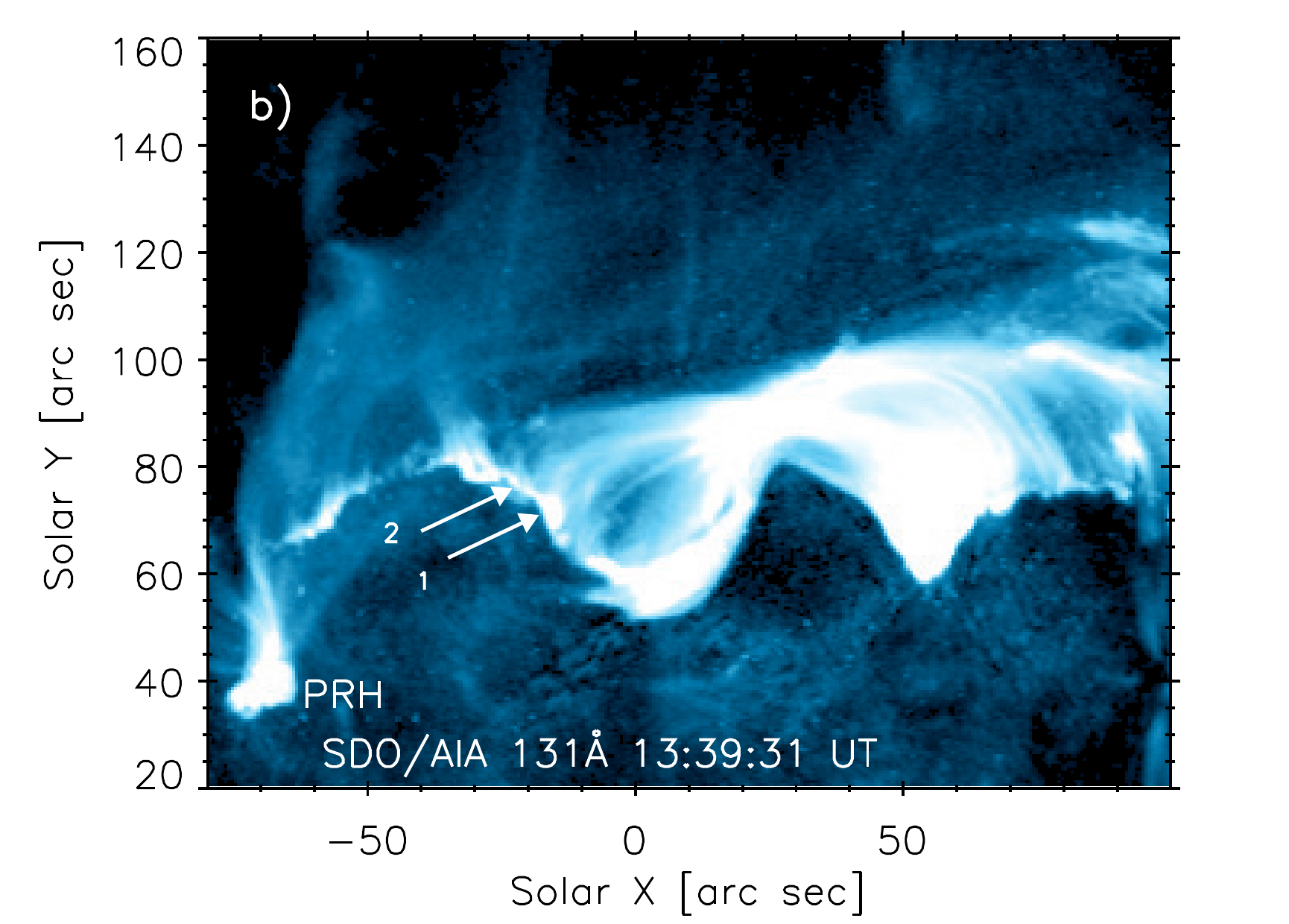}
        \includegraphics[width=8.29cm,clip,viewport=78 50 470 330]{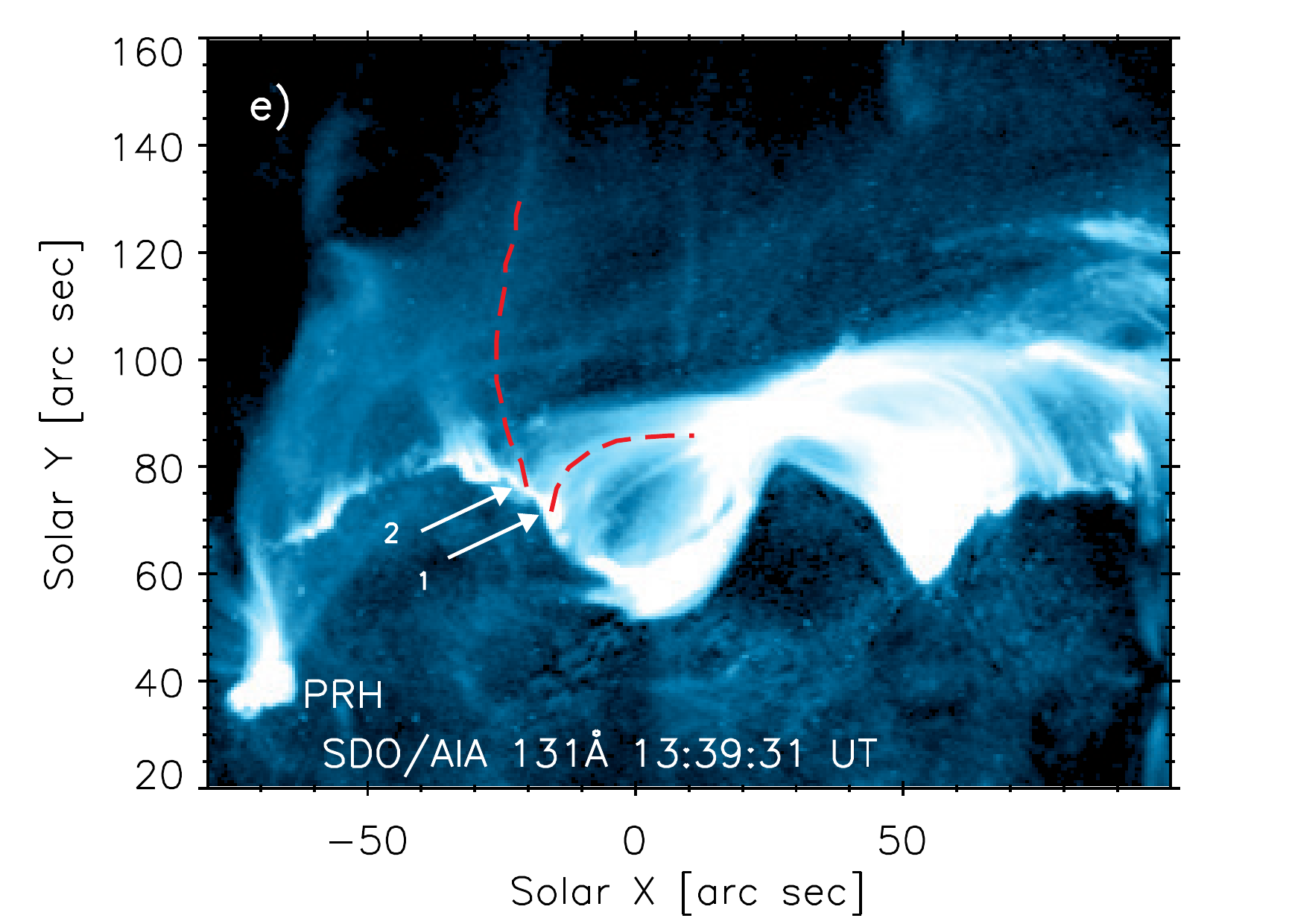}
       
       \includegraphics[width=9.625cm,clip,viewport=15 0 470 330]{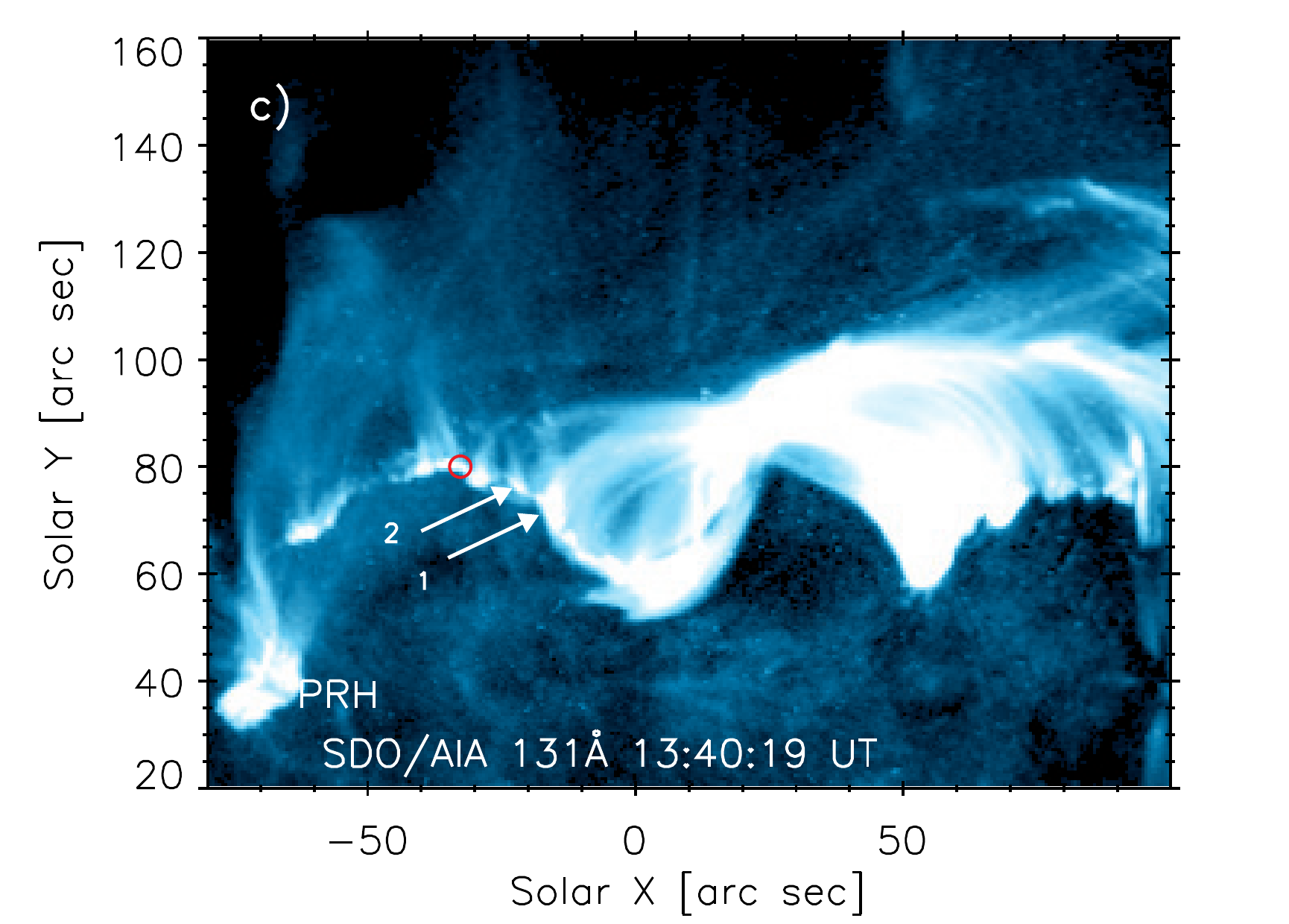}
       \includegraphics[width=8.29cm,clip,viewport=78 0 470 330]{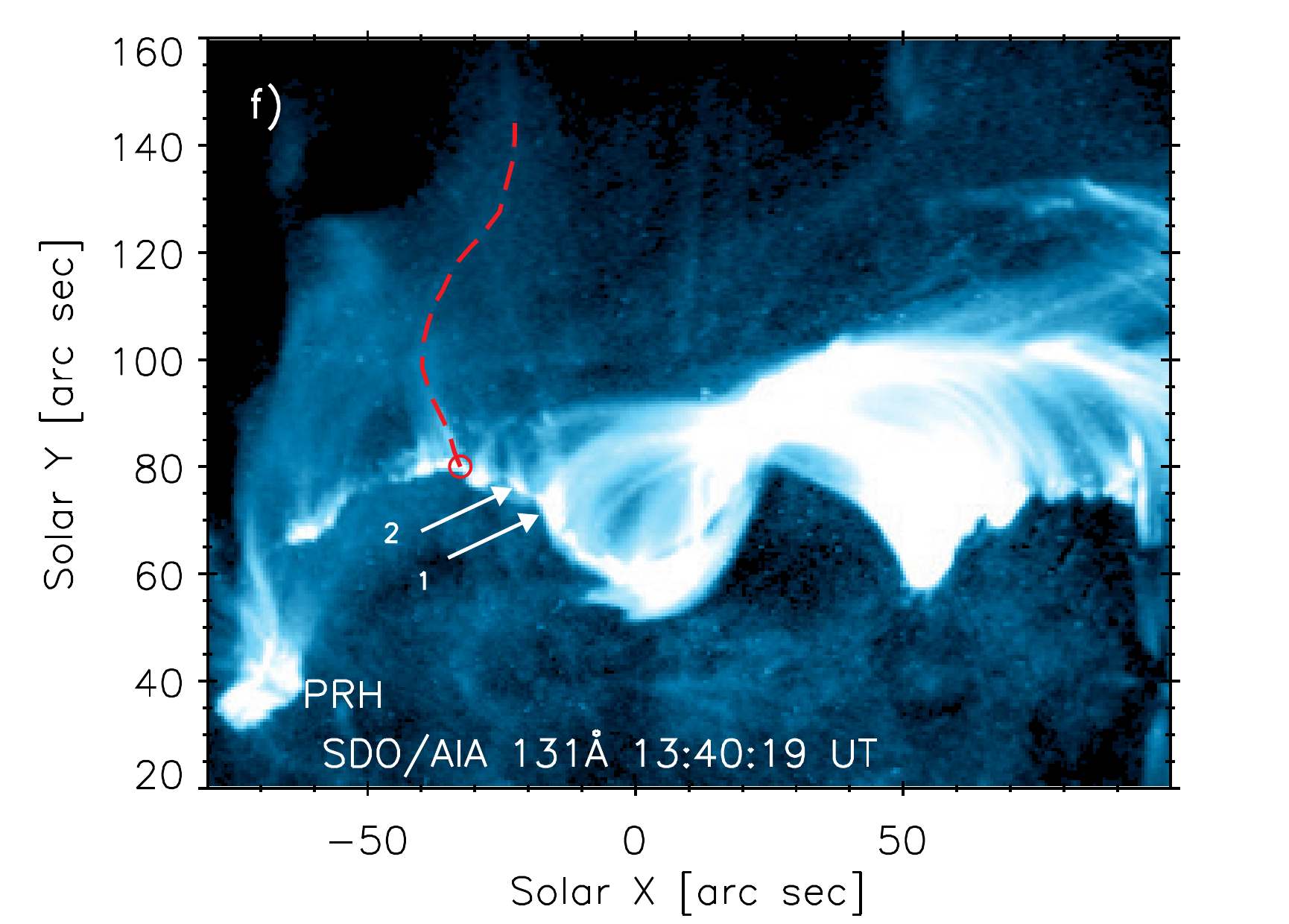}
        
\caption{Slipping reconnection of the loops during the elongation of the PR. 
White arrows show positions of loops which exchanged their connectivity. The original images in the left 
column show: (a) pre-reconnection state at 13:36:55\,UT, (b) post-reconnection state at 13:39:31\,UT
and (c) shows, in the red circle, the footpoint of the loop which slipped through the position 2 further 
away along PR.
Panels (d), (e) and (f) show the same images only the loops are highlighted by colours to better
visualise their state: pre-reconnection state in yellow and post reconnection state in red.
(A movie of this Figure is available (movie\_reco.mpg).)} \label{fig_reco}
\end{figure*}
Disappearance of SPRH and elongation of PR proceeded via slipping of hot loops seen in 131\AA. The loops, initially rooted in the curved part of SPRH (Figure~\ref{fig_sprh} and accompanying movie), slipped continually towards east of SPRH until the SPRH disappeared. At the former position of SPRH, only an elbow of newly elongated PR remained (Figures~\ref{fig_sprh}b and c). The new elongated part of PR first consisted of tiny bright 
kernels which later turned out to be footpoints of faint hot loops in 131\,\AA~(Figure~\ref{fig_sprh}b). Along the straight part of the original PR the footpoints of an arcade of flare loops were observed (Figure~\ref{fig_sprh}c, d). Finally, at the end of the new elongated part of PR a new hook PRH formed and started to quickly expand since the eruption of hot loops rooted in it (Figure~\ref{fig_sprh}d).

Slipping of the loops is presented in Figure~\ref{fig_reco} (and its accompanying movie). The first column shows the original {\it SDO}/AIA images in 131\,\AA~and the second one shows the same images, only the loops are highlighted: yellow colour and full lines show pre-reconnection state and the red colour and dashed lines show post-reconnection state. White arrows with numbers have same position at all six panels and point to the footpoints of two loops which we observed to exchange their connectivity. At 13:36:55\,UT (Figure~\ref{fig_reco}a, d), at the footpoint 1 we observed a loop which belonged to J1. As J1 was rising, the loop was stretching and its footpoint slipped along PR towards the position 2. Then, at 13:39:31\,UT (Figure~\ref{fig_reco}b, e), we observed a flare loop at the footpoint position 1 and at footpoint 2 we observed a stretched loop. Later, at the position of the red circle (Figure~\ref{fig_reco}c, f) we observed a footpoint of a loop which slipped further along PR and became a part of hot erupting loops.
The apparent slipping velocities of the loops in elongated part of PR were measured using the time-distance technique (not shown here) and were found to be 30--120 km\,s$^{-1}$, which is consistent with previous reports \citep{Dudik2014,Dudik2016,Li2015,Li2017}.
\begin{figure*}
        \centering
        \includegraphics[width=6.74cm,clip,viewport=15 50 310 350]{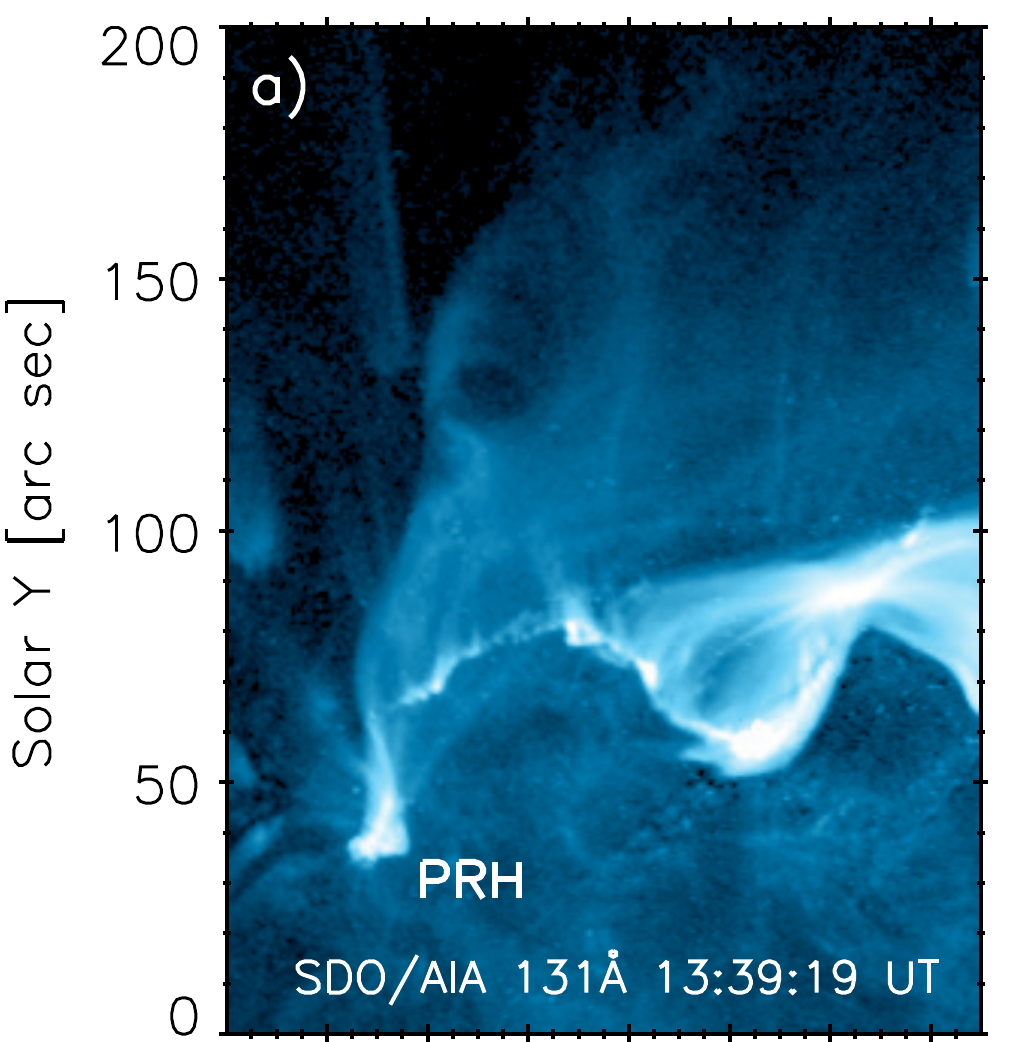}  
        \includegraphics[width=5.3cm,clip,viewport=78 50 310 350]{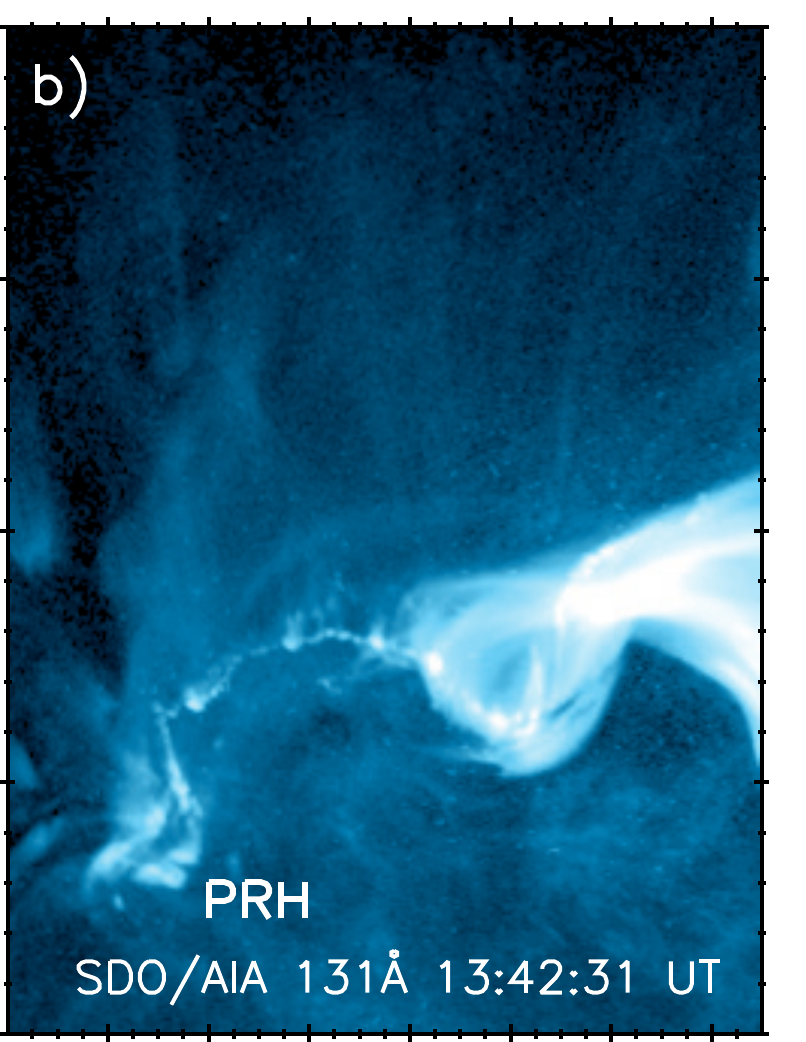}  
        \includegraphics[width=5.3cm,clip,viewport=78 50 310 350]{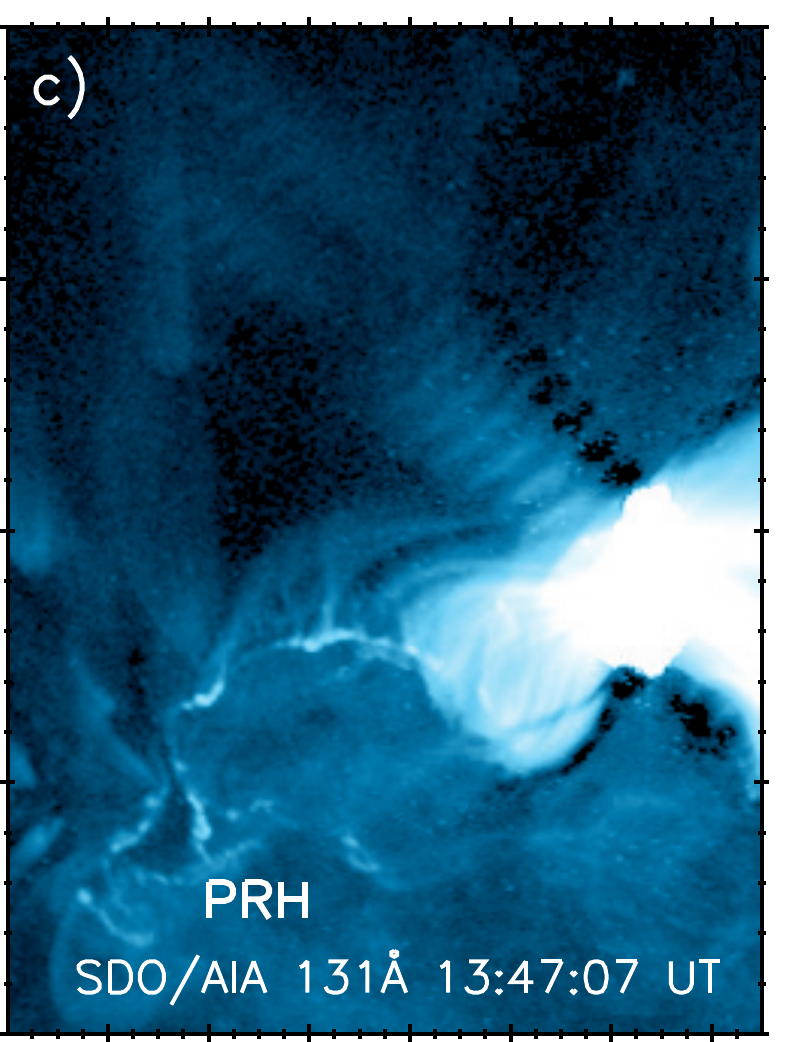}
        
        \includegraphics[width=6.74cm,clip,viewport=15 0 310 350]{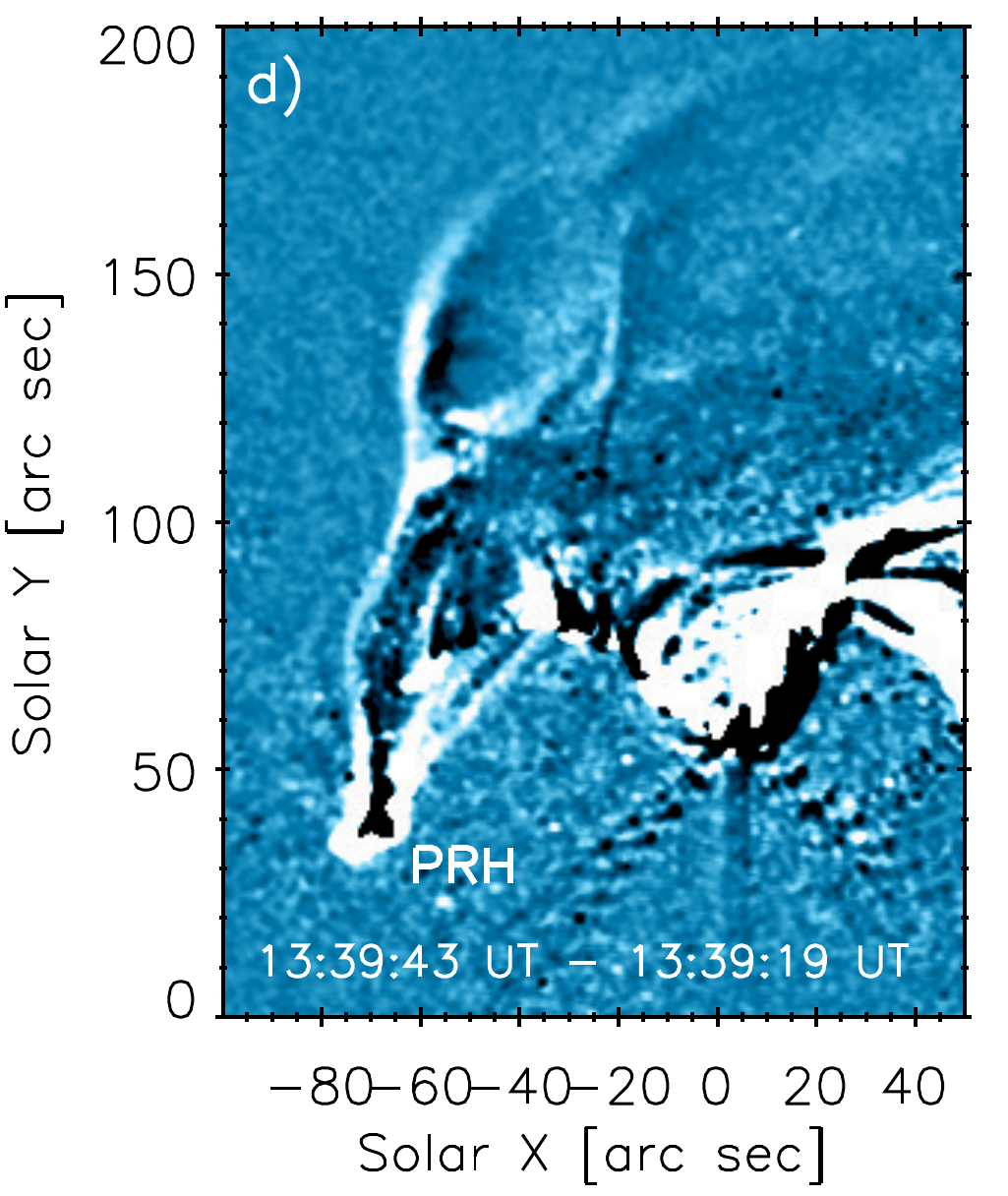}
        \includegraphics[width=5.3cm,clip,viewport=78 0 310 350]{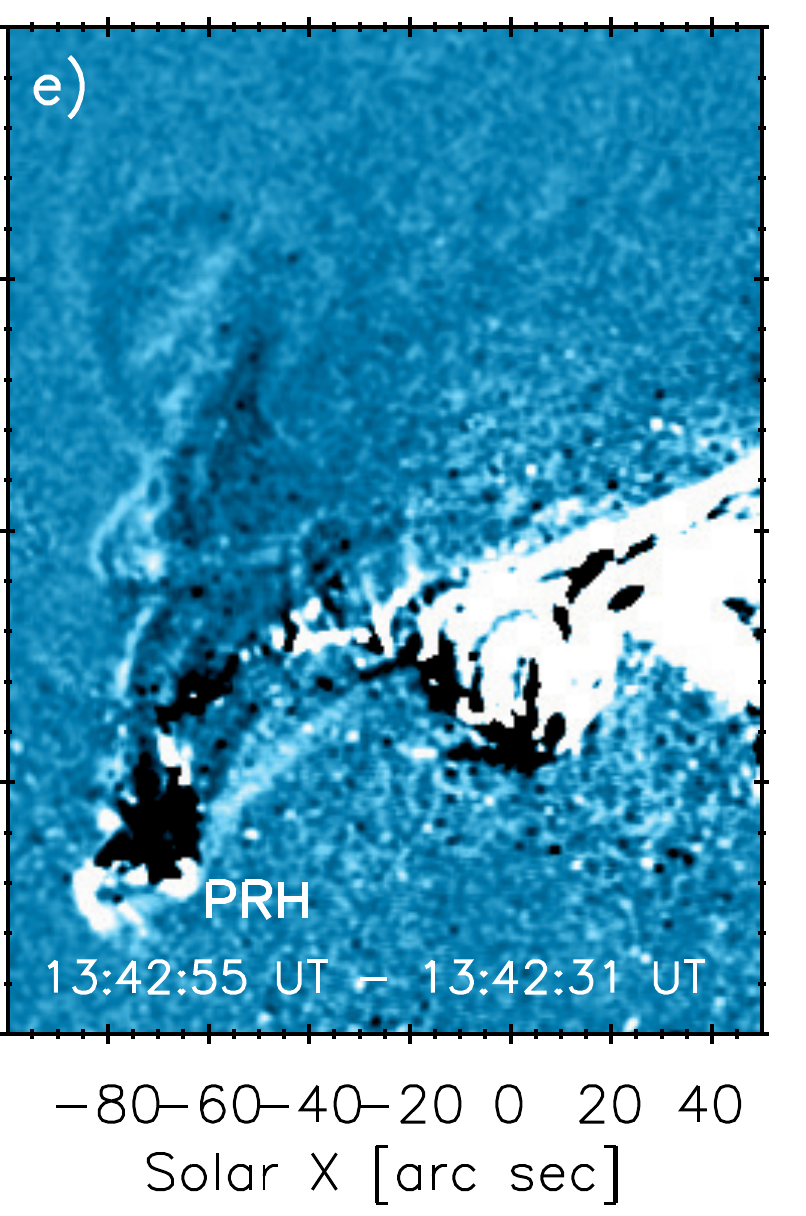}
        \includegraphics[width=5.3cm,clip,viewport=78 0 310 350]{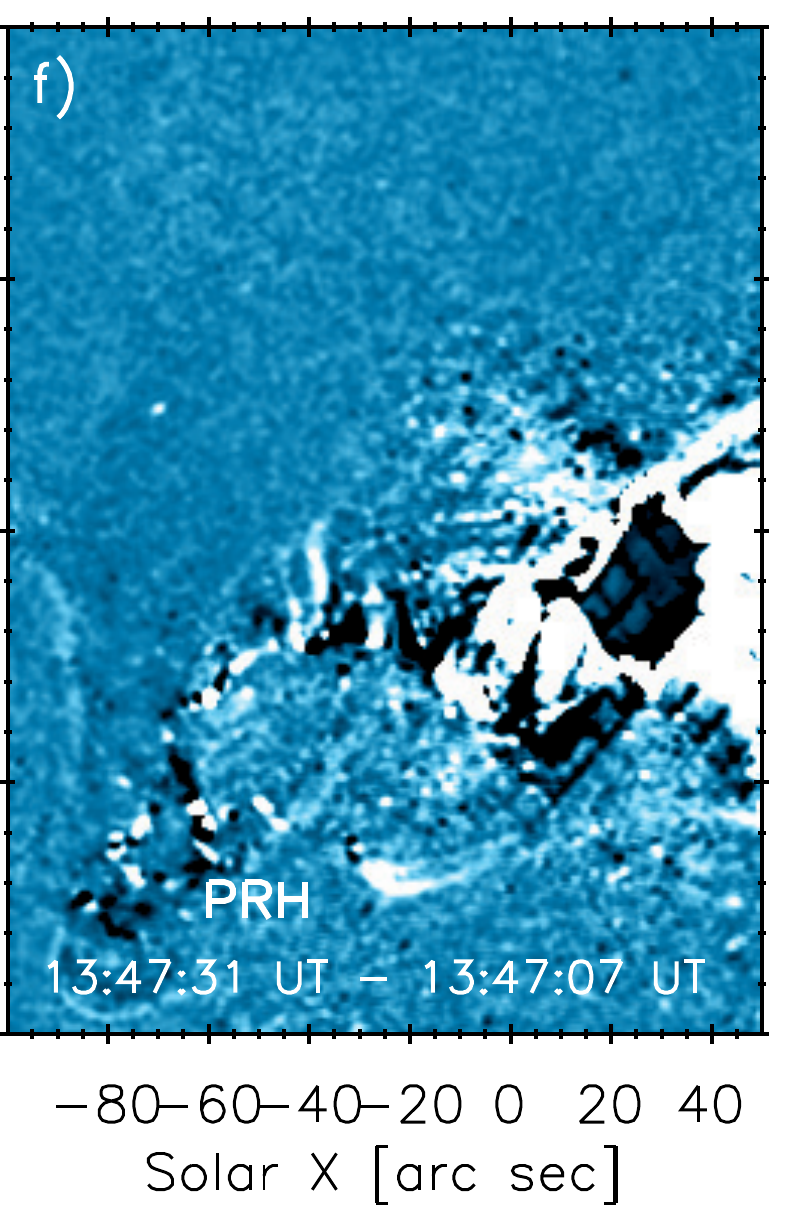}
        
\caption{Eruption of the hot loops rooted in PRH. Top row shows {\it SDO}/AIA 131\,\AA~images of: (a) hot loops 
approximately at the onset of their eruption, (b) rising hot loops and widening PRH and (c) shows dimming observed within 
the area of PRH after the escape of hot loops.
The running difference images at the bottom row document: (d) approximately onset of the eruption of hot loops rooted in PRH, 
(e) the rise of the hot loops and expansion of PRH and (f) escape of the hot loops and coronal dimming at PRH. Note, that 
dimmed area of PRH is not black in (f) because at that time the largest changes were detected especially at PRH extremity.
(A movie of this Figure is available (movie\_rope\_rd.mpg).)} \label{fig_rope}
\end{figure*}
The eruption of hot loops rooted in PRH occurred about 13:40\,UT (Figure~\ref{fig_rope} and its accompanying movie), and is in good temporal correspondence with drifting pulsating structures observed in the radio spectrum by \citet{Karlicky2018} during this flare. From about 13:39\,UT we observe the expansion of PRH (Figure~\ref{fig_rope} a--c ) and the rise and escape of the hot loops (Figure~\ref{fig_rope}b, c and e, f). 

Thus, during the impulsive phase of the flare we observed a large shift ($\approx$ 40\,$\arcsec$ as measured along solar $Y$ axis) 
between the position of original hook SPRH observed at the beginning of the flare and the position of new hook PRH with hot erupting loops rooted in it. 

\begin{figure*}
        \centering
        \includegraphics[width=3.71cm,clip,viewport=5 50 290 330]{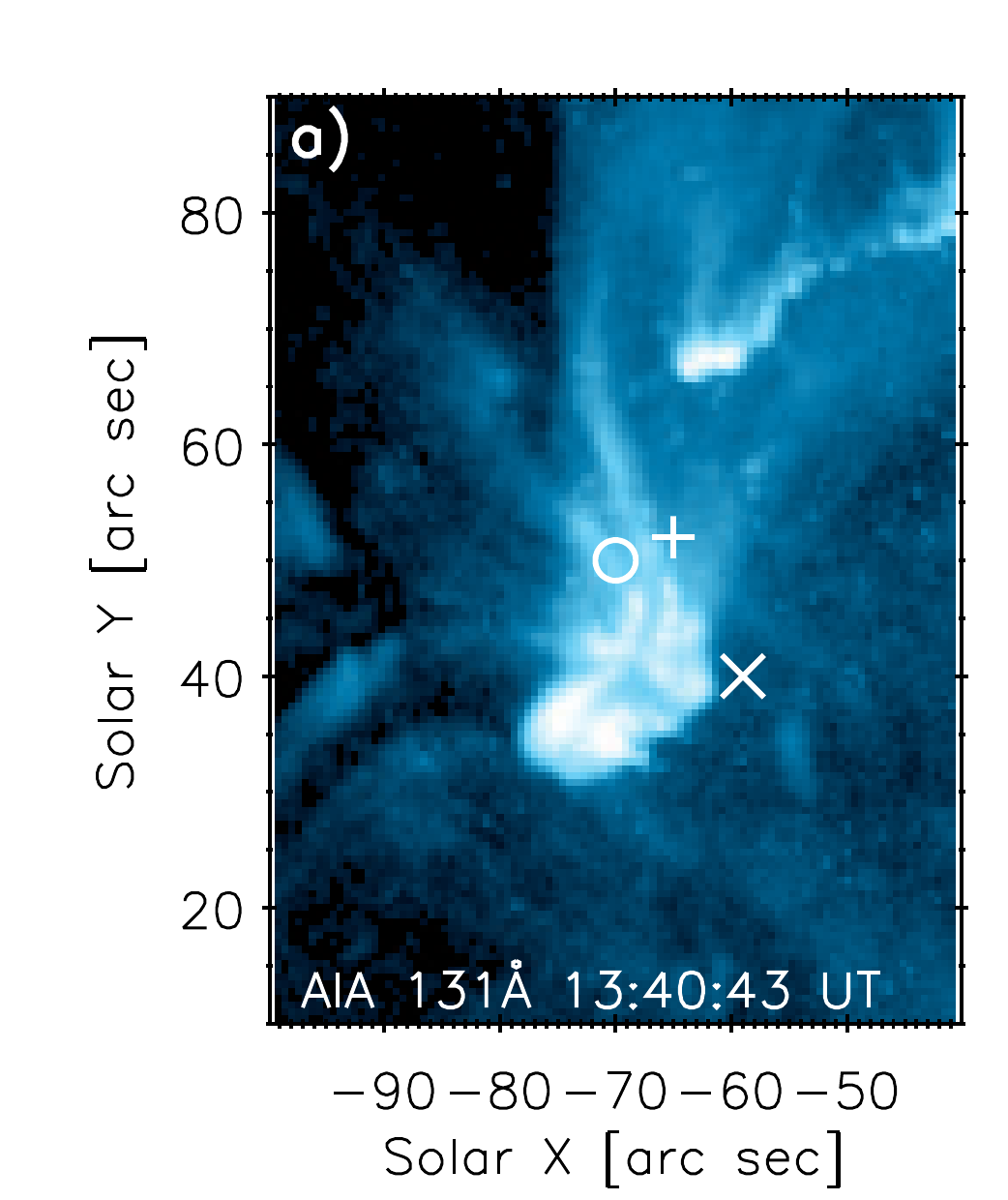}
        \includegraphics[width=2.76cm,clip,viewport=78 50 290 330]{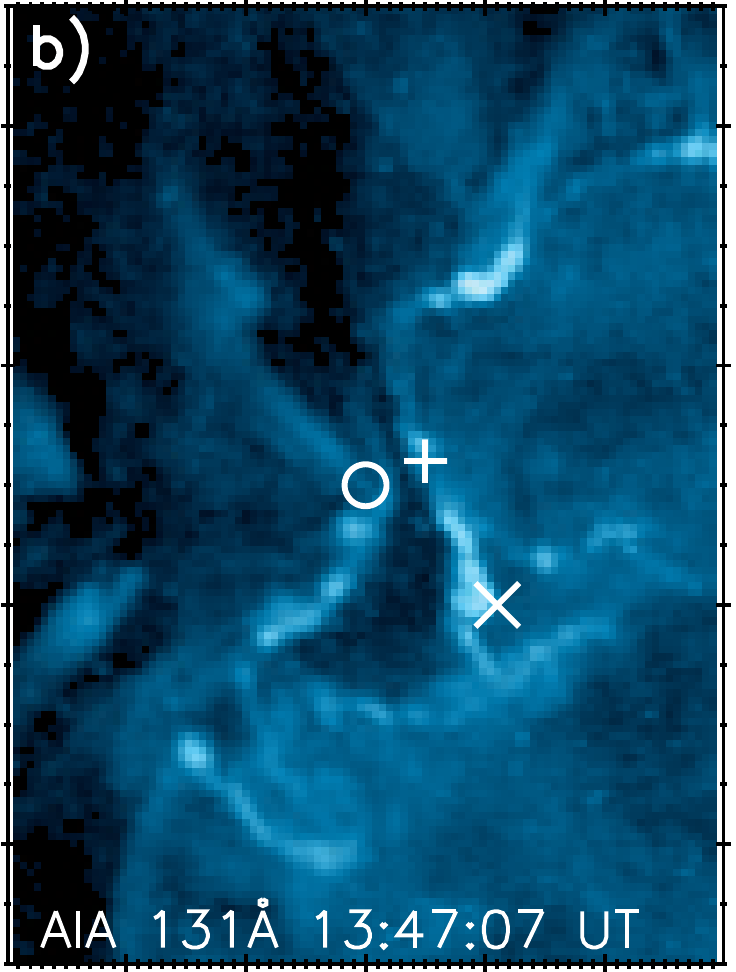}
        \includegraphics[width=2.76cm,clip,viewport=78 50 290 330]{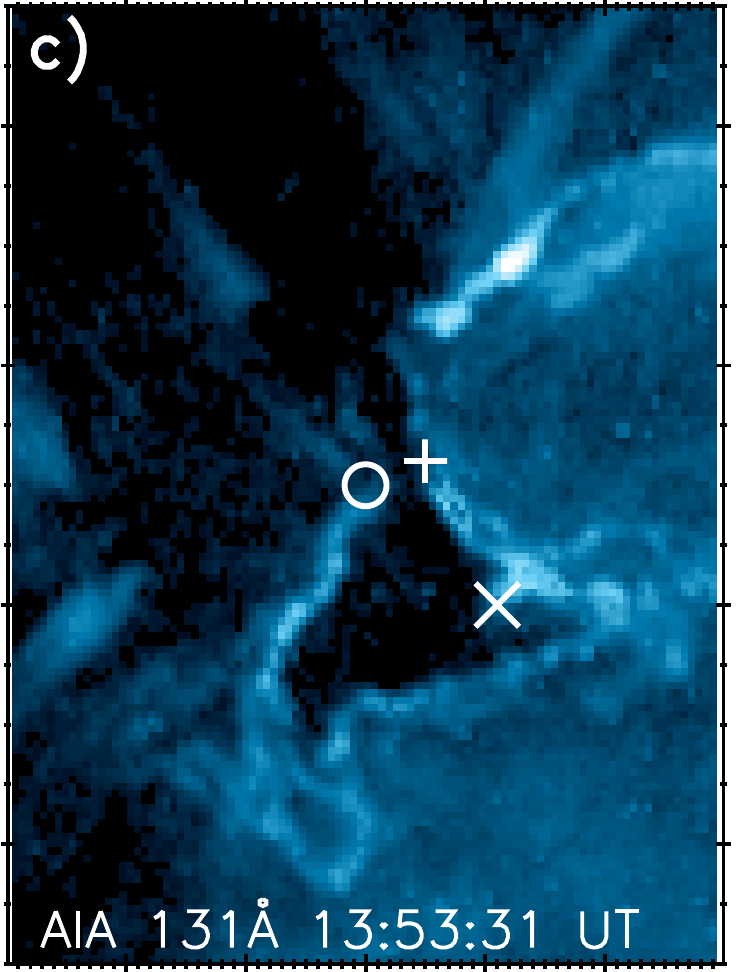}
        \includegraphics[width=2.76cm,clip,viewport=78 50 290 330]{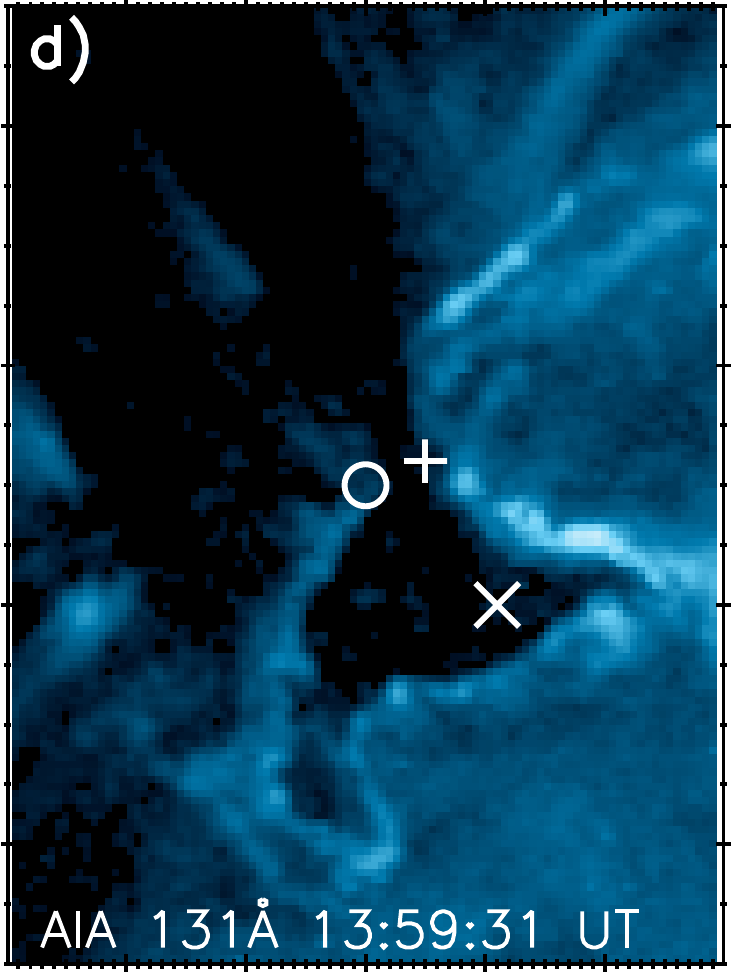}
        \includegraphics[width=2.76cm,clip,viewport=78 50 290 330]{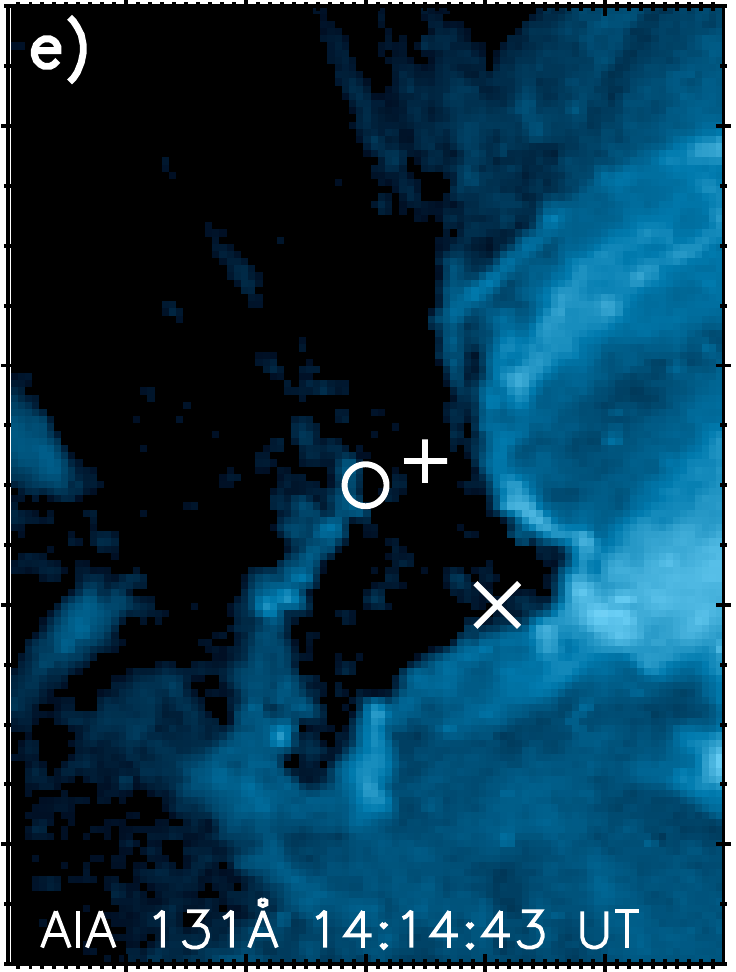}
        \includegraphics[width=2.76cm,clip,viewport=78 50 290 330]{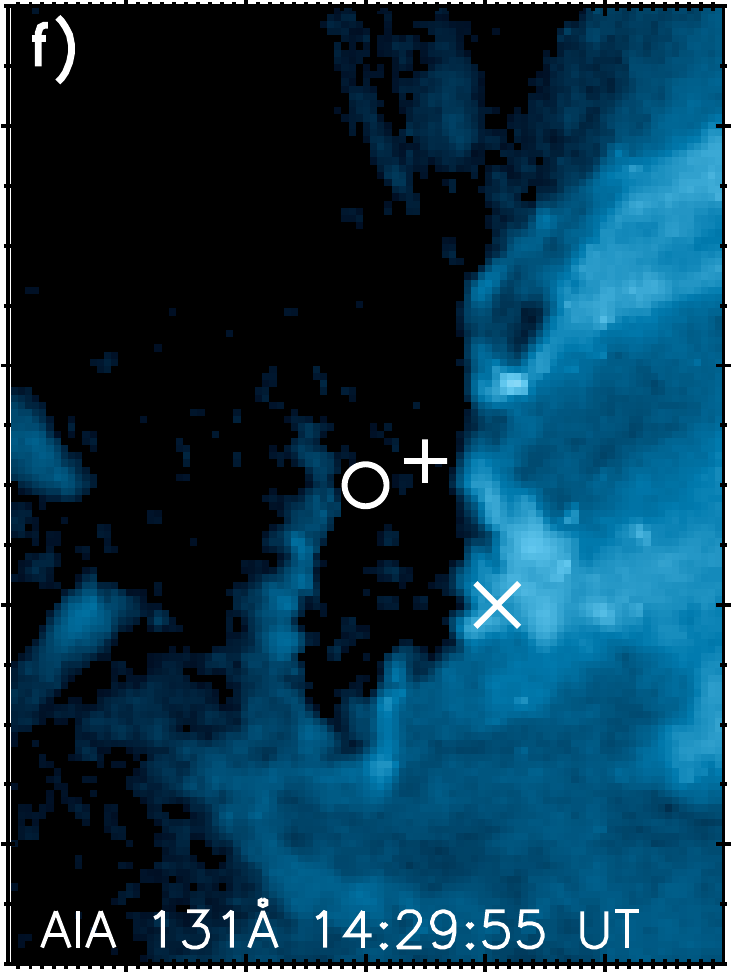}
 
	\includegraphics[width=3.71cm,clip,viewport=5 0 290 330]{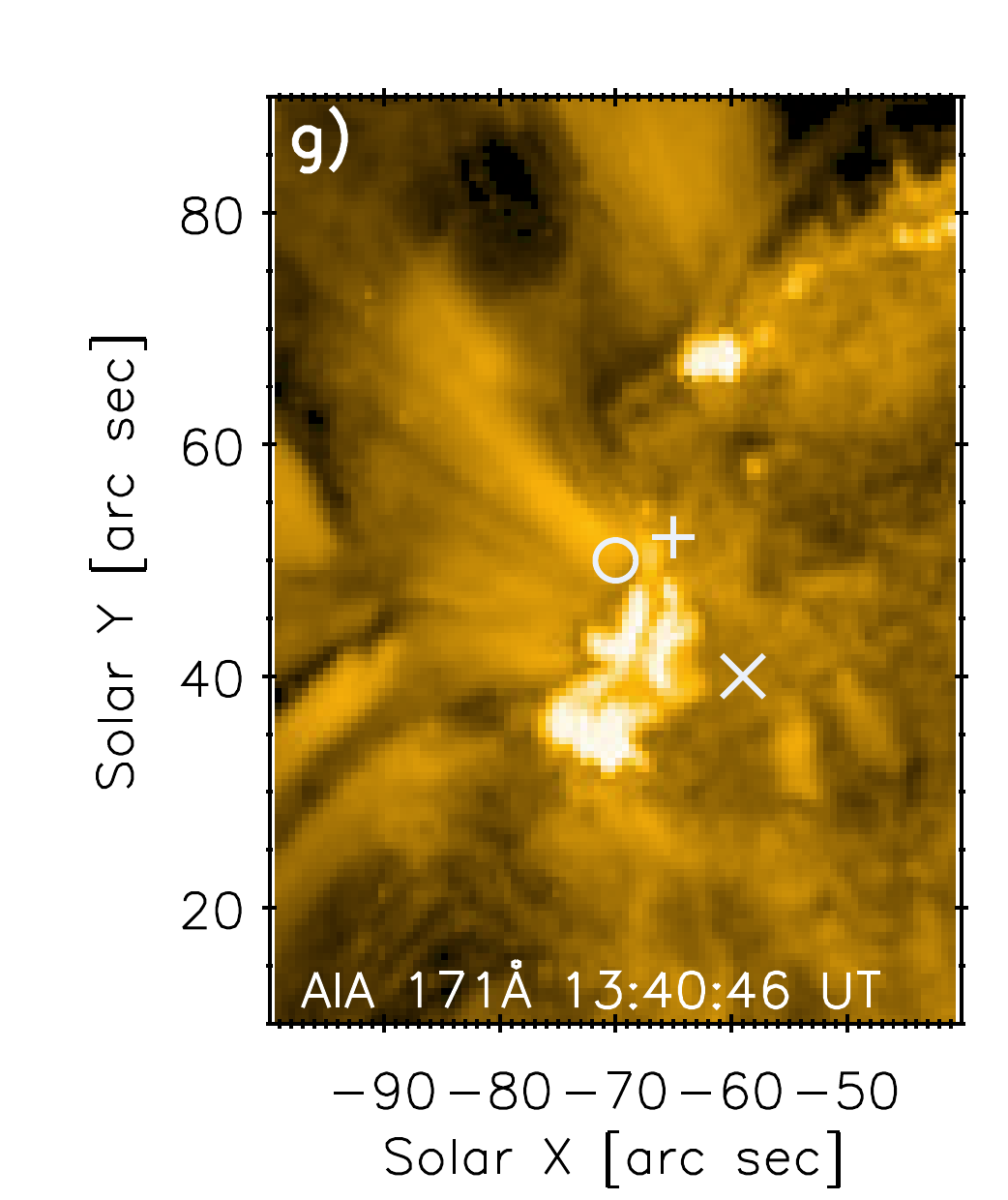}
        \includegraphics[width=2.76cm,clip,viewport=78 0 290 330]{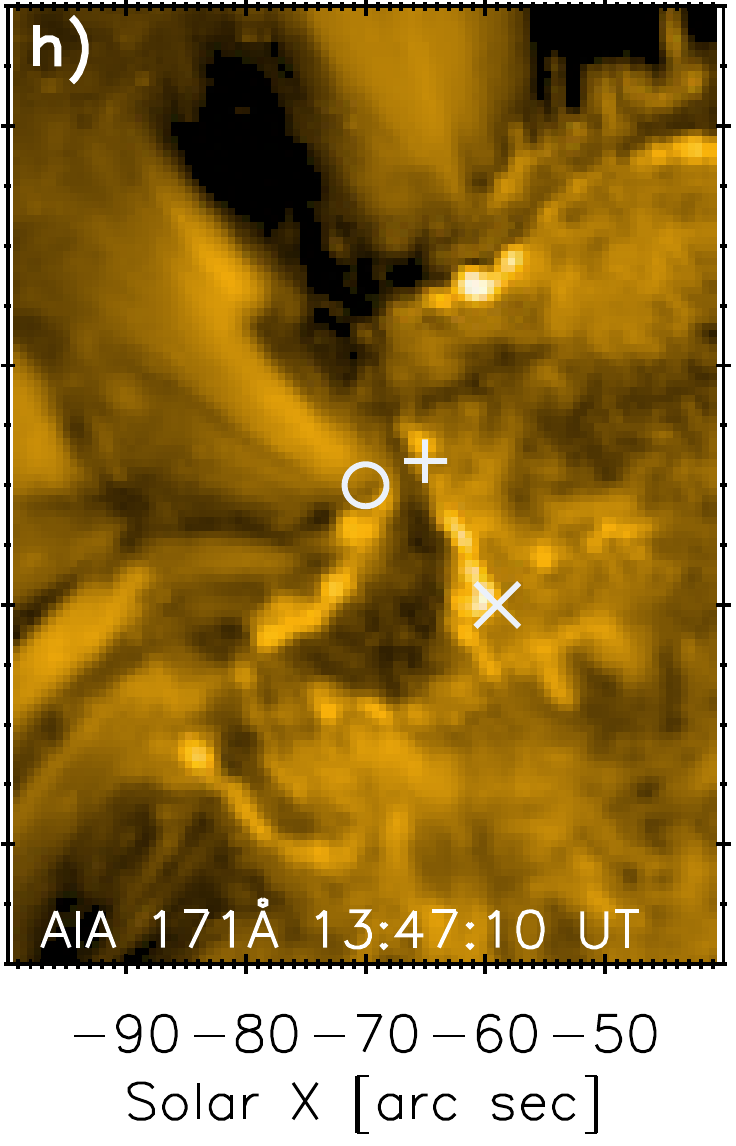}
        \includegraphics[width=2.76cm,clip,viewport=78 0 290 330]{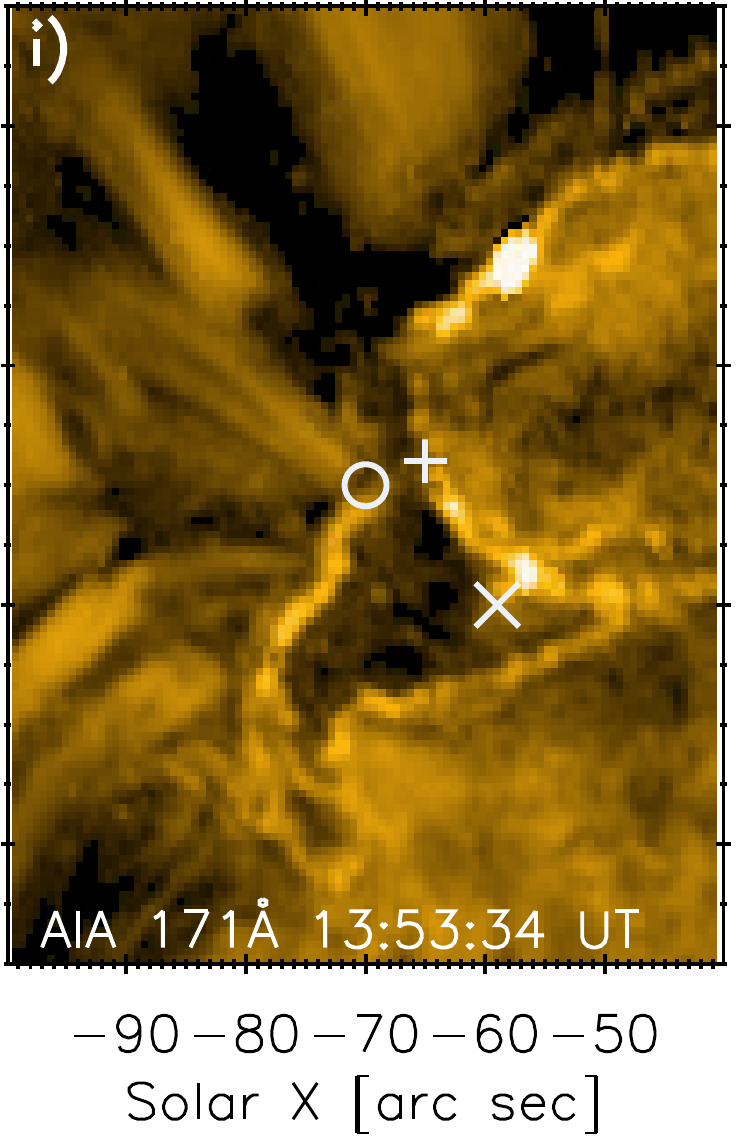}
        \includegraphics[width=2.76cm,clip,viewport=78 0 290 330]{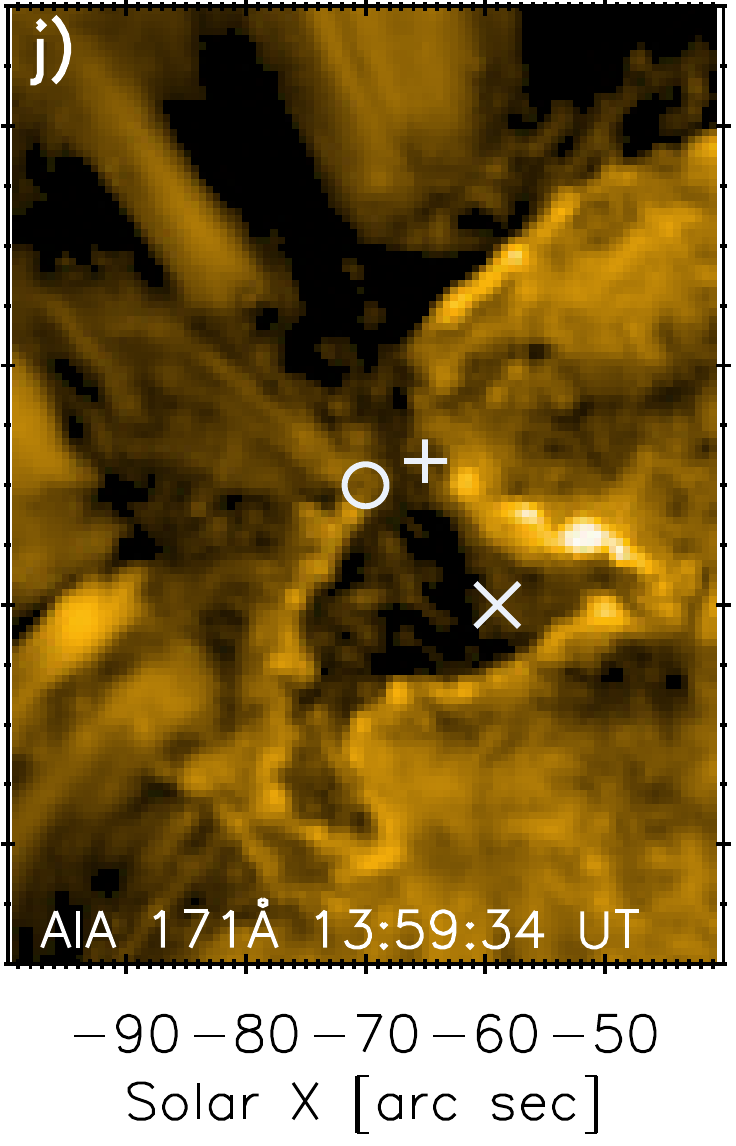}
        \includegraphics[width=2.76cm,clip,viewport=78 0 290 330]{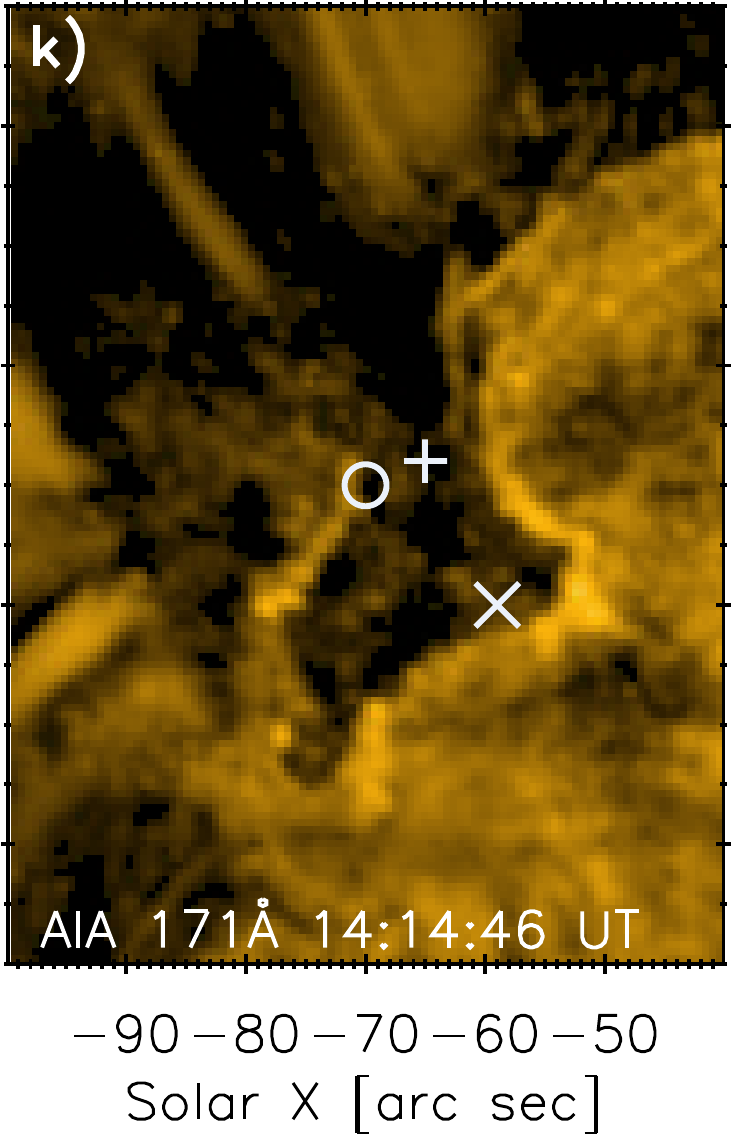}
        \includegraphics[width=2.76cm,clip,viewport=78 0 290 330]{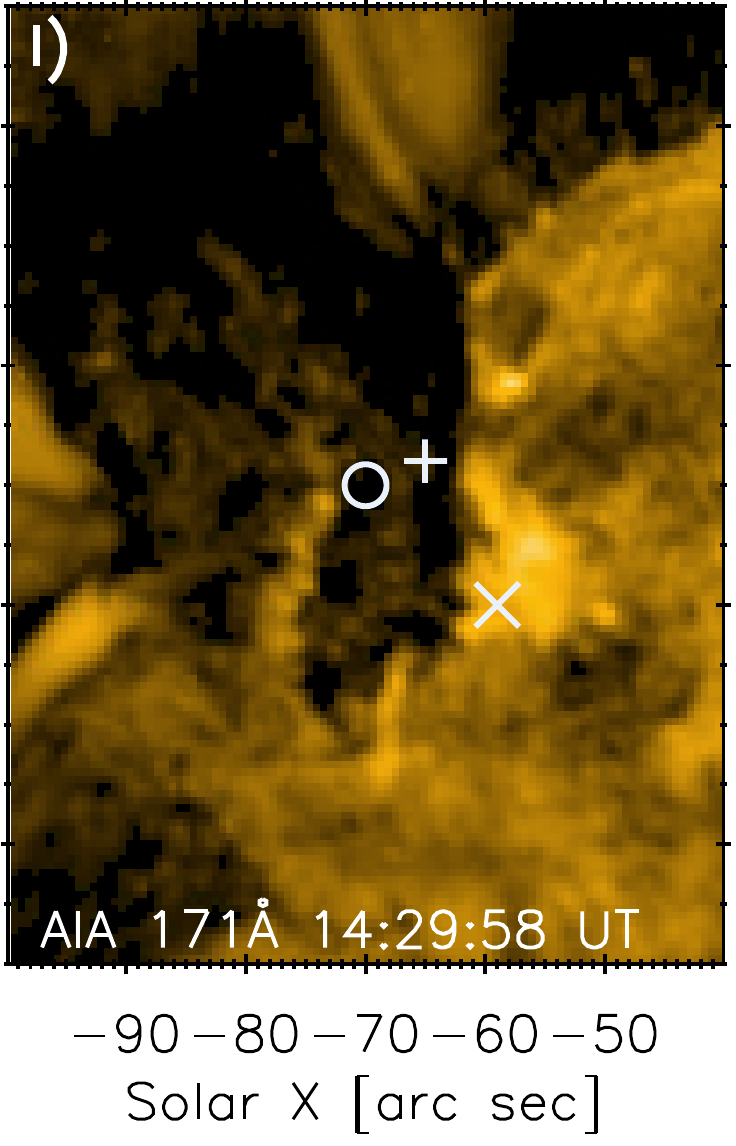}
        
\caption{Evolution of the positive ribbon hook PRH. The hook first expanded and later it contracted.
The circle, plus and $\times$ symbols mark the positions of the loops which reconnect at some stage 
of PRH evolution due to 3D magnetic reconnection. See explanation in the text. (Movies of this Figure are available (movie\_prh\_131.mpg, movie\_prh\_171.mpg).)} \label{fig_prh}
\end{figure*}

\subsection{Observation of expansion and contraction of the positive polarity ribbon hook}

Detailed evolution of the PRH can be seen in Figure~\ref{fig_prh} (and its accompanying movies) for two {\it SDO}/AIA filters: 131\,\AA~and 171\,\AA. As the hot erupting loops rose higher up into the corona, the PRH was expanding. This process continued approximately to 14:05\,UT, when expansion turned to contraction of the hook. To describe this observation, we put 3 symbols to each panel of Figure~\ref{fig_prh}: $\bigcirc$, $+$ and $\times$. Each of those symbols has its specific position which is the same throughout all panels of Figure~\ref{fig_prh}.

The symbol $\bigcirc$ is located at solar coordinates $[X,Y]=[-70\arcsec, 50\arcsec]$, where the footpoints of the loop system are visible both in 171\,\AA~and 131\,\AA~filters (Figure~\ref{fig_prh}a, g), the symbol $+$ is located at solar coordinates $[X,Y]=[-65\arcsec, 52\arcsec]$, at the footpoint of a faint loop visible in 171\,\AA~only (Figure~\ref{fig_prh}g), and the symbol $\times$ is located at solar coordinates $[X,Y]=[-59\arcsec, 40\arcsec]$, close to the outer boundary of PRH (Figure~\ref{fig_prh}g). Following the sequence of panels a--f or g--l of the Figure~\ref{fig_prh}, one can notice that these points change their positions relative to the PRH extremity. 

At 13:40\,UT the $\bigcirc$ (Figure~\ref{fig_prh}a, g) marks the footpoints of coronal loops (Figure~\ref{fig_goes}b) located to the left of PRH extremity. Expanding PRH approached the footpoints of loops (Figure~\ref{fig_prh}b, c and h, i), crossed the $\bigcirc$ position (Figure~\ref{fig_prh}d, e and j, k), and finally the $\bigcirc$ is located within the hook (Figure~\ref{fig_prh}f, l). When PRH swept the $\bigcirc$ position, the loops rooted in it disappeared (cf.~\ref{fig_prh}a, d and g, j). At 14:29\,UT the $\bigcirc$ was enclosing only the part of dimmed area inside the hook.

The symbol $+$ marks the footpoint of a faint loop at Figure~\ref{fig_prh}g. As the hook ribbon swept the $+$, the loop disappeared (Figure~\ref{fig_prh}g--j) and at 14:14\,UT the symbol $+$ resided inside the hook (Figure~\ref{fig_prh}e, f and k, j).

The $\times$ is situated to the right of PRH extremity (Figure~\ref{fig_prh}a and g). Later, the hook swept through its position as in previous examples (Figure~\ref{fig_prh}b, c and h, i). Although we did not observe any particular loop starting at the position of $\times$, the decrease of intensity was observed there (Figure~\ref{fig_prh}a--c and g--i). At 13:59\,UT the symbol $\times$ clearly resided in the dimmed area of PRH (Figure~\ref{fig_prh}d, j). But contrary to the other two symbols, the PRH extremity moved across the $\times$ again (Figure~\ref{fig_prh}d--f and j--l). Figure~\ref{fig_prh}f shows that, finally, at the position of the $\times$ we observed enhanced 131\AA~emission coming from the area where the footpoints of flare loops were located.

\section{INTERPRETATION}

We now focus on interpretation of the observed changes in positions of ribbon hooks.

\subsection{Drift of the ribbon hook}

The tether-cutting reconnection of J1 and J2 structures at the beginning of the flare led to a formation of sigmoidal loop structure S, as proposed e.g. by \cite{Moore2001, Aulanier2010}. The S was seen almost edge-on, with its eastern footpoint rooted at the small hook SPRH (Figure~\ref{fig_sprh}b) and with the other one connected to the negative polarity (Figure~\ref{fig_flevol}e, f).
The tether-cutting process led to enlarging and strengthening of the pre-eruptive flux rope observed as H$\alpha$ filament (Figure~\ref{fig_model}d, f). The footpoints of the enlarged flux rope S were thus displaced with respect to those of H$\alpha$ filament before its activation.
We interpret the SPRH as the hook of the J-shaped positive polarity ribbon PR, that exists due to the presence of  the flux rope S. The conjugate hook of negative polarity ribbon is difficult to identify due to the complex shape of the negative ribbon at this time (Figure~\ref{fig_flevol}e, f). 

\begin{figure}
        \centering
        \includegraphics[width=8.45cm,clip,viewport=25 50 320 310]{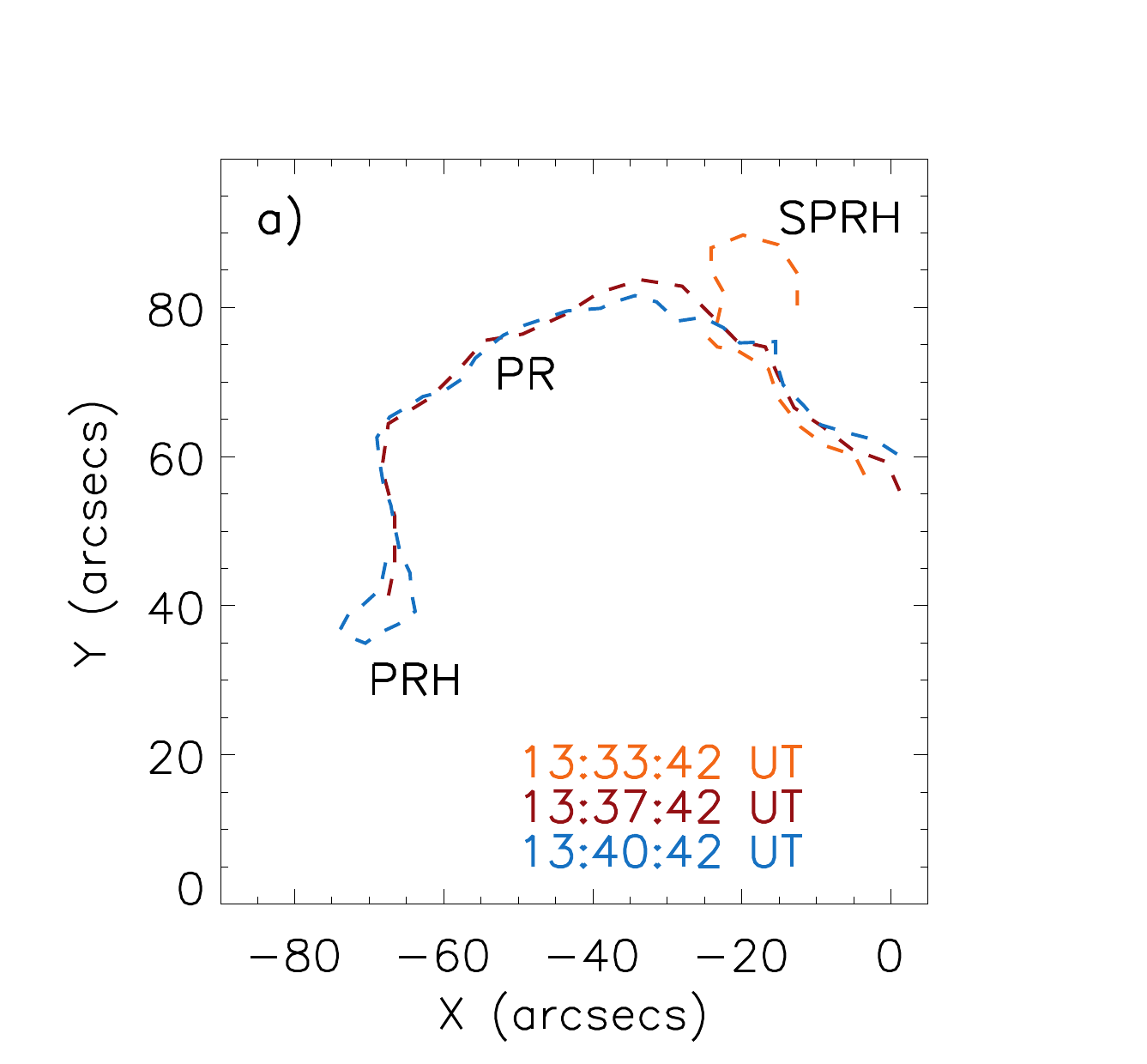}
        \includegraphics[width=8.45cm,clip,viewport=25 10 320 310]{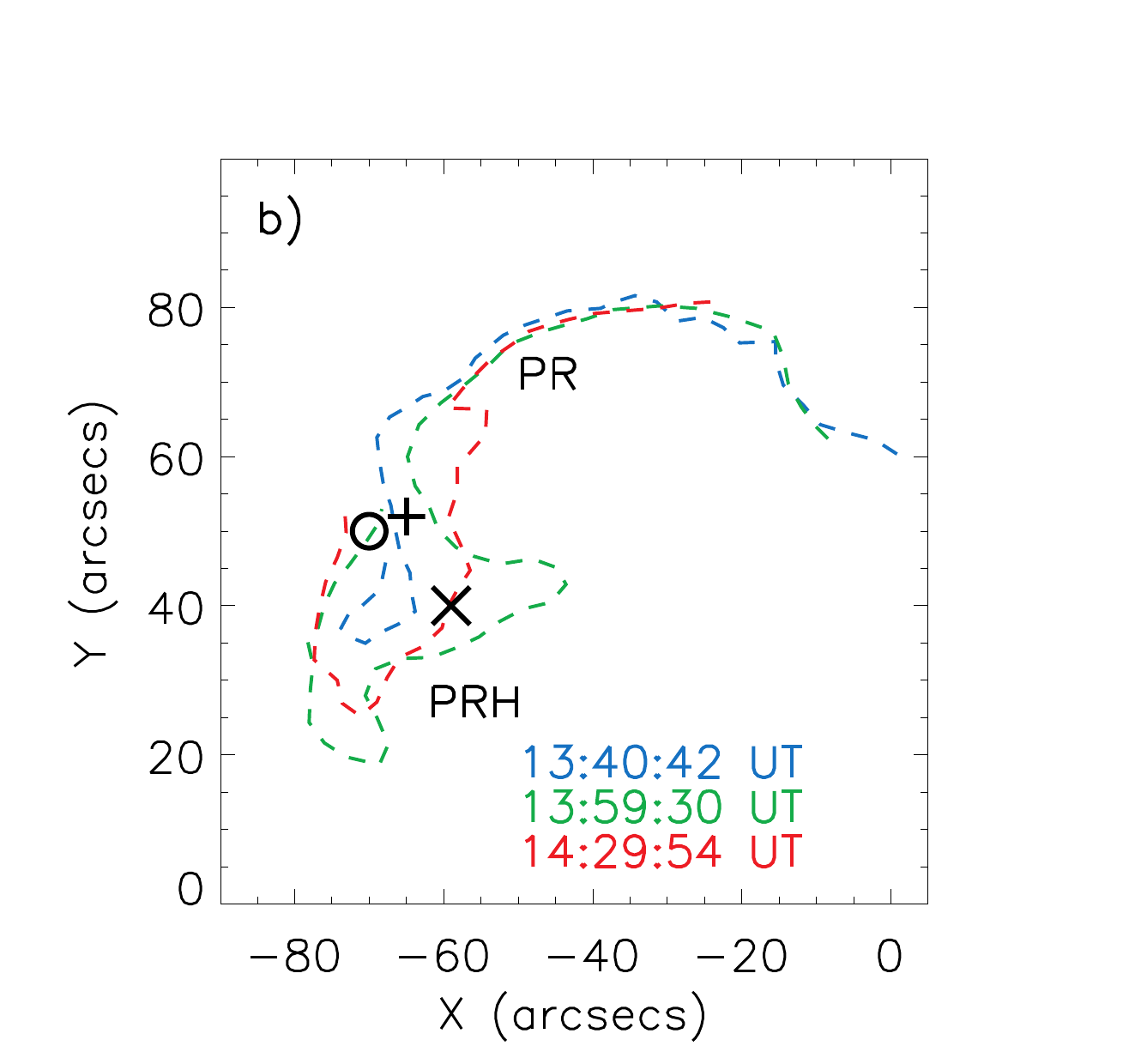}
        \caption{Manually traced-out ribbons from {\it SDO}/AIA images in 304\AA~filter at given times showing: (a) the drift of the SPRH (orange dashed line) through the elongation of the PR (brown dashed line) to the final PRH (blue dashed line) of the erupting flux rope and (b) the expansion and contraction of PRH. The three symbols, $\bigcirc$, $+$ and $\times$ on panel b), show the same positions as in Figure~\ref{fig_prh}.} \label{fig_drift_expcont}
\end{figure}

Figure~\ref{fig_drift_expcont}a shows the schematic time evolution of PR, SPRH and PRH. The dashed lines represent manually traced ribbon and its hook as observed in 304\AA~filter of {\it SDO}/AIA data for three different times. The SPRH was observed only for few minutes and then it disappeared. The loops rooted in it, i.e. along the hook part of orange dashed line in Figure~\ref{fig_drift_expcont}a, continually slipped to the east and formed a new elongated part of the positive ribbon PR, depicted by brown dashed line in Figure~\ref{fig_drift_expcont}a. In Figure~\ref{fig_reco} we showed that during this apparent slipping motion the loops exchanged their connectivity with its neighbours as was described by \cite{Aulanier2006,Aulanier2007, Janvier2013_III}. 

Elongation of PR took several minutes. Then, a new and larger hook PRH appeared at its end, showed by blue dashed line in Figure~\ref{fig_drift_expcont}a.
We interpret this evolution as the shift in the position of the flux rope footpoint, from the filament F (Figure~\ref{fig_model}f) to the flux rope S (Figure~\ref{fig_flevol}c--f, Figure~\ref{fig_sprh}a--c) and finally to the hot erupting flux rope with its eastern/positive footpoint rooted in PRH (Figure \ref{fig_rope}e) and western/negative foopoint rooted in NRH (Figure~\ref{fig_flevol}g). The  positive footpoint of the flux rope thus exhibited significant drift (Figure~\ref{fig_drift_expcont}a) while becoming the hook PRH of the erupting flux rope. The observed shift between the position of two hooks, SPRH and later PRH, along the PR can be estimated from Figure~\ref{fig_drift_expcont}a to about 80$\arcsec$.

The changes in positions of the flux rope footpoints were recently predicted by the MHD model of \cite{Aulanier2019} who explained it as a consequence of ar--rf reconnections. During these reconnections with the surrounding coronal arcades, the flux rope is eroded on its inner side (facing the PIL), while it is built up on the outer side (Figure~\ref{fig_cartoon}c--d). This leads to a gradual drift of a flux rope footpoints. We suggest that the disappearance of the SPRH and formation of new PRH could happen due to this process, although the drift of the flux rope footpoints in present observations is larger than in the model of \citet{Aulanier2019}. This interpretation of the drifting footpoints of the erupting flux rope also allows us to reconcile the footpoints of the pre-eruptive flux rope (filament and the corresponding NLFFF model), with the distant footpoints of the erupting loops rooted in PRH seen by AIA. 

\subsection{Expansion and contraction of the ribbon hook}

Since the eruption of the hot flux rope, observed during the impulsive phase of the flare, the ribbon hook PRH first widened and then it shrank. Schematic evolution of PRH is shown in Figure~\ref{fig_drift_expcont}b. Coloured dashed lines represent part of the PR together with PRH for three different times. Position of the ribbon and its hook has been traced again manually from {\it SDO}/AIA data observed in 304\AA~filter.

To illustrate this evolution we selected three positions around the curved PRH extremity. 
These positions are marked in Figure~\ref{fig_drift_expcont}b by the same symbols as shown in Figure~\ref{fig_prh}.
The first one was marked by the symbol $\bigcirc$. The footpoints within the $\bigcirc$ belonged to coronal arcades as can be seen in AIA 171\,\AA~(Figure~\ref{fig_model}c, $[X,Y]=[-70\,\arcsec, 50\,\arcsec]$). The expanding hook swept through the $\bigcirc$ position (green dashed line in Figure~\ref{fig_drift_expcont}b) and the 
coronal loop turned to the flux rope field line inside the dark area encircled by the ribbon hook (red dashed line in Figure~\ref{fig_drift_expcont}b). The decrease of EUV emission at the location of $\bigcirc$ can be explained by coronal dimming due to evacuation of plasma along the field line of erupted flux rope \citep[][]{Vanninathan2018}. The evolution of the location corresponding to the second symbol, the $+$, was similar. At first we observed a faint loop rooted in this position (the $+$ position relative to blue dashed line in Figure~\ref{fig_drift_expcont}b). When it was swept by PRH, the faint loop disappeared, and the location was within the dimmed area of PRH (Figure~\ref{fig_drift_expcont}b). We interpret this again as an arcade field line reconnecting to become a flux rope field line. Within the 3D model of \cite{Aulanier2019} such reconnection change can occur in two reconnection geometries: {\it a}a--{\it r}f and  {\it a}r--{\it r}f (Figure~\ref{fig_cartoon}), where the letter `a' denotes an arcade type field line, `r' a flux rope field line, and `f' a flare loop. The use of italics highlights the changes in the field line topology of the particular reconnection process considered, and corresponding to our observations. 

The aa--rf reconnection geometry is present in the 2D standard CSHKP model. However, the ar--rf reconnection geometry is purely 3D. Unfortunately, in both observed cases, at locations $\bigcirc$ and $+$, we observed only one loop of a pair of loops undergoing the reconnection process changing
the coronal loop \textit{a} to the flux rope \textit{r}. Therefore, neither aa--rf nor ar--rf reconnection geometry can be ruled out. Still, we may infer that aa--rf is more likely. This preference is supported by the fact that the $\bigcirc$ and $+$ symbols were located close to the tip of the ribbon hook and marked outer and inner arcade type field lines, respectively. In the model, field lines starting from similar locations with respect to the hook undergo aa--rf reconnection (see Figure 4 in \cite{Aulanier2019}).

The situation is different for the last position marked by $\times$ . This location was observed to change its position relative to the hook extremity more than once. Initially it was located out of the ribbon hook, then it moved into the hook, and later it moved out of the hook again (Figure~\ref{fig_drift_expcont}b). We did not observe any loop starting at the location of $\times$, we observed only this back-and forth drift which evidenced that hook was expanding and then contracting. 

Such behaviour of QSL hooks is addressed by \cite{Aulanier2019}. They show that this `non-monotonic drift' (expansion and contraction) of the QSL hooks during the eruption `implies that individual field lines sequentially move in-and-out of the flux rope'. Such drift of the flux rope footpoints then comes from sequential reconnections involving purely 3D ar--rf reconnection geometry.  

Therefore, let us imagine a field line starting at $\times$, and combine this assumption with the observed evolution at the position of $\times$. Since at the beginning this imaginary field line is outside of the hook and then inside it, it likely underwent reconnection from an arcade type field line into a flux rope field line. Finally, it turned into a flare loop when coming out of the hook again. Thus schematically, it would underwent {\it a--r--f} reconnection series. This sequence could be achieved by several combinations of 3D reonnection geometries: a) {\it a}a--{\it r}f followed by a{\it r}--r{\it f}, b) {\it a}a--{\it r}f followed by r{\it r}--r{\it f}, and c) {\it a}r--{\it r}f followed by a{\it r}--r{\it f}; where the code letters in italics again highlight the reconnection schemes. 

Considering the position of $\times$ symbol relative to the hook during the observations and with the help of the
assumed reconnection scheme we now attempt to make a comparison with the model. 

The reconnection scenario of a) was not identified in the model, so it is questionable if it can even exist. 

The reconnection series in case b) are shown in the Figure 4 of the model by \cite{Aulanier2019}: the arcade field lines which undergo 
{\it a}a--{\it r}f reconnection are located close to the tip of the QSL hook. The flux rope field line produced by this reconnection is also located close to the tip of the QSL hook and a flare loop appears at straight part of the J-shaped ribbons. The flux rope field lines rooted close to the tip of the hook can further undergo the r{\it r}--r{\it f} reconnection, producing a new flux rope field line with larger twist and with its footpoint rooted again at the tip of the QSL hook. The position of the symbol $\times$ relative to PRH in our observations is however away from the tip of the PRH. Therefore, we argue that the sequence of aa--rf reconnection followed by rr--rf reconnection is not consistent with observations.

Finally, the c) is the most likely scenario. The initial position of $\times$ symbol resembles the model situation in Figure 6 of \cite{Aulanier2019}. Its position relative to PRH extremity is similar to the position of the orange inclined arcade. This inclined arcade has its footpoint near the curved part of the QSL hook. It reconnects into a flux rope field line in the ar--rf reconnection geometry.  
The $\times$ symbol was located outside the curved part of the PRH, at the side of active region facing the PIL. 
When the ribbon hook swept the $\times$ for the first time, it appeared inside the PRH and we observed noticeable drop of intensity there. This is consistent with the orange arcade reconnecting into a flux rope field line in the model of \citet{Aulanier2019} via the ar--rf reconnection geometry. 
Finally, the fact that the location of the $\times$ symbol moved outside of the PRH can be considered as an evidence for the second ar--rf reconnection process of series in case c). At 14:30\,UT the $\times$ appears in area of enhanced emission both in 131 and 171\,\AA, with fainter 131\,\AA~loops originating from this position (Figure \ref{fig_prh}, panels f and l; see also animation accompanying Figure \ref{fig_flevol}). Thus the evolution observed at the location $\times$ likely provides us with the first observational candidate for series of two ar--rf reconnections.

We note that during this evolution of PRH, a pair of secondary ribbons appeared close to the PRH extremity (Figure~\ref{fig_rope}c). A small part of them can be noticed in Figures~\ref{fig_prh}b and c, in vicinity of $\times$. Evolution of these secondary ribbons was studied in detail by \cite{Li2017}. Since the location of $\times$ was swept by PRH and not by these secondary ribbons, the existence of secondary ribbons likely had no influence on the behaviour of reconnection at $\times$. Furthermore, \citet{Li2017} reported that the secondary ribbons involve neighbouring magnetic domains; i.e., reconnection there happens away from our studied location.

\section{Conclusions}

We report on imaging observations of a solar flare and eruption of 2015 November 4. The flare and eruption involved formation, disappearance of a ribbon hook, its re-appearance and deformation in a different location, connected to evolution of the hot flux rope and slipping motion of hot loops. This evolution is interpreted as the drift of the flux rope footpoints recently predicted by the 3D extensions to the standard solar flare model by \cite{Aulanier2019}.

Before the flare we observed the H$\alpha$ filament which was satisfactorily reproduced by a NLFFF model. The filament got activated due to the magnetic flux cancellation that occurred close to its central part. The following tether-cutting reconnection involving its activated parts produced a hot sigmoidal loop. We interpreted this sigmoidal loop structure, rooted in the small hook of the positive polarity ribbon, as part of the flux rope enlarged and strengthened by the tether-cutting reconnection. Later, the hot loops of this sigmoidal structure slip-reconnected to the south-east, while the small hook disappeared. The ribbon elongated due to this slipping reconnection, leaving only an elbow at the location of the former small hook. A new and larger hook formed at the end of elongated ribbon, at a distance of more than 40$\arcsec$ along Solar Y axis and the shift along the PR itself was estimated to 80$\arcsec$.
This new hook belonged already to the hot flux rope which erupted immediately. The observed large shift between the footpoints of pre-eruptive and erupting flux ropes is interpreted here as an observational evidence of purely 3D reconnections identified by \cite{Aulanier2019}. 

Following the evolution of the large hook after the eruption of the flux rope and the associated dimmings during the gradual phase of the flare, we showed that hook exhibited deformation of its extremity. This was observed as an expansion and contraction and can be interpreted again as a consequence of 3D reconnection series which ought to involve the ar--rf reconnection geometry \citep{Aulanier2019} as only this one can shift the flux rope footpoints back-and-forth. Thus according to our observation, the magnetic reconnection can proceed to a very late phase of the flare and it could contribute to diminishing the area of the observed dimmed regions. Drifting of a flux rope footpoints can also influence the estimation of mass of the material lifted up during the eruption. Finally, we showed that the hot erupting flux rope was longer than the pre-eruptive one and much longer than the H$\alpha$ filament observed before the flare and modelled by the NLFFF extrapolation.

Our results show that the footpoints of erupting flux rope can drift large distances during solar eruptions. This drift is important for our understanding of the evolution of coronal dimmings and mapping of the ICMEs back to their footpoints, as well as the origins of flares and eruptions in the lower solar atmosphere.\\

\acknowledgements
A.Z. and J.D. acknowledge the support from the Grant 17-16447S of the Grant Agency of the Czech Republic and RVO:67985815 support form their domestic institution. 
G.A. thanks the Programme National Soleil Terre of the CNRS/INSU for financial support, as well as the Astronomical Institute of the Czech Academy
of Sciences in Ond\v{r}ejov for financial support and warm welcome during his visits.
J.K.T. acknowledges support from the Austrian Science Fund (FWF): P27292-N20.
H$\alpha$ data were provided by the Kanzelh\"{o}he Observatory, University of Graz, Austria. 
P.G. acknowledges support from the project VEGA 2/0004/16.
AIA and HMI data are provided courtesy of NASA/SDO and the AIA and HMI science teams. Hinode is a Japanese mission developed and launched by ISAS/JAXA, 
with NAOJ as domestic partner and NASA and STFC (UK) as international partners. It is operated by these agencies in co-operation 
with ESA and NSC (Norway).

\bibliographystyle{aasjournal}
\bibliography{references}

\end{document}